\documentclass[12pt]{article}

\usepackage[table]{xcolor}	% table row shading
\usepackage{tocloft}
\usepackage{natbib}
\usepackage{relsize}
\usepackage{epsfig}
\usepackage{graphics}
\usepackage{color}
\usepackage{amsmath}
\usepackage[mathlines]{lineno}
\usepackage{longtable}
\usepackage{rotating}
\usepackage{multibib}

\usepackage{epstopdf}
\usepackage{amssymb}
 \usepackage{color}	    % colored text
\usepackage{rotating}		% landscape table
\usepackage{subcaption}		% subtables
\usepackage{multirow}
\usepackage{booktabs}
\usepackage{adjustbox}
\usepackage{ulem}
\usepackage{authblk} % affiliation to authors
\usepackage[absolute]{textpos}
\usepackage{geometry}
\usepackage{longtable}
\usepackage{booktabs}

\newcommand{\Var}{\mathrm{Var}}

\DeclareMathOperator*{\argmax}{arg\,max}

\graphicspath{
 Figures/
}

\begin{document}

\title{A guide to state-space modeling of ecological time series}
\author[1,2,*]{Marie Auger-M\'eth\'e}
\author[3,4]{Ken Newman}
\author[5]{Diana Cole}
\author[6]{Fanny Empacher}
\author[1,2]{Rowenna Gryba}
\author[7]{Aaron A. King}
\author[8,9]{Vianey Leos-Barajas}
\author[10]{Joanna Mills Flemming}
\author[11]{Anders Nielsen}
\author[12]{Giovanni Petris}
\author[6]{Len Thomas}

\affil[*]{ auger-methe@stat.ubc.ca}
\affil[1]{ Department of Statistics, University of British Columbia, Vancouver, British Columbia, V6T 1Z4, Canada}
\affil[2]{Institute for the Oceans and Fisheries, University of British Columbia, Vancouver, British Columbia, V6T 1Z4, Canada}
\affil[3]{Biomathematics and Statistics Scotland, Edinburgh,  Scotland, EH9 3FD, UK}
\affil[4]{School of Mathematics, University of Edinburgh,
Edinburgh, Scotland, EH9 3FD, UK}
\affil[5]{School of Mathematics, Statistics and Actuarial Science, University of Kent, Canterbury, Kent, CT2 7FS, UK}
\affil[6]{Centre for Research into Ecological and Environmental Modeling, University of St Andrews, Scotland}
\affil[7]{Center for the Study of Complex Systems and Departments of Ecology \& Evolutionary Biology and Mathematics, University of Michigan, Ann Arbor, Michigan, 48109, USA}
\affil[8]{Department of Statistics, University of Toronto, Toronto, Ontario, M5G 1X6, Canada}
\affil[9]{School of the Environment, University of Toronto, Toronto, Ontario, M5S 3E8, Canada}
\affil[10]{Department of Mathematics and Statistics, Dalhousie University, Halifax, Nova Scotia, B3H 4R2, Canada}
\affil[11]{National Institute for Aquatic Resources, Technical University of Denmark, 2920 Charlottenlund, Denmark}
\affil[12]{Department of Mathematical Sciences, University of Arkansas, Fayetteville, Arkansas, 72701, USA}

\date{}

\begin{textblock}{8}(5,1.4)
\noindent \textbf{Running head:} Ecological state-space models\end{textblock}

\maketitle

%%% Would remove 1 on title page
%\thispagestyle{empty}

% Abstract max length: 350 words

\begin{abstract}
      State-space models (SSMs) are an important modeling framework for analyzing ecological time series. These hierarchical models are commonly used to model population dynamics, animal movement, and capture-recapture data, and are now increasingly being used to model other ecological processes. SSMs are popular because they are flexible and they model the natural variation in ecological processes separately from observation error. Their flexibility allows ecologists to model continuous, count, binary, and categorical data with linear or nonlinear processes that evolve in discrete or continuous time. Modeling the two sources of stochasticity separately allows researchers to differentiate between biological variation and imprecision in the sampling methodology, and generally provides better estimates of the ecological quantities of interest than if only one source of stochasticity is directly modeled. Since the introduction of SSMs, a broad range of fitting procedures have been proposed. However, the variety and complexity of these procedures can limit the ability of ecologists to formulate and fit their own SSMs. We provide the knowledge for ecologists to create SSMs that are robust to common, and often hidden, estimation problems, and the model selection and validation tools that can help them assess how well their models fit their data. We present a review of SSMs that will provide a strong foundation to ecologists interested in learning about SSMs, introduce new tools to veteran SSM users, and highlight promising research directions for statisticians interested in ecological applications. The review is accompanied by an in-depth tutorial that demonstrates how SSMs can be fitted and validated in \texttt{R}. Together, the review and tutorial present an introduction to SSMs that will help ecologists to formulate, fit, and validate their models.
\end{abstract}


\noindent
\textbf{Type of contribution:} review

% ***************************************************
\section{\label{S.Intro}Introduction}

State-space models (SSMs) are a popular modeling framework for analyzing ecological time-series data. They are commonly used to model population dynamics \citep{Newman-etal-2014},
including metapopulation dynamics \citep{Ward-etal-2010}, they have a long history in fisheries stock assessment \citep{Aeberhard-etal-2018}, and have been recently proposed as a means of analyzing  sparse biodiversity data \citep
{Kindsvater-etal-2018}. Moreover, they have been a favored approach in movement ecology for more than a decade \citep{Patterson-etal-2008}, and are increasingly used with biologging data \citep{Jonsen-etal-2013}. In addition, the flexibility of SSMs is advantageous when modeling complex capture-recapture data \citep{King-2012}. SSMs are also used in epidemiology \citep{Dukic-etal-2012, Fasiolo-etal-2016} and disease ecology \citep{Hobbs-etal-2015}. These common uses of SSMs, and their many unique applications \citep[e.g., investigating animal health from photographs, \citealt{Schick-etal-2013}; plant invasion, \citealt{Damgaard-etal-2011}; and host-parasitoid dynamics,][]{Karban-and-deValpine-2010}, demonstrate their wide-spread importance in ecology.

\begin{figure}
\begin{center}
\includegraphics[width=470pt]{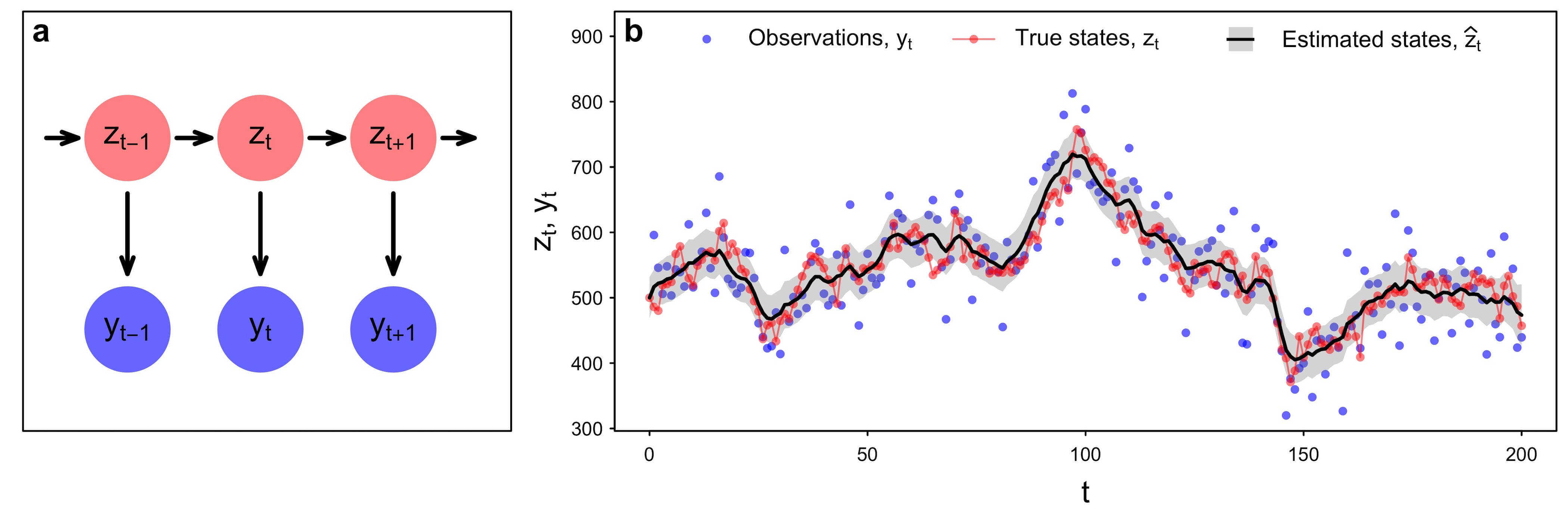}
\end{center}
\caption{The dependence structure and evolution of the two time series comprising a simple univariate SSM. Panel a represents the dependence relationships with arrows and demonstrates that once the dependence of the observations, $y_t$, on the states, $z_t$, is accounted for, the observations are assumed independent. Panel b represents our toy model (Eqs. \ref{E.state.NDLM}-\ref{E.obs.NDLM}). The blue and red dots are the simulated observations and states respectively. The black line and gray band are the estimated states and associated 95\% confidence intervals. The true states, but not the observations, usually fall in the 95\% confidence intervals. This demonstrates that the state estimates can be a closer approximation of the truth than the observations.}
\label{fig:structure}
\end{figure}

SSMs are popular for time series in part because they directly model temporal autocorrelation in a way that helps differentiate process variation from observation error. 
SSMs are a type of hierarchical model \citep[see Table \ref{t.terms} for definitions;][]{Cressie-etal-2009} and their hierarchical structure accommodates the modeling of two time
series: 1) a state, or process, time series that is unobserved and attempts to reflect the true, but hidden, state of nature; and 2) 
an observation time series that consists of observations of, 
or measurements related to, the state time series. For example,
actual fish population size over time would be the state time series,
while incomplete and imprecise counts of fish sampled in a survey,
or caught in a fishery, would be the observation time series. Process variation represents the stochastic processes 
that changes the population size of a fish stock through time (e.g., the 
birth and death processes), while observation error 
reflects differences between the hidden state and the observed data due to randomness or imprecision in the sampling or
survey methodology. These two stochastic components act 
at different levels of the model hierarchy, and
the SSM framework allows them to be modeled separately. The assumptions that the hidden states are autocorrelated (e.g., that a large population in year $t$ will likely lead to a large population in year $t+1$), and that 
observations are independent once we account for their dependence on the states (Fig. \ref{fig:structure}a), allow SSMs to separate these two levels of stochasticity. When we fit a SSM to time series, we can often estimate the process and observation parameters, as well as the hidden states. These estimates
of the hidden states generally reflect the true state of nature better than the original observations (Fig. \ref{fig:structure}b). For example, the estimates of the hidden states will generally reflect the true fish population size better than the survey- or fisheries-based counts. 

The first SSMs, often referred as  Normal Dynamic Linear 
Models (NDLMs), were a special case where the state and the observation time
series were modeled with linear equations and normal distributions. Two seminal papers on NDLMs, \citet{Kalman-1960}
and \citet{Kalman-and-Bucy-1961}, provided an algorithmic 
procedure, the now famous Kalman filter, for making inferences
about the hidden states given imperfect observations and known parameters. These papers led to developments that revolutionized 
aerospace engineering in the 1960s and allowed the Apollo mission 
to correct the trajectory of a spacecraft going to the moon, 
given inaccurate observations of its location through time
\citep{Grewal-Mohinder-2010}. The earliest applications of SSMs 
to ecological data, which used NDLMs and the Kalman filter,
were in the 1980-90s and focused primarily on 
fisheries \citep{Mendelssohn-1988, Sullivan-1992} and 
animal movement \citep{Anderson-Sprecher-1991}. 
The first animal movement SSMs were closely analogous 
to the original aerospace application in that they recreated
the trajectory of an animal based on inaccurate observations.
However, these ecological models required parameters to be estimated.
Unlike  a planned mission to the moon, we rarely have \textit{a 
priori} knowledge of the intended speed and direction of an animal. 
Developments in the time-series literature made use of the Kalman filter to evaluate the likelihood function for unknown parameters, thus allowing
calculation of maximum likelihood parameter estimates in addition to state estimates \citep{Harvey-1990}. NDLMs, however, are a restricted class of SSMs and
their applicability to many ecological time series, which have
nonlinear and non-Gaussian structure, is limited.

Since their initial development, there have been important advancements in SSMs, and in their
application in ecology. In the 1990s, the simultaneous popularization of 
Markov chain Monte Carlo methods \citep[MCMC,][]{Gilks-etal-1995}, including the freely 
available BUGS software  \citep{Lunn-etal-2009}, and high speed desktop computing considerably expanded the diversity of possible SSMs 
to include non-Gaussian and nonlinear formulations
\citep[e.g.,][]{Meyer-Millar-1999}. As a result, Bayesian ecological SSMs were developed for a variety of applications in the following decades, including capture-recapture models \citep[e.g.,][]{Dupuis-1995,Gimenez-etal-2007, Royle-2008} and formulations structured around matrix population models \citep{Buckland-etal-2004}. Further developments have
advanced fitting procedures in both Bayesian and frequentist 
frameworks \citep{deValpine-2004, Ionides-etal-2015, Kristensen-etal-2016, Monnahan-etal-2017}.
These methods provide the means to fit increasingly complex 
SSMs with multiple hierarchical levels
\citep[e.g.,][]{Jonsen-etal-2005} and integrate disparate datasets
\citep[e.g.,][]{Hobbs-etal-2015}.

However, while advancements in fitting SSMs have changed how we
model time series in ecology, the computational burden required 
to fit some of these models is often high enough that
comparisons between multiple SSMs can be difficult, and the 
complex structure of some SSMs complicates model validation
and diagnostics. In the ecological literature, there has been a recent interest in 
model comparison and validation for hierarchical 
models or models for datasets with complex dependence structure \citep{Hooten-Hobbs-2015, Roberts-etal-2017, Conn-etal-2018}. In line with this, new validation tools for SSMs are being developed
\citep{Thygesen-etal-2017}.

While SSMs are powerful tools for modeling ecological time series, 
the fitting procedures may seem prohibitively complex to
many practitioners. The variety
of inference procedures and tools that can be used to fit SSMs 
\citep{Harvey-1990, Doucet-etal-2001, Durbin-and-Koopman-2012, 
Ionides-etal-2015, Kristensen-etal-2016} 
may bewilder all but the most quantitative ecologists, thus limiting the ability of many researchers to formulate, fit, and evaluate
their own SSMs. While there are some popular application-specific \texttt{R} \citep{R} packages with functions to fit specialized SSMs
\citep[e.g., \texttt{MARSS} for multivariate
NDLMs, \citeauthor{Holmes-etal-2012}, \citeyear{Holmes-etal-2012}; \texttt{bsam}, for
animal movement,][]{Jonsen-etal-2005}, few ecologists are aware of the full range of SSMs that can be fitted with such packages. In addition, these packages may be inadequate for the data-at-hand, especially when using SSMs to answer novel questions or with new data types. A further complication with the application of SSMs
is the potential for estimability issues where some states or
parameters cannot be estimated well, or at all,
given the available data \citep{Dennis-etal-2006, Knape-2008, AugerMethe-etal-2016}.
For example, such estimability issues may arise because the formulation of a SSM is too complex for the data (e.g., the time resolution of the process model is too
fine relative to the time resolution of the observations). While 
there has been some effort to provide a general, and easy to use, 
set of tools for ecologists to fit SSMs to their data \citep[e.g.,][]{King-etal-2016, deValpine_et_al:2017}, the available tools and array of
choices may be overwhelming to those with little familiarity with SSMs. Given these challenges and the recent advancements in inference methods and model diagnostics, we believe the time is ripe to
provide a review of these developments
for scientists wanting to fit SSMs to ecological time series.

In this review, we first demonstrate the flexibility of SSMs through a set of examples (Section \ref{S.examples}) and discuss how ecologists should consider SSMs as a default modeling technique for many of their time series (Section \ref{S.OTHER}). Next, we review the different inference methods that can be used to fit a SSM to data (Section \ref{S.Fitting}). We then discuss how one can assess whether a SSM suffers from estimability
or identifiability issues (Section \ref{S.Estimability}). Lastly, we describe model selection procedures (Section \ref{S.Model.Comparison}) and diagnostic tools that can be used to
verify whether a model is adequate (Section \ref{S.Diagnostics}), crucial steps that are often
ignored. This review is accompanied by an in-depth tutorial that provides examples of how one can use \texttt{R} \citep{R} to fit, and validate, SSMs with various inference methods. We believe this review will give a strong foundation to ecologists interested in learning about SSMs and hope it will provide new tools to veteran SSM users interested in inference methods and model validation techniques.

% ***************************************

\section{\label{S.examples}Examples of ecological SSMs}

SSMs are flexible hierarchical models for time series, where observations are imperfect measures of temporally evolving hidden states. Through examples, we demonstrate that SSMs can model univariate or multivariate observations, as well as biological processes that evolve in discrete or continuous time steps. We also show that SSMs can be linear or nonlinear, and can use a variety of statistical distributions (e.g., normal, Poisson, multinomial). To show the structural flexibility of SSMs, we chose many examples from population and movement ecology, two fields that have been crucial in the development of these models. However, SSMs can be used to model time series from all branches of ecology.

%---------------------------------------------------------
\subsection{A toy example: normal dynamic linear model \label{S.M.toy}}

To formalize the description of SSMs, we start by describing a simple, toy example. It models a time series of univariate observations, denoted $y_t$, made at discrete and evenly-spaced points in
time $t$ $(t = 1, 2, \ldots, T)$. The time series of states, denoted $z_t$, is defined at the same time points $t$ as the observations. Our model is a simple normal dynamic linear model (NDLM), thus process variance and observation error are modeled with Gaussian distributions and both time series are modeled with linear equations.

SSMs make two main assumptions. First, SSMs assume that the state time series evolves as a Markov process \citep{Aeberhard-etal-2018}. This Markov process, which is generally of first-order, is a relatively simple way to incorporate temporal dependence. For our toy model, this means that the state at time $t$, $z_t$, depends only on the state at the previous time step, $z_{t-1}$. Second, SSMs assume that the observations are independent of one another once we account for their dependence on the states. More formally, we say that, given
the corresponding state $z_t$, each observation $y_t$ is conditionally independent of all other observations, $y_s$, $s \ne t$. Thus, any dependence between observations is the result of the dependence between hidden states \citep{Aeberhard-etal-2018}. For our toy model, this means that $y_t$ is independent of $y_{t-1}$, and all other 
observations, once we account for the dependence of $y_t$ on $z_t$ (Fig. \ref{fig:structure}a). In a population dynamics context, this could be interpreted to mean that the values of observations are autocorrelated because the process driving them (i.e., the true population size of the animal) is autocorrelated through time. In contrast, the discrepancy between the true population size and the observation is not correlated in time. We can see this structure in the equations for our toy SSM:
\begin{linenomath*}
\begin{align} 
  \label{E.state.NDLM}
   z_t &= \beta z_{t-1} + \epsilon_t,  & \epsilon_t \sim \text{N} \left( 0, \sigma^2_p \right ), \\
   \label{E.obs.NDLM}
   y_t & = \alpha z_t + \eta_t, & \eta_t \sim \text{N} \left( 0, \sigma^2_o \right ).
 \end{align}
\end{linenomath*}
The autocorrelation in the states is captured by the parameter $\beta$. The observations are a function of the states only and the parameter $\alpha$ allows the observation at time $t$ to be a biased estimate of the state at time $t$. The process  variation ($\epsilon_t$) and observation error ($\eta_t$) are both modeled with normal distributions but have different standard deviations ($\sigma^2_p$ and $\sigma^2_o$). We have not defined the state at time $0$, $z_0$, and many authors will provide an additional equation, often referred as the initialization equation, which describes the probability of different values of $z_0$ (e.g., $z_0 \sim \text{N}(0, \sigma_{z_0}^2)$). For our toy example, we view $z_0$ as a fixed and unknown parameter. 

The terminology used to refer to the process and observation equations varies in the literature. A process equation can be referred as a process model, state equation, state model, or transition equation. An observation equation can be referred as an observation model,
measurement equation, or measurement model. In this paper, we generally use the terms `process equation' and `observation equation' respectively, and we often describe SSMs with equations that combine a deterministic function with a stochastic component (e.g., Eqs. \ref{E.state.NDLM}-\ref{E.obs.NDLM}).

To further reveal the dependence structure and understand how to fit SSMs to data, it can help to additionally characterize a SSM in terms of probability distributions
for the states and the observations, e.g.:
\begin{linenomath*}
\begin{align}
 \label{E.state.general}
  & f(z_t|z_{t-1},\boldsymbol{\theta}_p), & t=1,\ldots, T, \\
 \label{E.obs.general}
  & g(y_t|z_t,\boldsymbol{\theta}_o), & t=1,\ldots, T.
 \end{align}
\end{linenomath*}
In the case of our toy model, $f$ and $g$ are two normal probability density functions, while $\boldsymbol{\theta}_p$ and $\boldsymbol{\theta}_o$ are
vectors of parameters associated with each equation (i.e., $\boldsymbol{\theta}_p = (\beta, \sigma_p^2)$, $\boldsymbol{\theta}_o = (\alpha, \sigma_o^2)$). Eq. \ref{E.state.general} describes the autocorrelation in state values as a first-order Markov process, while Eq. \ref{E.obs.general} describes how observations depend simply on the states. This definition also demonstrates that states are random variables and thus that SSMs are a type of hierarchical model.

One of the goals of fitting a SSM to data is typically to estimate unknown parameters. Here, to contrast them with the states, we refer to these as the fixed parameters and denote them together as $\boldsymbol{\theta}$. For example, here $\boldsymbol{\theta} =  (\boldsymbol{\theta}_p, \boldsymbol{\theta}_o, z_0)$, thus $\alpha$, $\beta$, $\sigma^2_p$, 
$\sigma^2_o$, $z_0$ in Eqs. \ref{E.state.NDLM}-\ref{E.obs.NDLM}. A second important goal is to estimate the unobserved states, $\mathbf{z}_{1:T} = (z_1, z_2, ..., z_T)$, where $T$ is the length of the time series. The notation $1{:}t$, which we use throughout, refers to the sequence 1, 2, $\ldots$, $t$. Fig. \ref{fig:structure}b shows how close estimates of the states ($\hat{\mathbf{z}}_{1:T}$) can be to their true values.

SSMs can be fitted using frequentist or Bayesian approaches to statistical inference. When using a Bayesian approach, a third level is added to the model hierarchy: the prior distribution(s) for the fixed parameters denoted by the probability density function, $\pi(\boldsymbol{\theta} |\boldsymbol{\lambda})$, where $\boldsymbol{\lambda}$ are known values called hyperparameters. While we refer to $\boldsymbol{\theta}$ as fixed parameters to differentiate them from the states, in Bayesian inference $\boldsymbol{\theta}$ is a vector of random variables. In Section \ref{S.Fitting} and Appendix S2, we discuss how we can use these probabilistic descriptions of the model for inference.

This simple linear and normal model is a toy example that we will use throughout to explain the concepts associated with fitting and validating SSMs. We will also use this model in Appendix S1 to demonstrate how to use \texttt{R} to fit a SSM to data. While the simplicity of this toy example makes it a helpful teaching tool, it is not a particularly useful model for ecology. We now turn to the description of a set of ecological SSMs.

%---------------------------------------------------
\subsection{Handling nonlinearity \label{S.M.simple.pop}}

We use a set of simple univariate SSMs for population dynamics to demonstrate that even simple ecological models can rarely be blindly modeled as NDLMs. \citet{Jamieson-Brooks-2004} applied multiple SSMs to abundance estimates from North American ducks obtained through annual aerial counts. We start with one of their simplest models, for which the process equation is the stochastic logistic model of \citet{dennis1994}. This model allows for density dependence, i.e., a change in growth rate dependent on the abundance in the previous year:
\begin{linenomath*}
\begin{align}
z_t &= z_{t-1} \exp\left(\beta_0 + \beta_1 z_{t-1} + \epsilon_t \right),
& \epsilon_t \sim \text{N}(0, \sigma_p^2). \label{eq:ducks_process}
\end{align}
\end{linenomath*}
As in the toy example above, $z_t$ denotes the true hidden state, in this case the number of ducks in year $t$. The parameter $\beta_0>0$ determines the median rate of population growth when population size is 0. The parameter $\beta_1\leq 0$ determines how much the growth rate decreases with an increase in population size, with $\beta_1 = 0$ indicating no density dependence. The process variation, $\epsilon_t$, is normally distributed and represents the random change in growth rate each year. The observations $y_t$ are modeled as unbiased estimates of the true population size with a normally distributed error:
\begin{linenomath*}
\begin{align}
y_t &= z_t + \eta_t,
& \eta_t \sim \text{N}(0, \sigma_o^2).
\end{align}
\end{linenomath*}
Even though the observation equation is linear with a Gaussian error, the SSM is not a NDLM because of the exponent in the process equation (Eq. \ref{eq:ducks_process}). \citet{Jamieson-Brooks-2004} also modeled the population size on a logarithmic scale, $w_t=\log(z_t)$, which resulted in the following formulation:
\begin{linenomath*}
\begin{align}
    w_t &= w_{t-1} + \beta_0 + \beta_1 \exp(w_{t-1}) + \epsilon_t,
    & \epsilon_t \sim \text{N}(0, \sigma_p^2),
    \\
    y_t &= \exp(w_t) + \eta_t,
    & \eta_t \sim \text{N}(0, \sigma_o^2) \label{eq:untrans1Pop}.
\end{align}
\end{linenomath*}
While such reconfiguration can sometimes linearize the model, in this case the model remains nonlinear. 
\citet{Jamieson-Brooks-2004} use a Bayesian framework to fit this model, see their original paper for the description of the priors.

The modeling of density dependence has been extensively debated in the literature, and \citet{Jamieson-Brooks-2004} also explored an alternative 
process equation, a stochastic Gompertz model:
\begin{linenomath*}
\begin{align}
    z_t &= z_{t-1} \exp\left(\beta_0 + \beta_1 \log z_{t-1} + \epsilon_t \right),
& \epsilon_t \sim \text{N}(0, \sigma_p^2),
\end{align}
\end{linenomath*}
which assumes that the per-unit-abundance growth rate depends on the log abundance, $\log(z_{t-1})$, instead of the abundance, $z_{t-1}$ \citep{dennis1994}. Such a model is often linearized as follows:
\begin{linenomath*}
\begin{align}
w_t &= \beta_0 + (1 + \beta_1) w_{t-1} + \epsilon_t,
& \epsilon_t \sim \text{N}(0, \sigma_p^2),
\\
g_t &= w_t + \eta_t,
& \eta_t \sim \text{N}(0, \sigma_o^2),
\end{align}
\end{linenomath*}
where $w_t = \log (z_t)$ and $g_t = \log (y_t)$ are the logarithms of the states and observations, respectively \citep[e.g.,][]{Dennis-etal-2006}. The linear version of this model is a NDLM that can be fitted with tools such as the Kalman filter \citep{Dennis-etal-2006}. This statistical convenience may have contributed to the uptake of the stochastic Gompertz SSM in the literature. However, it may not always be adequate to assume that the growth rate depends logarithmically on population density \citep{dennis1994}.

Many papers have extended these models to incorporate external covariates \citep[e.g.][]{viljugrein2005,Saether-etal-2008,Linden-Knape-2009}. For example, one could account for the influence of fluctuating availability of wetlands on the population size of ducks by including the number of ponds in year $t$, $p_t$, as a covariate in the process equation and modifying the Gompertz stochastic model as follows: 
\begin{linenomath*}
\begin{align}
w_t &= \beta_0 + (1 + \beta_1) w_{t-1} + \beta_2 p_t + \epsilon_t,
& \epsilon_t \sim \text{N}(0, \sigma_p^2).
\end{align}
\end{linenomath*}

This set of examples shows how easy it is to adapt and extend models in the SSM framework. While even simple ecological models may only be linear with transformations and assumptions, \citet{Jamieson-Brooks-2004}, \citet{viljugrein2005}, and \citet{Linden-Knape-2009} showed that accounting for observation error improved the inference regardless of the process equation. For example, \citet{viljugrein2005} demonstrated that using a SSM, rather than a model that ignores observation error, decreased the size of the bias in the estimates of density dependence. This decreased bias, and a better quantification of uncertainty, reduced the cases where one would erroneously conclude the presence of density dependence.

%-----------------------------------------
\subsection{Joining multiple data streams \label{S.M.stock.assessment}}

Integrating multiple sources of data, often referred as data streams, into a single model can offset their individual limitations and reveal more complex ecological relationships \citep{Mcclintock-etal-2017}. To showcase how SSMs can extract the information provided by multiple data streams, we present a simplified version of a state-space stock assessment model described by \citet{Nielsen-Berg-2014}. SSMs are often used in fisheries stock assessments \citep{Aeberhard-etal-2018}, where the first data stream, $C_{a,t}$, represents how many fish from each age class $a$ are caught in the commercial fishery in each year $t$, and the second data stream, $I_{a,t,s}$, includes age-specific indices from distinct scientific surveys, $s$, which can occur in different years and only capture some portion of the age classes.

The hidden state in each year $t$ is a vector combining the log-transformed stock sizes, $N_{a,t}$, and fishing mortality rates, $F_{a,t}$, for each age class: $\mathbf{z}_t$ $=$ $(\log{N_{1,t}},\ldots,\log{N_{{\textsc{a}},t}},$ $\log{F_{1,t}},\ldots, \log{F_{\textsc{a},t}})'$, where $A$ represents the oldest age class. Just as for the toy example, the process equations describe the state in year $t$ as a function of the state in year $t-1$. However, unlike the toy model, we no longer have a single process equation. We have instead a set of equations describing recruitment, survival, and mortality:
\begin{linenomath*}
\begin{align}
 \log (N_{1,t}) &=\log (N_{1,t-1}) + \epsilon_{N_{1, t}},  \label{E.fsa.p.N1}\\
 \log (N_{a,t}) &=\log(N_{a-1,t-1}) - F_{a-1,t-1} - M_{a-1,t-1} + \epsilon_{N_{a,t}},  \hspace{3.5cm}  2\leq a \leq A, \label{E.fsa.p.Na}\\
 \log (F_{a,t}) &=\log (F_{a,t-1}) + \epsilon_{F_{a,t}}, \hspace{7.8cm} 1\leq a \leq A, \label{E.fsa.p.F}
\end{align}
\end{linenomath*}
where age and year specific log fishing mortality rates, $\log F_{a,t}$, are considered states that evolve as a random walk through time, but the equivalent natural mortality rate, $\log M_{a,t}$, is assumed known from outside sources. While the main equation describing the population growth (Eq. \ref{E.fsa.p.Na}) is based on demographic processes, the other equations (Eqs. \ref{E.fsa.p.N1} and \ref{E.fsa.p.F}) are simply assuming that recruitment and fishing mortality are each correlated across years. The formulation of Eq. \ref{E.fsa.p.Na}, and Eqs. \ref{E.fsa.obs.C}-\ref{E.fsa.obs.I} below, are based on the well-known Baranov catch equation, which states that a cohort continuously decreases in size through time according to two sources of mortality \citep[i.e., fishing and natural, see][for detail]{Aeberhard-etal-2018}. Derived from a continuous-time equation, the Baranov equation maps the surviving cohort size as depending exponentially on fishing and natural mortality rates. Thus, as shown here, the SSM can be modeled by expressing the age-specific stock size and mortality rates on the logarithmic scale.

The process variation for all of these equations are assumed to be Gaussian with zero mean, but they differ in their variance and covariance parameters. For recruitment and survival, the variation is assumed to be uncorrelated, i.e.,  $\epsilon_{N_{1, t}} \sim \text{N}(0, \sigma^2_{N_{a=1}})$, and $\epsilon_{N_{a,t}}$ and $\epsilon_{N_{\textsc{a},t}} \sim \text{N}(0,\sigma^2_{N_{a>1}})$. However, for fishing mortality, the yearly variation is assumed to be correlated across age classes (i.e., $\boldsymbol{\varepsilon}_{F_t} = (\epsilon_{F_1,t}, ..., \epsilon_{F_{\textsc{a}},t})' \sim \text{N} (\mathbf{0}, \boldsymbol{\Sigma}_F)$) due to age/size correlations in capture probability. The covariance matrix, $\boldsymbol{\Sigma}_{F}$, is assumed to have an auto-regressive order 1, AR(1), correlation structure (i.e., each element $\Sigma_{a,\tilde{a}} = \rho^{|a-\tilde{a}|} \sigma_a \sigma_{\tilde{a}}$, where $\rho^{|a-\tilde{a}|} = \mbox{cor}(\epsilon_{F_{a_t}}, \epsilon_{F_{\tilde{a} _t}})$).

The two different sets of data streams (i.e., the observed age-specific log-catches, $\log C_{a,t}$, and the age-specific log-indices from scientific surveys, $\log I_{a,t,s}$) are related to the time series of the unobserved states, $\mathbf{z}_t$, with the following observation equations:
\begin{linenomath*}
\begin{align}
 \log C_{a,t} &= 
\log\left(\frac{F_{a,t}}{K_{a,t}}(1-e^{-K_{a,t}})N_{a,t}\right)+\eta_{a,t,c} \label{E.fsa.obs.C},\\
 \log I_{a,t,s} &= \log\left(Q_{a,s}
 e^{-K_{a,t}\frac{D_s}{365}}N_{a,t}\right)+\eta_{a,t,s} \label{E.fsa.obs.I},
\end{align}
\end{linenomath*}
where $K_{a,t}$ is the total mortality rate of age class $a$ in year
$t$ (i.e., $K_{a,t}=M_{a,t}+F_{a,t}$), $D_s$ is the number of days
into the year when the survey $s$ is conducted, and each $Q_{a,s}$ is a model parameter describing the catchability coefficient. The observation error terms, $\eta_{a,t,c}$ and $\eta_{a,t,s}$, are assumed to be Gaussian distributed and their variances are designed such that the catch data, and each scientific survey have their own covariance matrix. We can use different covariance structures for each matrix \citep[e.g., independent catches across ages, but each survey index has an AR(1) correlation structure across ages; see][for other examples]{Berg-Nielsen-2016}. 

This example depicts how to harness more information from independent data streams. The observation equations (Eqs. \ref{E.fsa.obs.C}-\ref{E.fsa.obs.I}) account for the differences in how each data stream is related to a common of set of states (i.e., stock sizes, $N_{a,t}$). In addition, the potentially more biased data stream (i.e., the fisheries catch data) provides direct information on the other set of states (i.e., fishing mortality rate, $F_{a,t}$), which would otherwise be difficult to estimate. This type of structure provides the opportunity to model more complex ecological mechanisms (e.g., Eqs. \ref{E.fsa.p.N1}-\ref{E.fsa.p.F}). SSMs that integrate multiple data streams have been used in other fields of ecology, including movement ecology \citep{Mcclintock-etal-2017} and disease ecology \citep{Hobbs-etal-2015}.

%---------------------------------------------------
\subsection{Accounting for complex data structure \label{S.M.movement.model}}

SSMs are well suited to handle the complex structure of many ecological datasets. For example, the first difference correlated random walk model \citep[DCRW,][]{Jonsen-etal-2005}, one of the earliest SSMs for animal movement, was developed to account for the peculiarities of Argos doppler shift location data \citep{Jonsen-etal-2005}. Argos tags are often used to track marine animals because they overcome some of the challenges associated with using conventional GPS units in an aquatic environment. However, unlike most GPS datasets, Argos locations, $\mathbf{y}_i = \left[\begin{smallmatrix} y_{i,lon}\\ y_{i,lat}  \end{smallmatrix}\right]$, data have large observation errors \citep[mean error ranging from 0.5-36km,][]{Costa-etal-2010}, including large outliers. In addition, they are collected at irregular time intervals, $i$, (i.e., when the animal is at the surface and the satellites are overhead), and have a quality rating that classifies each location into one of six categories, $q_i$. All of these aspects of the data are incorporated in the simplified version of the DCRW presented below. 

While the observations are taken at irregular time intervals, the process equation models the true locations of the animal at regular time intervals $t$, $\mathbf{z}_t = \left[\begin{smallmatrix} z_{t,lon} \\ z_{t,lat}\end{smallmatrix}\right]$, for $T$ time steps. The process equation assumes that the animal's location at time $t$ is not only dependent on the previous location, $\mathbf{z}_{t-1}$, but also on the animal's previous displacement in each coordinate, $\mathbf{z}_{t-1} - \mathbf{z}_{t-2}$:
\begin{linenomath*}
	\begin{align}
        \mathbf{z}_t &= \mathbf{z}_{t-1} + 
        \gamma(\mathbf{z}_{t-1} - \mathbf{z}_{t-2}) +
        \boldsymbol{\epsilon}_t, & \boldsymbol{\epsilon}_t &\sim \text{N}(0,\mathbf{\Sigma}), & 1 \leq t \leq T, \label{e.dcrw.p}
	\end{align}
\end{linenomath*}
where
\begin{linenomath*}
	\begin{align}
        \boldsymbol{\Sigma} &=  \begin{bmatrix}
	  		\sigma_{\epsilon,lon}^2 & \rho\sigma_{\epsilon,lon}\sigma_{\epsilon,lat} \\
	  		\rho\sigma_{\epsilon,lat}\sigma_{\epsilon,lon} & \sigma_{\epsilon,lat}^2
		\end{bmatrix}.
	\end{align}
\end{linenomath*}
The parameter $\gamma$ can take values between 0 and 1 (i.e., $0 \le \gamma \le 1$), and controls the degree of correlation between steps. Values close to 0 mean that the movement only depends on the previous location. Values close to 1 reflect strong correlation in both latitudinal and longitudinal displacements, and thus mean that the animal has a tendency to move at the same speed and in the same direction as the previous step. The covariance matrix for the process variation, $\boldsymbol{\Sigma}$, allows for covariance between longitude and latitude, but in many instances it is simpler to assume that $\rho = 0$.

\begin{figure}
\begin{center}
\includegraphics[width=470pt]{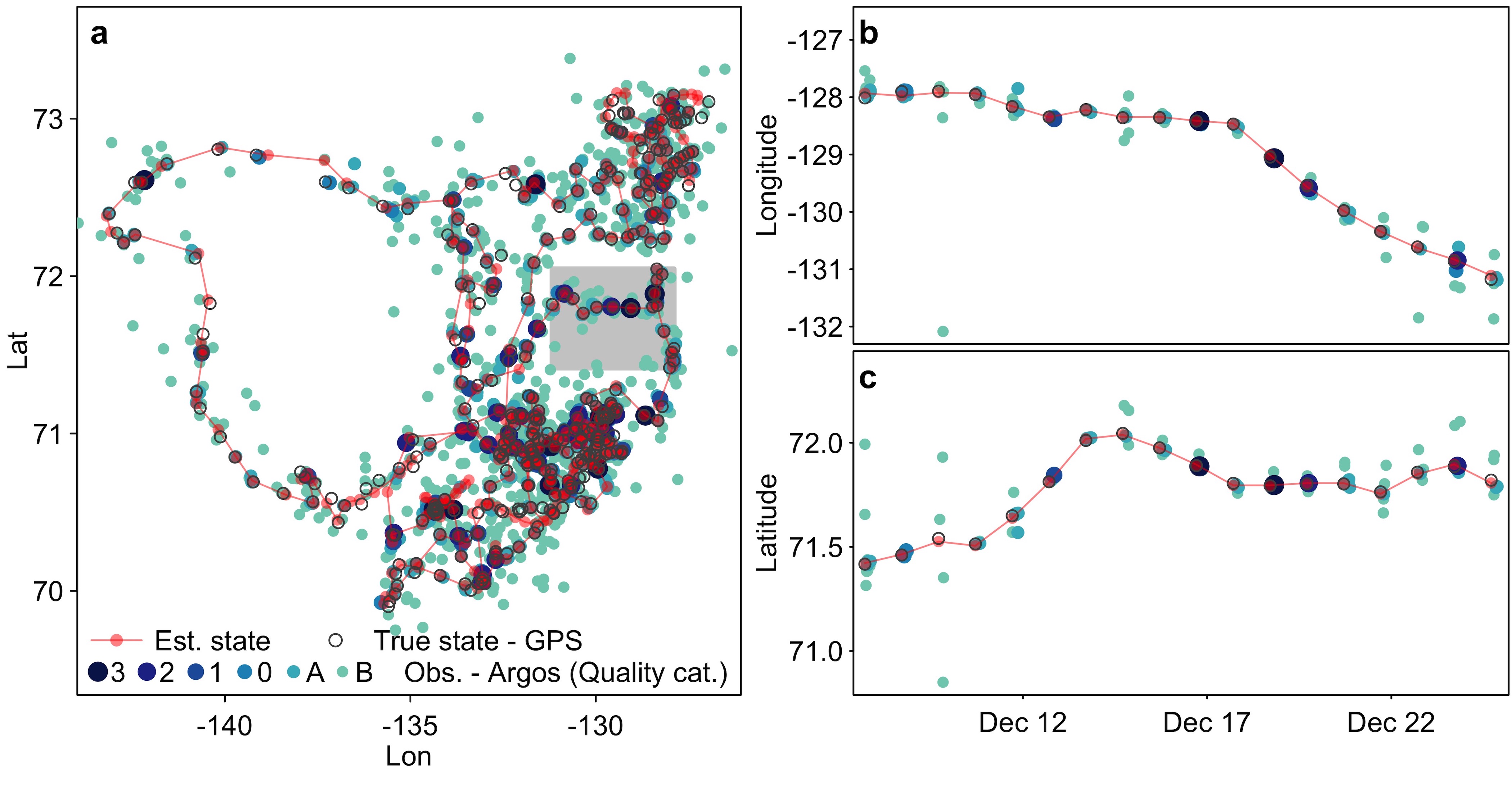}
\end{center}
\caption{The DCRW (Eqs. \ref{e.dcrw.p}-\ref{e.dcrw.v}) fitted to a polar bear Argos track and validated with GPS data. Panel a maps the observed Argos data with points in shades of blue and green (darker colors representing higher quality observations), the estimated true locations in red, and the true locations of the bear (GPS data) with open circles. Panels b and c show the longitudes and latitudes of a small subset of the time series (indicated by a grey box in the map). These panels highlight the temporal clustering of observations, which likely helped the state estimation procedures.}
\label{fig:polarbear}
\end{figure}

The observation equation accounts for various characteristics of the Argos data:
\begin{linenomath*}
	\begin{align}
        \mathbf{y}_i &= (1-j_i)\mathbf{z}_{t-1} + j_i\mathbf{z}_t + \boldsymbol{\eta}_i, & \boldsymbol{\eta}_i &\sim \text{T}(\boldsymbol{\Psi} \circ \mathbf{S}_i, \mathbf{D}_i), & 1 \leq i \leq N,
        \label{e.dcrw.o}
	\end{align}
\end{linenomath*}
where
\begin{linenomath*}
	\begin{align}
		\boldsymbol{\Psi} &=  \begin{bmatrix}
	  		\psi_{lon}\\
	  		\psi_{lat}
		\end{bmatrix},\\
		  \mathbf{S}_i &=  \begin{bmatrix}
	  		s_{lon,q_i}\\
	  		s_{lat,q_i}
		\end{bmatrix},\\
	  \mathbf{D}_i &= \begin{bmatrix}
	  		df_{lon,q_i}\\
	  		df_{lat,q_i}
		\end{bmatrix},
		\label{e.dcrw.v}
	\end{align}
\end{linenomath*}
and $N$ is the number of observed Argos locations.
Because data are taken at irregular time intervals, the true location of the animal is linearly interpolated to the time of the observation, with $j_i$ representing the proportion of the regular time interval between $t-1$ and $t$ when the observation $\mathbf{y}_i$ was made. Because the data often have outliers, the measurement errors are modeled with t-distributions, which have fat tails. Finally, to model the differences in error size between the six quality categories, each category, $q_i$, is associated with unique bivariate t-distributions: $\text{T}(\boldsymbol{\Psi} \circ \mathbf{S}_i, \mathbf{D}_i)$. In particular, each category is associated with a unique scale parameter, $s_{c,q_i}$, and degrees of freedom, $df_{c,q_i}$, for each coordinate (i.e., $c = lon$ or $lat$). Instead of estimating these 24 parameters, many researchers fix them to known values derived from field experiments \citep[e.g.,][]{Jonsen-etal-2005}. To allow for discrepancies between these fixed values and the ones that may fit the data best, we can add a correction factor for each coordinate, $\psi_c$. Note that the Hadamard product, $\circ$, simply states that we perform entrywise multiplication of the correction factors to the scale parameters, i.e., $\psi_c s_{c,q_i}$, for $c= (lon,lat)$. Fig. \ref{fig:polarbear} shows the DCRW fitted to a polar bear track, and Appendix S1: Section S1 2.3 provides the code to fit the model. 

Datasets with unexplained outliers and data points with differing quality ratings are common in ecology, and the flexibility of SSMs allow to directly account for these characteristics in the model, rather than arbitrarily discarding data.

%---------------------------------------------------
\subsection{Accommodating continuous-time processes }\label{S.cont.mov.model}

So far we have only described SSMs where the hidden state evolves in discrete time steps. However, many biological processes occur in continuous time and modeling them as such can facilitate the use of irregularly-timed observations \citep{Dennis-etal-2014, McClintock-etal-2014}. Using a simplified version of the movement model of \citet{Johnson-etal-2008}, we showcase how SSMs can accommodate continuous-time process equations.

The SSM of \citet{Johnson-etal-2008} models the movement of an animal with a continuous-time correlated random walk. The process equation is formulated in terms of how changes in velocity $v$ through time affect the location $\mu$ of an animal. While the model describes an animal moving in two dimensions (e.g., latitude and longitude), for simplicity, we assume the velocity processes in each coordinate to be independent and only describe the process for one coordinate. Velocities at time $t$, denoted $v(t)$, are the first set of states. Change in velocity over time is described using a type of diffusion model called an Ornstein-Uhlenbeck (OU) process. At time $t + \Delta$, velocity is:
\begin{linenomath*}
\begin{align}
v(t+\Delta) &= e^{-\beta\Delta} v(t) + \zeta(\Delta), & \zeta(\Delta) \sim \text{N}\left(0, \sigma^2_{OU} (1-e^{-2\beta\Delta})/2\beta\right), &\;\;\;\;\;\; \beta > 0
\label{eq:OU}
\end{align}
\end{linenomath*}
where $\beta$ represents how fast the temporal correlation in velocity tends towards 0, and $\zeta(\Delta)$ is a random perturbation. As both increases, the autocorrelation in velocity decreases. In addition, as the time difference ($\Delta$) increases, the velocity value at time $t+\Delta$ depends less on the previous velocity value and more on the random perturbation. This assumption is often reasonable as we expect an animal to continue at the same speed during a short period of time and be more likely to change speed over long time periods.

While the core of the process model describes changes in velocity, the observations are locations. Thus, we have a second set of states, the locations $\mu(t)$, which are related
to velocities as follows:
\begin{linenomath*}
\begin{align}
\mu(t+\Delta) &= \mu(t) + \int_{t}^{t+\Delta}v(u) du.
\label{eq:muInt}
\end{align}
\end{linenomath*}
Integrating the rate of change, here speed, over the time interval is often key to link continuous-time processes to ecological observations \citep[e.g., to model oxygen concentration,][]{Appling-etal-2018}. Such integration can be difficult to handle, but \citet{Johnson-etal-2008} solved Eq. \ref{eq:muInt} to show that the change in location in time $\Delta$ is simply:
\begin{linenomath*}
\begin{align}
\mu(t+\Delta) &= \mu(t) + v(t)\left(\frac{1-e^{-\beta \Delta}}{\beta}\right) + \xi(\Delta), & \xi(\Delta) \sim \text{N}\left(0, \frac{\sigma^2_{OU}}{\beta^2}\right).
\label{eq:mu}
\end{align}
\end{linenomath*}

Because $\Delta$ can take any non-negative value, we can keep track of the states at any time intervals, thus easily accommodating observations, $y_i$, collected at irregular-spaced times, $t_i$. For the state, $\mathbf{z}_i$, the final process equations in matrix notation form are:
\begin{linenomath*}
\begin{align}
    \mathbf{z}_i &= \begin{bmatrix}
    \mu_i \\ v_i
    \end{bmatrix}
    = \begin{bmatrix}
    1 & (1-e^{\beta\Delta_i})/\beta \\ 0 & e^{-\beta\Delta_i}
    \end{bmatrix} \begin{bmatrix}
    \mu_{i-1} \\ v_{i-1}
    \end{bmatrix} + \boldsymbol{\eta}_i,
    &  \boldsymbol{\eta}_i \sim \text{N}\left(\mathbf{0}, \boldsymbol{\Sigma}^2_p\right),
\label{eq:process}
\end{align}
\end{linenomath*}
where $u_i$ and $v_i$ are $u(t)$ and $v(t)$ at the time when the $i^{th}$ observation occurred, $\Delta_i = t_i-t_{i-1}$, and the variance-covariance matrix was solved to be:
\begin{linenomath*}
\begin{align}
    \boldsymbol{\Sigma}^2_p &= \begin{bmatrix}
 \frac{\sigma^2_{OU}}{\beta^2}&\frac{\sigma^2_{OU}}{2\beta^2}\left(1-e^{-\beta\Delta_i}\right)^2 \\
\frac{\sigma^2_{OU}}{2\beta^2}\left(1-e^{-\beta\Delta_i}\right)^2&\sigma^2_{OU} (1-e^{-2\beta\Delta_i})/2\beta
\end{bmatrix}.
\end{align}
\end{linenomath*}

The observation equation can be chosen as usual, for example as simply adding normal error to the true location:
\begin{linenomath*}
\begin{align}
y_i &= \mu_i + \epsilon_i, & \epsilon_i \sim \text{N}\left(0, \sigma^2_o\right).
\label{eq:obs}
\end{align}
\end{linenomath*}
The SSM defined by Eqs. \ref{eq:process}-\ref{eq:obs} is a linear Gaussian SSM and can therefore be fitted with a Kalman filter.

This model allows various extensions to include different aspects of animal movement. For example, \citet{Johnson-etal-2008} show how haul-out behavior of tagged seals can be incorporated using data on how long the tag has been dry (e.g., by making $\beta$ an increasing function of dry time). To account for the large outliers associated with Argos data one can use a t-distribution (see Section \ref{S.M.movement.model}), in which case the Kalman filter will no longer be adequate and other fitting methods will be required  \citep{Albertsen-etal-2015}. While continuous-time models can be more complex to understand, they are useful in a variety of contexts where data is collected at unequal time intervals and when ecological processes are intrinsically continuous \citep[e.g., population dynamics,][]{Dennis-etal-2014}.

%----------------------------------------
\subsection{Integrating count and categorical data streams}

The SSM framework provides the flexibility to create joint models that integrate different data types and link biological processes. Here, we use the model of \citet{Schick-etal-2013} to demonstrate how count and categorical data can be integrated in a single SSM for the health, monthly movement, and survival of North Atlantic right whales (\textit{Eubalaena glacialis}).

\citet{Schick-etal-2013} extracted two types of data from photographic observations of individual whales. The first type, denoted $y_{i,t,k}$, is the number of sightings of individual $i$ in geographic zone $k$ and month $t$. The second type, denoted $H_{q,i,t}$, is the value for the $q^{th}$ visual health metric for individual $i$ in month $t$. The six visual health metrics (e.g., skin condition, entanglement status) are on ordinal scales, each with two or three levels. In addition, ancillary data (e.g., search effort, whale age) are used. 

Three process equations model the health, survival, and monthly movement of each individual whale. Whale $i$ in month $t$ is characterized by its age $a_{i,t}$, health status $h_{i,t}$ (defined on an arbitrary, but positive, scale: $(0,100)$), and location $k_{i,t}$ (one of nine geographic zones). Health status, $h_{i,t}$, is modeled as a function of previous health status and age:
\begin{linenomath*}
\begin{align}
h_{i,t} &= \beta_0 + \beta_1 h_{i,t-1} + \beta_2 a_{i,t-1} + \beta_3 a_{i,t-1}^2 + \epsilon_{i,t}, & \epsilon_{i,t} \sim \text{N}(0, \sigma^2).
\end{align}
\end{linenomath*}
When $\beta_2 > 0$ and $\beta_3 < 0$, the quadratic age term allows for the fact that health status, and thus survival probability, initially increases but declines with advanced age. Survival from month $t$ to $t+1$ is modeled as a 
Bernoulli process, with survival probability modeled with a logit link function:
\begin{linenomath*}
\begin{align}
    \text{logit}(s_{i,k,t}) & = \alpha_{0,k} + \alpha_1 h_{i,t}.
\end{align}
\end{linenomath*}
Here, $\alpha_{0,k}$ denotes the fixed effect for zone $k$ and $\alpha_1$ the relationship with health. Hence, survival probability depends on health and on the occupied zone, allowing researchers to identify the geographic zones associated with reduced survival. 

While a whale is assumed to stay in a single zone during the month, it can move between zones each month. The monthly location of each individual, $z_{i,t}$, is only known when the individual is sighted that month. The subscript $t$, throughout, represents the number of months since the beginning of the time series. For each month of the year (January, ..., December), denoted $t^{(u)}$, the movement between zones is modeled with a transition matrix, where each element, $m_{j,k,t^{(u)}}$, describes the probability of moving from zone $j$ to zone $k$ (i.e. $m_{j,k,t^{(u)}} = \text{Pr}(z_{i,t^{(u)}+1} = k |z_{i,t^{(u)}} = j)$). As the complete geographic range of the whales is assumed to be covered by these zones, a living whale will be in one of the nine distinct zones at time $t+1$, $\sum_{k=1}^9 m_{j,k,t^{(u)}} = 1$. The changes in transition probabilities between the months of the year, $t^{(u)}$, allow for the modeling of seasonal migration.

The model has two sets of observation equations. First, the number of sightings of whale $i$ in location $k$ and month $t$ is modeled as a Poisson random variable.
\begin{linenomath*}
\begin{align}
y_{i,k,t} &\sim \text{Pois}(\lambda_i E_{k,t}),
\end{align}
\end{linenomath*}
where $E_{k,t}$ denotes the search effort in zone $k$ and month $t$ and $\lambda_i$ denotes the expected number of sightings of individual $i$ per unit effort. The number of sightings of whale $i$ is only modeled in months where the individual is alive based on state $s_{i,k,t}$ and in the appropriate monthly geographical zones according to state $z_{i,t}$. Second, each visual health metric is modeled as coming from a multinomial logit distribution. The probability of being in each level of the $q^{th}$ health metric depends on the true health status, $h_{i,t}$, and the model is structured so as to ensure that the ordinal aspect of the variables is respected (i.e., that lower values means lower health). For example, if the health metric $ H_{q,i,t}$ has a three-level ordinal scale, the observation equations for this metric are: \begin{linenomath*}
\begin{align}
H_{q,i,t}& \sim \text{Multinom}(1, \boldsymbol{p}_{q, i,t})
\\
\text{logit}(p_{q, i,t,1})&= \log\left(\frac{p_{q, i,t,1}}{p_{q, i,t,2}+p_{q,i,t,3}}\right) =  c_{q,0,1} + c_{q,1,1} h_{i,t}
\\
\text{logit}(p_{q,i,t,1} + p_{q, i,t,2})&= \log\left(\frac{p_{q,i,t,1} + p_{q, i,t,2}}{p_{q,i,t,3}}\right) =  c_{q,0,2} + c_{q,1,2} h_{i,t} \label{eq:summultnomlink}
\\
p_{q,i,t,3} &= 1- p_{q,i,t,1}-p_{q,i,t,2}.
\end{align}
\end{linenomath*}
The vector $\boldsymbol{p}_{q,i,t}$ contains the probabilities with which an individual with true health $h_{i,t}$ is assigned a specific health level. Since $h_{i,t}$ is positive, forcing the parameters $c_{q,0,1} < c_{q,0,2}$ and $c_{q,1,1} < c_{q,1,2}$, and modeling cumulative probabilities (Eq. \ref{eq:summultnomlink}) ensure that the order of the levels is accounted for. The probabilistic nature of the model allows health metrics to depend on the true health status but to be observed with error.

By integrating different data types, this SSM allows inference about various aspects of North Atlantic right whales. For example, we can learn which visual health metrics show the strongest links to underlying health, whether geographic regions (and thereby human activity) have an impact on survival, and at which times of the year movement to certain zones occurs. While this joint model may seem complex at first sight, each of the individual hierarchical levels are relatively straightforward.

%----------------------------------------
\subsection{\label{S.CRM}Capturing heterogeneity with random effects}

SSMs can account for additional dependencies and heterogeneity in parameter values with random effects. This feature has been used to incorporate individual variation in capture-recapture models \citep{Royle-2008, King-2012}. Capture-recapture models, such as the Cormack-Jolly-Seber model,  are often used to estimate survival probabilities and gain insight on the factors that may affect survival. They model data, where individuals are uniquely identifiable via artificial (e.g., rings) or natural marks (e.g., coloring) \citep{King-2012}. One of the first applications of SSMs to such data was by \citet{Royle-2008} to demonstrate how to model variation in survival and capture probabilities.   \citet{Royle-2008} applied the model to a 7-year study of European dippers (\textit{Cinclus cinclus}).

\citet{Royle-2008} presents a SSM parametrization of a Cormack-Jolly-Seber model, where the observation $y_{i,t}$ represents whether individual $i$ was capture during the $t^{th}$ sampling occasion (i.e., $y_{i,t} = 1$ means the individual was captured at time $t$) and the state $z_{i,t}$ describes whether individual $i$ is dead or alive at time $t$ (i.e., $z_{i,t} = 1$ means the individual was alive at time $t$). At the time of first capture, $f_i$, the state is considered fixed: $z_{i,f_i} = 1$. Afterward, the process and observation equations are both Bernoulli trials, representing the survival and capture processes for each individual $i$:
\begin{linenomath*}
\begin{align}
    z_{i,t} &\sim \text{Bernoulli}(\phi_{i,t-1}z_{i,t-1}), & f_i < t \leq T, \label{E.crmp}\\
    y_{i,t} &\sim \text{Bernoulli}(p_{i,t} z_{i,t}), & f_i < t \leq T, \label{E.crmo}
\end{align}
\end{linenomath*}
where $T$ is the total number of sampling occasions, and $\phi_{i,t}$ is the probability of survival of individual $i$ over the interval $(t, t+1)$ if the individual was alive at time $t-1$, and $p_{i,t}$ is the probability of capturing individual $i$ during the $t^{th}$ sampling occasion if it is alive. These probabilities are multiplied to the state values. Thus, an individual's probability of surviving to time $t$ becomes 0 if the animal was dead at time $t-1$ (i.e., $z_{t-1}=0$) regardless of the value of $\phi_{i,t-1}$, which means that once the animal is dead it remains dead for the rest of the time series. Similarly, the probability of being captured at time $t$ becomes 0 if the individual is dead at that time.

We could simplify the model by having a single overall survival probability (i.e., $\phi_{i,t} = \phi$) and a single capture probability ($p_{i,t} = p$). However, differences between sampling occasions and individuals (e.g., due to variations in environmental and body conditions) often warrant for temporal and individual variations in survival and capture probabilities. \citet{Royle-2008} modeled the variations in these probabilities as follows:
\begin{linenomath*}
\begin{align}
    \text{logit}(\phi_{i,t}) &= b_t + \beta_i, & \beta_i &\sim \text{N}(0, \sigma^2_{\beta}) \\
    \text{logit}(p_{i,t}) &= a_t + \alpha_i,  & \alpha_i &\sim \text{N}(0, \sigma^2_{\alpha})
\end{align}
\end{linenomath*}
where $a_t$ and $b_t$ are the fixed temporal effects (i.e., effects associated with each sampling occasion), $\alpha_i$ and $\beta_i$ are the latent individual effects, and $\sigma_{\alpha}^2$ and $\sigma_{\beta}^2$ are the variances for the random effects. Just as for generalized linear model, the logit link function ensures that probability parameters stay between 0 and 1. The fixed temporal effects require that we estimate $(T-1) + (T-2)$ additional parameters \citep[for details, see][]{Royle-2008}. In contrast, the individual random effects allow one to model heterogeneity in survival and capture probabilities with only two additional parameters.

SSMs are now commonly used to model capture-recapture data because their mechanistic structure allows one to incorporate additional complexity \citep{King-2012}. Using random effects to model variation in parameter values can be used in many other ecological applications.

%----------------------------------------
\subsection{\label{S.HMM}Modeling discrete state values with hidden Markov models}

Hidden Markov models (HMMs) are a special class of SSMs, where the states are discrete rather than continuous \citep[generally categorical with a finite number of possible values;][]{Langrock-etal-2012}. HMMs have gained popularity in ecology, where they are used to model capture-recapture data \citep[e.g.,][]{Choquet-Gimenez-2012, Johnson-etal-2016} and animals that switch between distinct behavioral modes \citep{Langrock-etal-2012}. Recently, \citet{McClintock-etal-2020} have demonstrated that HMMs are widely applicable in ecology. Having discrete states in a SSM becomes important when choosing fitting procedures (see Section \ref{S.discrete.states}), and thus we provide a few examples.

The two main characteristics of HMMs are: 1) each observation is assumed to be generated by one of $N$ distributions, and 2) the hidden state sequence that determines which of the $N$ distributions
is chosen at time $t$ is modeled as a Markov chain, where the
probability of being in each mode at time $t$ depends only on the
state value at the previous time step \citep{Langrock-etal-2012}. The capture-recapture model presented in Section \ref{S.CRM} is an HMM, because state $z_{i,t}$ can only have one of two discrete values: 0 if the individual is dead or 1 if alive. The state value directly affects the observation equation (Eq. \ref{E.crmo}), and the observation, $y_{i,t}$, is  generated by one of two distributions: $y_{i,t} \sim \text{Bernoulli}(0)$ if $z_{i,t} = 0$ or $y_{i,t} \sim \text{Bernoulli}(p_{i,t})$ if $z_{i,t} = 1$. As seen in the process equation (Eq. \ref{E.crmp}), the probability of being in each state at time $t$ depends only on the state value at the previous time step. This process can be viewed  as a Markov chain with the following transition probability matrix:
\begin{linenomath*}
\begin{align}
\boldsymbol{\Gamma} & = \begin{bmatrix}
    1 & 0 \\
    1-\phi_{i,t-1} & \phi_{i,t-1}
\end{bmatrix}, \label{E.tpm}
\end{align}
\end{linenomath*}
where for each individual, the probability of staying dead (i.e., $\text{Pr}(z_{i,t} = 0 |z_{i,t-1} = 0)$) is 1, that of resurrecting (i.e., $\text{Pr}(z_{i,t} = 1 |z_{i,t-1} = 0)$) is 0, that of dying (i.e., $\text{Pr}(z_{i,t} = 0 |z_{i,t-1} = 1)$) is the probability that it did not survive (i.e., $1-\phi_{i,t-1}$), and that of surviving (i.e., $\text{Pr}(z_{i,t} = 1 |z_{i,t-1} = 1)$) is $\phi_{i,t-1}$. 

In other contexts, the transition probabilities may be more flexible, allowing for transition between all states, and the SSM may include both discrete and continuous states. For example, the model presented in Section \ref{S.M.movement.model} was originally developed to model the movement of animals tracked with Argos data that switched between two behavioral modes \citep{Jonsen-etal-2005}. Instead of having a single $\gamma$ parameter that controls how correlated the steps are (e.g., Eq. \ref{e.dcrw.p}), this model has two parameters, $\gamma_{b_t}$, each one associated with one of the behavioral modes, $b_t = 1$ or $b_t = 2$. When $\gamma_1$ is close to 0 and $\gamma_2$ is close to 1, the movement path switches between tortuous and directed movement. The switch between the behavioral modes is modeled with a simple Markov chain (i.e., $\text{Pr}(b_t = j | b_{t-1} = i) = \alpha_{ij}$). While here we could allow the animal to switch back and forth between the behavioral modes (i.e., no $\alpha_{ij}$ is set to 0), the transition out of a given mode always need to sum to one  (i.e., $\sum_{j=1}^2 \alpha_{ij}= 1$ for $i, j = 1,2$).

%----------------------------------------
\section{\label{S.OTHER}SSMs as a framework for ecological time series}

Time is one of the fundamental axes that shape ecological systems \citep{Wolkovich-etal-2014} and time series are crucial to understand the complex processes and interactions that govern all aspects of ecology \citep[e.g.,][]{Boero-etal-2015,Damgaard-2019}. While SSMs have a long history in only a few fields of ecology, the breath of their applications has been expanding and could be extended to most ecological time series. SSMs provide a framework that can be used to understand the mechanisms underlying complex ecology systems and handle the large uncertainties associated with most ecological data and processes.

SSMs have been increasingly used in plant ecology. \citet{Damgaard-2012} showed the usefulness of using SSMs to analyze plant cover data collected through quadrats (specifically through pin-point methods). \citet{Bell-etal-2015} demonstrated how SSMs could be used to estimate canopy processes (e.g., conductance and transpiration) using imperfectly monitored stem sap flux data. \citet{Clark-etal-2011} used the SSM approach to model the growth, fecundity, and survival of more than 27,000 individual trees. They showed how these processes are linked with light competition and spatiotemporal variation in climate.

SSMs are now used for paleoecological research. For example, \citet{Tome-etal-2020} used SSMs to identify the drivers of changes in the mass and diet of a small mammal during the late Pleistocene. They use three separate linear Gaussian SSMs to model temporal changes in mass (as estimated from molar size) and in two stable isotopes (extracted from jaw bone collagen) as responses to each other and of a set of covariates related to climate (e.g., maximum temperature) and community structure (e.g., species richness). \citet{Einarsson-etal-2016} developed a SSM for sediment core data. One process equation modeled the change in abundance of midges egg capsule. The other modeled the change in pigment concentration characterizing potential resources (e.g., diatoms). These process equations modeled the abundance of each group with a Gompertz population model (similar to Section \ref{S.M.simple.pop}), modified to add the effect of the other group's abundance at the previous time step. The measurement equations modeled sediment mixing and its associated uncertainty. They use their model to show that the cyclic fluctuations in midges are likely driven by consumer-resource (a.k.a exploiter-victim or predator-prey) interactions. 

Detecting cyclicity in ecological time series can be challenging due to temporal autocorrelation, and \citet{Louca-Doebeli-2015} showed that SSMs can outperform statistical tests for cyclicity. For example, they showed that simpler models would often lead to erroneous conclusions that cycles are present; while the SSM generally had an appropriate 5\% rate of Type I errors, simpler tests had rates as high as 79\%. The midge example of \citet{Einarsson-etal-2016} further demonstrates the usefulness of SSMs to identify the mechanisms behind cycles in ecological time series. Similar SSMs have been used to investigate fluctuations in other ecological fields \citep[e.g., host-parasitoid systems,][]{Karban-and-deValpine-2010}.

SSMs have been used in ecosystem ecology and biogeochemisty. For example, \citet{Appling-etal-2018} used a SSM to model changes in oxygen concentration in an aquatic ecosystem as a function of three important processes: ecosystem's gross primary production, respiration, and gas exchange rate with the atmosphere. The process equation of \citet{Appling-etal-2018} predicts the oxygen concentration at time $t$ as a function of its previous value and its instantaneous rate of change (similar to the model in Section \ref{S.cont.mov.model}). The rate of change is modeled through a mechanistic equation, which sums the three processes of interest. The study compared various versions of the model, including versions that were not SSMs (i.e., versions without measurement error or process stochasticity), and showed that the best SSM formulation significantly improved the accuracy and reduced the bias of estimates of gross primary productivity, respiration, and gas exchange. In some cases, the magnitude of the bias of the SSM was half as large as that of simpler models.  The study of \citet{Jia-etal-2011} is one of the many examples of applications of SSMs in soil science. \citet{Jia-etal-2011} used linear SSMs with normally distributed error to model the effects of elevation and the physical and chemical properties of soil (e.g., clay content and organic carbon) on the total net primary productivity of managed grasslands. They showed that the SSMs described the spatial patterns of soil total net primary productivity better than classical regression methods.

The term SSM has been used broadly in ecology to represent various types of hierarchical models with complex dependence structure. In particular, the term has been used for occupancy models that are based on capture-recapture SSMs similar to the one described in Section \ref{S.CRM} \citep[e.g.,][]{Kery-etal-2009, Mordecai-etal-2011}. While they have similar structure, many of them lack the specific temporal autocorrelation in process equation that we generally ascribe to SSM (Fig. \ref{fig:structure}a) and may be better thought as a related, but different, type of hierarchical model. For some of these models, it may be worth adding the Markovian dependence of the state in the process equation. However, to our knowledge, there is no studies that compare these related hierarchical models to SSMs.

The ubiquity of SSMs in ecology may have been obscured as some complex SSMs that combine various statistical techniques have not been identified as SSMs. For example, \citet{Thorson-etal-2016} present a joint species distribution model that has temporal dynamics. Although not called a SSM, their model has the essential structure of a SSM (Fig. \ref{fig:structure}A). We view their model as a Gompertz SSM (similar to Section \ref{S.M.simple.pop}) combined to dynamic factor analysis to reduce dimensions and Gaussian random fields to account for spatial autocorrelation. This complex multi-species model was used to demonstrate that the spatiotemporal patterns of butterfly from the same genus were significantly correlated and to identify dominant patterns in community dynamics of marine fish.

The complexity of SSMs may incite ecologists to ask: could we use a simpler alternative? In fisheries science, early papers on SSMs showed that they were particularly superior to simpler models when both the process variance and the observation error are large \citep[e.g.,][]{deValpine-Hastings-2002}. When one of the sources of stochasticity is small, and the model dynamics are not too complex, simpler models that account for either just the process variance or observation error give adequate results \citep{deValpine-Hastings-2002}. A key point, however, is that the simpler model performs adequately only if the model is well specified with regards to which source of stochasticity is most important. Thus, using a process variance-only model, is only suitable if we are certain that the observation error is negligible. Similarly, using an observation error-only model, is only suitable if we are certain that the process variation is small. While simpler alternatives can be adequate in some contexts \citep[see Chapter 11 of][]{Bolker-2008}, many studies have shown that SSMs provided  better inference than easier models \citep[e.g.,][]{Jamieson-Brooks-2004, Jia-etal-2011, Louca-Doebeli-2015, Appling-etal-2018}. For example, \citet{Linden-Knape-2009} showed that, unlike SSMs, simpler models often had unreliable point and uncertainty estimates for environmental effects, and that the 95\% confidence intervals excluded the true simulated value much more than 5\% of the time (up to 30\%). They showed that the SSMs always outperformed the simpler alternatives. As such, we believe that SSMs, and their extensions, should be a default statistical modeling technique for many ecological time series. In the rest of the paper, we provide the tools that allow ecologists to apply these complex models adequately.

% ***************************************************
\section{\label{S.Fitting}Fitting SSMs}

The goals of fitting a SSM to data include estimating the parameters, $\boldsymbol{\theta}$, the states, $\mathbf{z}$, or both. In ecology, we regularly need to
estimate both, as we rarely know the value of $\boldsymbol{\theta}$ \textit{a priori} and estimating the states is often a primary goal of the analysis. In movement
ecology, researchers often fit SSMs similar to that described in
Section \ref{S.M.movement.model} because the states provide better
estimates of the true locations of the animal than the data. In the
SSM literature, a  distinction is often drawn between three different
types of state estimation processes based on the amount of observations used to inform the estimates \citep{Shumway-Stoffer-2016}.
Using all of the observations, $y_{1:T}$, to estimate the states is
referred to as `smoothing'. Smoothing is common with ecological SSMs,
as we often have the complete dataset in hand when we start the
analysis. We denote the smoothed state estimate as
$\hat{z}_{t|1:T}$, with the subscript $t|1{:}T$ identifying that 
the state at time $t$ is estimated using the observations from time
$1$ to $T$. In the original engineering application and in other fields, states are often estimated while data continues to be collected, so only observations
up-to and including time $t$, $y_{1:t}$, are used to estimate the
state $\hat{z}_{t|1:t}$. This ubiquitous estimation procedure is
referred to as `filtering'. Finally, we can use a subset of the
observations that ends $s$ time steps before time $t$, $y_{1:t-s}$,
to predict the state at time $t$, $\hat{z}_{t|1:t-s}$, a procedure 
we refer to as `forecasting'. A common forecast is the
one-step-ahead prediction, $\hat{z}_{t|1:t-1}$, which is also
used within fitting algorithms (Appendix S2) and to validate models (Section \ref{S.Diagnostics}). While these three types of state estimation processes are useful, the uncertainty associated with the state estimates tend to decrease for processes that use more observations \citep[e.g.,][]{Shumway-Stoffer-2016}. 

The states are random variables, and thus have probability distributions. The states are sometimes referred to as random effects or latent variables. The fundamental differences in the procedures used to estimate states, as opposed to parameters (see Section \ref{S.Freq}), means that although we use estimation as an all-purpose term for both states and parameters, state estimation procedures are often referred as prediction, even when smoothing and filtering are used. The inferences about the states can include a variety of summary measures of their probability distributions. Above, the state estimates (e.g., $\hat{z}_{t|1:T}$) referred to point estimates such as the expected value. However, one can also calculate interval estimates (e.g., 95\% confidence intervals) and single measures of uncertainty (e.g., standard deviations or variances).

Methods for fitting SSMs can be divided into the two main
inferential approaches: frequentist and Bayesian. These approaches differ in their philosophies, see \citet{Bolker-2008} for a discussion. In brief, frequentist methods determine the probability of the data for a set of particular conditions (i.e., the hypothesis is fixed, but the data have a probability distribution). In contrast, Bayesian methods determine the probability that particular conditions exist given the data-at-hand (i.e., the data are fixed, but the hypothesis/parameters have probability distributions). The Bayesian approach requires the specification of prior beliefs for these distributions. Because of the early development of Bayesian computational methods for hierarchical models, historically it was easier to fit complex SSMs with a Bayesian approach and frequentist methods were limited to simple models \citep{DeValpine-2012}. As we will show, this is no longer true. There are now many accessible methods that allow to fit complex SSMs with a frequentist approach (e.g., see Sections \ref{S.Laplace}-\ref{S.Iterative.Filtering}). Thus, researchers can choose to work with their favored philosophical approach and/or based on the advantages of the algorithms available within each approach (see Section \ref{S.method.comp}). 

In terms of fitting procedures, frequentists maximize the likelihood, while Bayesians focus on the posterior density. As we show below, despite these differences, both approaches involve high-dimensional integration, which is at the crux of the difficulties associated with fitting SSMs to data. The many tools developed for fitting SSMs are essentially different solutions to this high-dimensional integration problem.

%---------------------------------------------------------
\subsection{Frequentist approach}
 \label{S.Freq}
 
When we fit a SSM with a frequentist approach, we search for the parameter values that maximize the likelihood, a method called maximum likelihood estimation with the resulting estimates called maximum likelihood estimates (MLEs). For our toy SSM (Eqs. \ref{E.state.NDLM}-\ref{E.obs.NDLM}), the joint likelihood for $\boldsymbol{\theta}$ and $\mathbf{z}_{1:T}$ would be defined as:
\begin{linenomath*}
\begin{align}
  L_{\textsc{j}}(\boldsymbol{\theta}, \mathbf{z}_{1:T} | \mathbf{y}_{1:T})=\prod_{t=1}^T g(y_t | z_t, \boldsymbol{\theta}_o) f(z_t | z_{t-1}, \boldsymbol{\theta}_p),
  \label{E.Joint.Likelihood}
\end{align}
\end{linenomath*}
where $T$ is the length of our time series and $\boldsymbol{\theta}$ is a vector of (unknown) model parameters that contains the parameters for the process equation, $\boldsymbol{\theta}_p$, and the observation equation, $\boldsymbol{\theta}_o$, and in this example the initial state, $z_0$. Maximizing the joint likelihood with respect to both parameters and the states is challenging. Instead, one can use a process with two interrelated steps, each focused on estimating either the parameters or the states.

To estimate the parameters, we maximize the marginal likelihood, $L_{\textsc{m}}(\boldsymbol{\theta} | \mathbf{y}_{1:T})$:
\begin{linenomath*}
\begin{align}
  \boldsymbol{\hat{\theta}} = \argmax_{\boldsymbol{\theta} \in \boldsymbol{\Theta}} L_{\textsc{m}}(\boldsymbol{\theta} | \mathbf{y}_{1:T}),
  \label{E.argmax.par}
 \end{align}
\end{linenomath*}
 where
\begin{linenomath*}
 \begin{align}
  L_{\textsc{m}}(\boldsymbol{\theta} | \mathbf{y}_{1:T})=\int  L_\textsc{j}(\boldsymbol{\theta}, \mathbf{z}_{1:T} | \mathbf{y}_{1:T}) \text{d} \mathbf{z}_{1:T}.
  \label{E.Marginal.Likelihood}
 \end{align}
\end{linenomath*}
Here, the key is that we integrate out the hidden states and thus have a function that only depends on the observations. The parameter estimates that result from maximizing the marginal likelihood have desired statistical properties \citep[consistency and asymptotic normality, see][]{Douc-etal-2011, DeValpine-2012}, where the estimates are anticipated to improve with increasing sample size. Such properties would be hard to achieve when maximizing the joint likelihood, because the number of states to estimate generally increase with the number of observations.

To estimate the hidden states, we can use the conditional distribution of the states given the observations and the estimated parameter values, for example:
\begin{linenomath*}
 \begin{align}
 p(\mathbf{z}_{1:T}|\mathbf{y}_{1:T},\boldsymbol{\hat{\theta}}) = \frac{L_\textsc{j}(\mathbf{z}_{1:T} | \mathbf{y}_{1:T},\boldsymbol{\hat{\theta}})}{\int  L_\textsc{j}(\mathbf{z}_{1:T} | \mathbf{y}_{1:T},\boldsymbol{\hat{\theta}}) \text{d} \mathbf{z}_{1:T}},
 \label{E.cond.dist.state}
 \end{align}
\end{linenomath*}
where $L_\textsc{j}(\mathbf{z}_{1:T} | \mathbf{y}_{1:T},\boldsymbol{\hat{\theta}})$ is similar to the right-hand side of Eq. \ref{E.Joint.Likelihood}, except that we use the MLEs for the parameters. Conditional distributions of the states, in particular the filtering distributions ($p(z_{t}|\mathbf{y}_{1:t}, \boldsymbol{\hat{\theta}})$, see Appendix S2 for an example), are at the base of filtering methods, such as the Kalman filter (Section \ref{S.Kalman.Filter}) and particle filter (Section \ref{S.Iterative.Filtering}). The means and variances of filtering densities can provide good point estimates and measures of uncertainty for state values (Appendix S2). As an approximation of the state estimates, one can also maximize $L_\textsc{j}(\mathbf{z}_{1:T} | \mathbf{y}_{1:T},\boldsymbol{\hat{\theta}})$ with respect to $\hat{\mathbf{z}}_{1:T}$:
\begin{linenomath*}
 \begin{align}
 \hat{\mathbf{z}}_{1:T} = \argmax_{\mathbf{z}_{1:T} \in \mathbf{Z}^T}L_\textsc{j}(\mathbf{z}_{1:T} | \mathbf{y}_{1:T}, \boldsymbol{\hat{\theta}}),
 \label{E.argmax.states}
 \end{align}
\end{linenomath*}
where $\mathbf{Z}^T$ is the set of all possible values
for the states. This maximization treats the states as if they were equivalent to parameters in an ordinary likelihood \citep[see][for more details]{Aeberhard-etal-2018} and is often used when the marginal likelihood is estimated with the Laplace approximation (see Section \ref{S.Laplace}). While Eq. \ref{E.argmax.states} treats the parameters as known when estimating the states, one can propagate the estimation variability when reporting the state estimate variance \citep[e.g., see \texttt{TMB} function \texttt{sdreport},][]{Kristensen-etal-2016}.

The marginal likelihood used to estimate the parameters, and thus the states, requires the computation of the high-dimensional integral found in Eq. \ref{E.Marginal.Likelihood}. This computation is difficult to achieve for most SSMs and the frequentist inference methods discussed below are different ways to either evaluate the marginal likelihood (e.g., Kalman filter) or to approximate it (e.g., Laplace and simulation-based approximations).

\subsubsection{Kalman filter \label{S.Kalman.Filter}}

For simple linear SSMs with Gaussian errors (i.e., NDLMs), the state estimates and marginal likelihood can be directly calculated using the Kalman filter \citep{Kalman-1960}. The Kalman filter provides an algorithm that, using only elementary linear algebra operations, sequentially updates the filtering mean and variance of the states \citep{Harvey-1990,Durbin-and-Koopman-2012}. While the Kalman filter was developed to estimate the state values for models with known parameter values, its output can be used to evaluate the marginal likelihood and thus to find the MLE. The Kalman smoother is an analogous algorithm that uses backward recursion in time to obtain the mean and variance of each smoothing distribution \citep[i.e., distribution of $z_{t|1:T}$;][]{Harvey-1990,Durbin-and-Koopman-2012}. See Appendix S2 for a detailed example of the Kalman filter as applied to our toy model.

In Appendix S1: Section S1 1.3.1, we demonstrate how to use the \texttt{R} package \texttt{dlm} \citep{petris2010} to perform Kalman filtering and smoothing, as well as forecasting. It can also be used to find MLEs of unknown fixed parameters. The package is flexible enough to allow univariate and multivariate NDLMs, accounting for constant or time-varying distributions of states and observations. More details about Kalman filter and smoother and \texttt{dlm} can be found in \citet{DLMwR} and \citet{petris2010}. See also Chapter 6 of  \citet{Shumway-Stoffer-2016} for description of filtering, smoothing, forecasting and maximum likelihood estimation.

The Kalman filter is among the most broadly used algorithms to fit SSMs to ecological data. For example, it has been used in population ecology \citep[e.g.,][]{Dennis-etal-2006}, movement ecology \citep[e.g.,][]{Johnson-etal-2008}, community ecology \citep{Ives-etal-2003}, and plant ecology \citep{Hooten-etal-2009}. The main advantage of the Kalman filter is that it is fast and easy to calculate \citep{DeValpine-2012}. In addition, unlike most other methods that provide an approximation of the likelihood, the Kalman filter provides an exact evaluation of the marginal likelihood for linear and Gaussian SSMs \citep[e.g., toy model;][]{DeValpine-2002}.

While the Kalman filter is an important algorithm for fitting SSMs to data, it does not work with nonlinear and non-Gaussian SSMs. Approximate
techniques based on the Kalman filter are available for linear models whose observations follow an exponential family distribution \citep[e.g., Poisson, see][Ch.\ 9]{Durbin-and-Koopman-2012}. Other approximate filtering and smoothing methods based on the Kalman filter, such as the extended Kalman filter and the unscented Kalman filter \citep[e.g.,][Ch.\ 10]{Durbin-and-Koopman-2012} are useful for some nonlinear and/or non-Gaussian SSMs. Such related methods have been used in ecology \citep[e.g.,][]{Einarsson-etal-2016}. However, for more complex, nonlinear, and non-Gaussian models, one must use one of the methods described below.

\subsubsection{Laplace approximation methods}
\label{S.Laplace}

The Laplace approximation is a commonly used tool for obtaining an approximation of the marginal likelihood of a SSM \citep{Fournier-etal-2012,Kristensen-etal-2016}. The general idea is that if the marginal likelihood (Eq. \ref{E.Marginal.Likelihood}) is a well-behaved unimodal function, it can be approximated with a Normal density function. We can use this approximation to find the MLE. For a given set of parameter values, the Laplace approximation of the marginal likelihood requires the maximization of the joint likelihood (Eq. \ref{E.Joint.Likelihood}) with respect to the states. Thus, the parameter estimation process also returns an approximation of the state estimates. See Appendix S2 for detail.

This method is flexible, and a variety of SSMs can be fitted using the Laplace approximation. However, the method assumes that the states can be locally-approximated with a Gaussian distribution, which means that the states are assumed to have an unimodal distribution. Because the method uses the second derivative of the log likelihood (Appendix S2), we cannot use the Laplace approximation with categorical states or other state distributions that are not twice differentiable. An important advantage of the Laplace approximation, over the simulation-based approaches described below, is the speed at which SSMs are fitted to data \citep[see][]{AugerMethe-etal-2017}. Many software use the Laplace approximation approach \citep[e.g.,][]{Fournier-etal-2012}. We demonstrate in Appendix S1: Section S1 1.3.2 how \texttt{TMB} \citep{Kristensen-etal-2016} is a particularly useful \texttt{R} package for SSMs. The Laplace approximation has been used in ecology, including in movement ecology \citep[e.g.,][]{AugerMethe-etal-2017} and fisheries science \citep[e.g.,][]{Aeberhard-etal-2018}.

%-------------------------------------------------------------
\subsubsection{Sequential Monte Carlo methods}
\label{S.Iterative.Filtering}

Monte Carlo methods can be used to estimate the states and evaluate the integral needed to obtain the marginal likelihood. Monte Carlo methods are computer intensive sampling procedures that generate random samples from specific probability distributions, which can then be used to evaluate integrals. While in this section we discuss Monte Carlo methods in the context of a frequentist inference approach, we will see in Section \ref{S.Bayesian} that Monte Carlo methods are commonly used for Bayesian inference. 

Sequential Monte Carlo methods, also referred to as particle filters, approximate the filtering distribution through simulated sampling \citep{DeValpine-2012}. In the context of SSMs fitted with a frequentist approach, these Monte Carlo methods generally sample the state space by generating samples using the process equation and weighting the samples with the observation equation. Sequential importance sampling \citep{Doucet-etal-2001} is a general procedure that can be used to generate $N$ time series of the states, referred to as particles, and using their weighted average as the state estimates (see Appendices S1-S2 for detail). However, sequential importance sampling is impractical for even moderately long time series (e.g., $T$=20) because only a small proportion of the $N$ randomly generated particles are generally supported by the observations. The reduced support for many of the particles, known as particle depletion,
is a serious problem with sequential importance sampling that leads to state estimates with unacceptably large variances.

The bootstrap filter \citep{Gordon-etal-1993}
is a procedure designed to remedy particle depletion. The bootstrap filter assesses the weight of a particle through time and iteratively removes
particles with low weights and replaces them with duplicates of
particles with higher weights. There are various algorithms for the bootstrap filter, see Appendices S1 and S2 for an example. While simple bootstrap filters can reduce particle depletion, they do not completely solve the problem particularly for long time series. There are various remedies aimed at reducing
particle depletion \citep{DeValpine-2012}, including more sophisticated importance sampling distributions that include information from the observations
\citep{Pitt-Shephard-1999} or changing the resampling methods
\citep{Liu-Chen-1998}. Sequential Monte Carlo methods are also used for Bayesian inference and these methods are often built to reduce particle depletion \citep[e.g., particle Markov chain Monte Carlo methods,][]{Andrieu-etal-2010, Michaud-etal-2020}.

Sequential Monte Carlo methods, such as sequential importance sampling, can be used to estimate the likelihood. However, the likelihood maximization required for frequentist inference comes with additional challenges (e.g., to maximize the likelihood, one must explore what is often a complicated likelihood surface). In principle, it is possible to use a general-purpose optimization algorithm such as Nelder-Mead to maximize the likelihood computed by a simple particle filter. However, such an approach is usually prohibitively expensive. In addition, the stochastic ingredients of a particle filter make each of its runs different, making it hard to identify the precise peak of the likelihood surface \citep{DeValpine-2012}. Several methods have been proposed to overcome this difficulty \citep{DeValpine-2012, Michaud-etal-2020}.

Iterated filtering is an attractive method for maximizing the likelihood using particle filter \citep{Ionides-etal-2015}. This method repeatedly applies the particle filter but perturbs the fixed parameters of the model at each observation time step.
These random perturbations enhance performance and forestall particle depletion by continually re-injecting random variability into the filter. However, because it applies artificial perturbations to parameters, iterative filtering is not learning about the model of interest (i.e., model with fixed parameters), but about a modified model (i.e., model where fixed parameters have been transformed into state variables).
Therefore, as filtering iterations proceed, one gradually cools (i.e., reduces the magnitude of) the artificial perturbations, so that the modified model approaches the model of interest as the iterations proceed.
Because statistical inference hinges on identification of the global likelihood maximum, it is usually advisable to perform many independent iterative filtering computations, starting from widely dispersed starting points.  See Appendix S2 for more detail.
Iterative filtering, and other similar sequential Monte Carlo methods, can be easily implemented using the \texttt{R} packages \texttt{pomp} and \texttt{nimble} \citep[see Appendix S1: Section S1 1.3.4;][]{King-etal-2016, Michaud-etal-2020, deValpine_et_al:2017}. 

The main advantage of sequential Monte Carlo methods is that they are flexible, thus can be used to conduct inference on any SSM \citep{Michaud-etal-2020}. Frequentist sequential Monte Carlo methods have been used in ecological fields such as bioenergetics \citep[e.g.,][]{Fujiwara-etal-2005} and movement ecology \citep[e.g.,][]{Breed-etal-2012}. The main disadvantage of sequential Monte Carlo methods is that they can be computationally expensive.

\subsubsection{Other methods}

The methods described above represent, in our view, the most
commonly used methods to fit SSMs to ecological data in a frequentist
framework. These methods are associated with comprehensive \texttt{R}
packages that facilitate their implementation. However, many other
methods exist \citep[see][for a review of frequentist methods]{DeValpine-2012}. Of note, \citet{Kitagawa-1987} provided a general algorithm for non-Gaussian SSMs similar to the Kalman
filter, but that approximates the non-normal distributions by discretizing them (e.g., through piecewise linear functions). It can be viewed as discretizing the continuous
state space and reformulating the model as a HMM
\citep{Pedersen-etal-2011}. \citet{deValpine-Hastings-2002}
demonstrated how flexible this approach was to fit nonlinear
non-Gaussian population dynamics models. The main advantages of this approach are that it can be computationally efficient for models with a few state dimensions and does not require Monte Carlo methods \citep{DeValpine-2012}. This approach appears particularly promising for population modeling, where the states are counts, and thus the state space is already discretized \citep{Besbeas-Morgan-2019}. \citet{Pedersen-etal-2011} demonstrated that while this method is general and can provide results similar to the Laplace approximation and Bayesian methods, it is computationally limited to problems with only a few state dimensions. This limitation arises from the curse of dimensionality, where even if each dimension has manageable number of cells (e.g., 1,000 cells), the number of values needed to be stored become impractical as the number of dimension increases \citep[e.g., three-dimension would results in $1,000^3 = 10^9$ cells, see][]{DeValpine-2012}.

%---------------------------------------------------------
\subsection{Bayesian framework}
\label{S.Bayesian}

When we fit a SSM with a Bayesian approach, the function of interest (also known as the target distribution) is the posterior distribution for the states and parameters given the observations:
\begin{linenomath*}
\begin{align}
   p(\boldsymbol{\theta},\mathbf{z}_{1:T}|\mathbf{y}_{1:T},\boldsymbol{\lambda}) = \frac{ L_{\textsc{j}}(\boldsymbol{\theta}, \mathbf{z}_{1:T} | \mathbf{y}_{1:T}) \pi(\boldsymbol{\theta} |\boldsymbol{\lambda})}{\int \int L_{\textsc{j}}(\boldsymbol{\theta}, \mathbf{z}_{1:T} | \mathbf{y}_{1:T})\pi(\boldsymbol{\theta} |\boldsymbol{\lambda}) \text{d}\mathbf{z}_{1:T} \text{d}\boldsymbol{\theta}},
   \label{E.posterior}
\end{align}
\end{linenomath*}
where $L_\textsc{j}(\boldsymbol{\theta}, \mathbf{z}_{1:T} | \mathbf{y}_{1:T})$ is the joint likelihood (i.e., $p(\mathbf{y}_{1:T} | \boldsymbol{\theta}, \mathbf{z}_{1:T})$, see for example Eq. \ref{E.Joint.Likelihood}), and $\pi(\boldsymbol{\theta}|\boldsymbol{\lambda})$ is the prior distribution(s) for the parameters with fixed hyperparameters, $\boldsymbol{\lambda}$. Eq.  \ref{E.posterior} is an application of Bayes' theorem ($p(\theta | \mathbf{y}) = \frac{p(\mathbf{y}| \theta) p(\theta)}{p(\mathbf{y})}$) and the denominator of Eq. \ref{E.posterior} represents the probability of the data (i.e., the marginal likelihood, which is the probability of the data for all possible values of the states and parameters). In Bayesian analyses, both the states, $\mathbf{z}_{1:T}$, and what we have been referring to as fixed parameters, $\boldsymbol{\theta}$, are considered random variables. The posterior distribution is a complete characterization of these random variables given the data and prior information. As such, the first inferential goal of a Bayesian analysis is often to evaluate the posterior distribution. While point estimates for the parameters and the states are not necessarily the primary goal of a Bayesian analysis, they can be obtained by summarizing the center of the posterior distribution (e.g., mean or mode of the posterior distribution). Similarly we can use the posterior distribution to obtain interval estimates and single measures of variation.

As for the frequentist framework, the fitting procedures are complicated by high-dimensional integrals and it is common to avoid calculating the integral and the posterior distribution explicitly. Instead, quantities of interest are generally approximated using Monte Carlo methods (see also Section \ref{S.Iterative.Filtering}), where large samples of states and parameters are randomly drawn from the posterior distribution. For example, one can approximate the point estimate of a parameter with the sample mean of the draws from the posterior distribution (often referred as the posterior mean). Simulating independent draws from Eq. \ref{E.posterior} is typically impossible. However, there are various algorithms that can approximate the posterior distribution with large samples of dependent draws. In particular, Markov Chain Monte Carlo (MCMC) methods are a broad class of algorithms that obtain samples from the target distribution (here the posterior distribution Eq. \ref{E.posterior}), by sampling from a Markov chain rather than sampling from the target itself. This Markov chain needs to have an invariant distribution (i.e., the probability distribution remains unchanged as samples are drawn) equal to the target distribution \citep{Geyer:2011}, a quality which is dependent on the initial condition of the chain and the transition probabilities, and relates to the importance of chain convergence as a diagnostic in MCMC sampling. MCMC algorithms fall into two broad families: Metropolis-Hastings samplers (which include Gibbs samplers) and Hamiltonian Monte Carlo.

\subsubsection{Metropolis-Hastings samplers}
\label{S.Metropolis-Hastings}

Metropolis-Hastings samplers are at the base of most MCMC algorithms used to sample the posterior distribution in a Bayesian analysis. Metropolis-Hastings samplers are iterative algorithms that construct an appropriate Markov chain to sample the target distribution. The general idea is that for each step $j$ of the chain, we use a proposal distribution to generate a candidate value for the variable of interest (e.g., a parameter value). The probability that this candidate value is used for that step rather than the previous value of the chain is based on the relative fit of the model with that candidate value compared to the previous value of the chain (see Appendix S2 for detail). 

In the context of SSMs, we have a multivariate posterior distribution for the states and the parameters. Using Metropolis-Hastings algorithms to sample for more than one random variables is complex, but there are various implementation tools to do so. For example, for each iteration $j$ of the chain, one can first sample sequentially all parameter values, and then sequentially sample the state values \citep[][see also Appendix S2]{Newman-etal-2014}. If groups of variables are related, they can be sampled simultaneously from a multivariate distribution rather than sequentially. In practice, states and parameters are often correlated, and thus it may be difficult to implement an efficient MCMC sampler that does not require very long simulations before convergence \citep{Newman-etal-2014}.

Gibbs samplers are commonly-used Metropolis-Hastings samplers for multivariate distributions, where the proposal distributions are conditional distributions of the target distribution and thus the candidate values are always accepted \citep[][see also Appendix S2]{Geyer:2011}. For NDLMs, the entire sequence $\mathbf{z}_{0:T}$ can be simulated at once from its conditional distribution, given the data $\mathbf{y}_{1:T}$ and the time-invariant parameter $\boldsymbol{\theta}$, using the Forward Filtering Backward Sampling algorithm described in \citet{Carter+Kohn:1994}. The Forward Filtering Backward Sampling algorithm can also be used to conduct inference for the SSMs that are conditionally linear and Gaussian. However, Gibbs samplers for nonlinear and non Gaussian models often require sampling from each conditional distribution sequentially  \citep[see chapter 4.5 of][for an overview]{Prado-West-2010}. A drawback of this particular Gibbs sampler design is that consecutive draws of $z_{0:T}^{j}$ and $z_{0:T}^{j-1}$ tend to be highly correlated, slowing the convergence and deteriorating the quality of the Monte Carlo approximations. Despite these drawbacks, Metropolis-Hasting samplers, including Gibbs samplers, are commonly used to fit ecological SSMs because they are flexible and freely available software to implement these algorithms have been available since the 1990s \citep{Meyer-Millar-1999}. They have been used to fit many of the original models described in Section \ref{S.examples}, including the population models of \citet{viljugrein2005}, the movement model of \citet{Jonsen-etal-2005}, the health and survival model of \citet{Schick-etal-2013}, and the capture-recapture model of \citet{Royle-2008}. 

Combining sequential Monte Carlo methods (Section \ref{S.Iterative.Filtering}) within MCMC algorithms can help alleviate some of the efficiency problems produce by generic MCMC algorithms \citep{Michaud-etal-2020}. In these combined algorithms, a sequential Monte Carlo algorithm draws the states, while a MCMC algorithm draw the parameters. Particle MCMC methods \citep{Andrieu-etal-2010} are particularly useful for SSMs \citep{Michaud-etal-2020}. Some particle filters, such as the bootstrap filter (Section \ref{S.Iterative.Filtering} and Appendix S2), can return unbiased estimates of the marginal likelihood (Eq. \ref{E.Marginal.Likelihood}). At each iteration $j$, a particle MCMC algorithm will estimate the marginal likelihood and use it to draw a full state sequence (i.e., one sample particle will be used for $\mathbf{z}_{1:T}^j$). While particle MCMC may still suffer from poor mixing when the likelihood estimates are highly variable, these algorithms tend to reduce the correlations between successive draws of the states \citep{Michaud-etal-2020}. Custom-made particle MCMC algorithms have been used to fit different ecological SSMs, including population models \citep[e.g.,][]{Knape-deValpine-2012, White-etal-2016} and complex models for range expansion \citep{Osada-etal-2019}. The recent implementation of such algorithms in \texttt{R} packages such as \texttt{pomp} and \texttt{nimble} will facilitate their uptake \citep{Michaud-etal-2020}.

There are a few important general Bayesian software and \texttt{R} packages that can be easily used to fit ecological SSMs using Metropolis-Hastings samplers. Generating draws from the posterior distributions can done using software from the BUGS \citep[Bayesian analysis Using Gibbs Sampling, see][]{Lunn_et_al:2013} project and their associated \texttt{R} packages: \texttt{WinBUGS} can be called in \texttt{R} via \texttt{R2WinBUGS} \citep{Lunn-et-al-2000}, \texttt{OpenBUGS} via \texttt{BRugs} \citep{Lunn-etal-2009}, while \texttt{MultiBUGS} \texttt{R} interface is in development \citep{Goudie-etal-2017}. \citet{gimenez2009} provide a tutorial on how to fit ecological models (including some of the SSMs of Section \ref{S.M.simple.pop}) with \texttt{WinBUGS} in \texttt{R}. \texttt{JAGS} \citep[Just Another Gibbs Sampler;][]{Plummer:2003} is an alternative to BUGS project software that is written for UNIX, thus preferred by Mac and Linux users. \texttt{JAGS} is available through the \texttt{R} package
\texttt{rjags} \citep{rjags}. The \texttt{R} package \texttt{nimble} \citep{deValpine_et_al:2017} is a recent alternative to \texttt{JAGS} and BUGS software that is more transparent in how the sampling is performed. \texttt{nimble} allows users to write custom Gibbs samplers that perform block updating or implement a variety of other techniques including particle MCMC \citep{deValpine_et_al:2017, Michaud-etal-2020}. All these software allow one to write general models in a language based on BUGS. The user can set up the sampler in \texttt{R}, and once compiled, can use it to simulate draws to make inference about states and parameters. See Appendix S1: Sections S1 1.3.5 and S1 1.3.6 for detailed examples in \texttt{JAGS} and \texttt{nimble}.  

\subsubsection{Hamiltonian Monte Carlo}
\label{S.HMC}

An efficient alternative to Metropolis-Hastings sampling is provided by Hamiltonian Monte Carlo (HMC) methods, which have gained popularity in recent years thanks in part to their
implementation in the \texttt{Stan} software \citep{STAN:2012}. These methods are inspired by analogies drawn from physics and rely heavily on deep differential geometric
concepts, which are beyond the scope of this review. HMC can be a more efficient sampler than Metropolis-Hastings as fewer iterations are typically required and fewer rejections occur. This is achieved by the addition of a momentum variable that helps the Markov chain to remain within the typical set of the target distribution, rather than conducting random walk to explore the target distribution as is frequently done by Metropolis-Hastings samplers. Interested readers can read the introduction for ecologists by \citet{Monnahan-etal-2017} and explore the statistical details in \citet{Neal:2011} or \citet{Betancourt:2017}. Conducting inference for general SSMs via HMC is possible when all parameters and states are continuous or when the posterior distribution can be marginalized over any discrete parameters or states. Continuous distributions are required because density gradients of the target distribution are required to direct the sampling through the typical set of the target distribution \citep{Betancourt:2017, Monnahan-etal-2017}. Unlike Metropolis-Hastings samplers, HMC methods draw samples from the joint posterior distribution directly and can scale well to high dimensional spaces. General SSMs can be fitted either by defining the posterior as in Eq. \ref{E.posterior} or by marginalization over the state process to derive the posterior distribution of the time-invariant parameters only, $p(\boldsymbol{\theta} | \mathbf{y}_{1:T}, \boldsymbol{\lambda})$.  

One of the most popular software that uses Hamiltonian Monte Carlo is \texttt{Stan}, available in \texttt{R} through the package \texttt{rstan} \citep{rstan:2018}. See Appendix S1: Sections S1 1.3.7 and S1 2.3.2 for detailed examples using \texttt{rstan}. \citet{Monnahan-etal-2017} showed that \texttt{Stan} can fit ecological SSMs more efficiently than Gibbs software like \texttt{JAGS}. Although the parameterization of the SSM affects \texttt{Stan}'s efficiency, it can reduce computing time by orders of magnitude \citep{Monnahan-etal-2017}. Other advantages of \texttt{Stan} over \texttt{JAGS} include better diagnostics for when the algorithms is unable to explore the entire posterior, which could results in biased inference \citep{Monnahan-etal-2017}. The main disadvantage of HMC is that one cannot easily work with discrete parameters, which makes it harder to have SSMs with discrete latent states \citep[e.g., counts, categories;][]{Monnahan-etal-2017}. We discuss methods to work around this limitation in Section \ref{S.discrete.states}. The use of HMC is increasing in ecology \citep{Monnahan-etal-2017}, and HMC has been recently used to fit ecological SSMs \citep[e.g., in ecosystem ecology,][and in fisheries science, \citealt{Best-Punt-2020}]{Appling-etal-2018}.

\subsubsection{Other algorithms}

The algorithms and software discussed above are the most commonly used to fit SSMs to ecological data in a Bayesian framework. For a more general introduction on how to develop statistical algorithms to fit Bayesian ecological models, please refer to \citet{Hooten-Hefley-2019}. However, the development of Bayesian sampling algorithms is an active field of research. New methods, such as variational inference, appear particularly promising for fitting SSMs \citep[e.g.][]{Ong-etal-2018}. 

\subsubsection{Convergence diagnostics}
\label{S.Convergence.Diagnostic}

Regardless of the sampling method, it is important to assess whether it has reached the target posterior distribution. Convergence between multiple chains usually indicates that they have reached the invariant distribution. As such, multiple approaches have been developed to assess whether chain convergence has been achieved. In general, samples from the first iterations are discarded, as these likely occurred before the chain has reached the target distribution \citep[][but see \citealt{Geyer:2011}]{Gelman-Shirley-2011}. In the Metropolis-Hastings setting, this period is referred to as `burn-in'. A somewhat similar initial period, referred as the `warm-up', is discarded with HMC. Then, as a first step, convergence within and between chains can be assessed visually via traceplots (see Appendix S1). More formal metrics exist. The Gelman-Rubin metric, $\hat{R}$ \citep[][see \citealt{Brooks-Gelman-1998} for the multivariate analogue]{Gelman-Rubin-1992}, is one of the most popular multi-chain diagnostics. Although $\hat{R} <1.1$ generally indicates convergence \citep{Gelman-etal-2013}, recent research indicates that a threshold closer to one may be more suitable in some scenarios \citep{Vats-Knudson-2018}. Note that pseudo-convergence can occur in many different scenarios. For example, the sampler can get caught in one mode if the target distribution has multiple modes that are not well connected by the Markov chain dynamics \citep{Geyer:2011}. Running the chain for a long period can help limit these pseudo-convergence problems \citep{Geyer:2011}. A detailed summary of convergence methods is available in \citet{Cowles-Carlin-1996} and further research on convergence diagnostics includes \citet{Boone-etal-2014}, \citet{VanDerwerken-Schmidler-2017}, and \citet{Vats-Knudson-2018}. Both \texttt{JAGS} and \texttt{BUGS} project software, as well as the \texttt{R} package \texttt{coda} \citep{coda}, provide several methods to assess convergence.

\subsubsection{Priors}
\label{S.priors}

Selection of priors is a significant part of a Bayesian analysis because priors affect the resulting posterior distribution \citep{Robert-2007}. Several approaches can be taken depending on the information available about the model parameters and the philosophy of the modeler. Ecologists often use `noninformative' priors. These priors (e.g., a uniform distribution over the parameter space) are often thought to be objective and are generally chosen with the goal of maximizing the influence of the data on the posterior. However, noninformative priors may still have important effects on the posterior, and they should not be used naively \citep{Gelman-etal-2017, Lemoine-2019}. For example, \citet{Lele-2020} showed that noninformative priors could significantly influence the parameter and state estimates of ecological SSMs. Alternatively, ecologists can use informative priors, which are created using knowledge of the parameters or previously collected data  \citep[e.g.,][]{Meyer-Millar-1999, Dunham-Grand-2016}. As there are many advantages to using informative priors, they are increasingly used in ecological models \citep{Hooten-Hobbs-2015}. For example, informative priors can be used to supplement SSMs with limited time-series data \citep{Chaloupka-Balazs-2007} and can improve state estimates \citep{Dunham-Grand-2016}. In most cases, noninformative and informative priors are used in the same model on different parameters. For technical reasons, it can be sometime advantageous to use conjugate priors (i.e., priors with the same distribution as the conditional posterior distribution or the posterior distribution). \citet{Kass-Wasserman-1996} and \citet{Millar-2002} have summarized priors typically used in fisheries models, including many SSMs. \citet{Lemoine-2019} advocates for the use of weakly informative priors as default in ecology and provides a guide to their implementation. More generally, \citet{Robert-2007} and \citet{Gelman-etal-2013} provide a thorough review of available priors, selection and examples for a variety of models.

%---------------------------------------------------
\subsection{Information reduction approaches}

While uncommonly used with SSMs, information reduction approaches, such as synthetic likelihood or Approximate Bayesian Computation (ABC), appear promising to fit complex, highly nonlinear, ecological SSMs \citep{Fasiolo-etal-2016}. These methods bypass the calculation of the exact likelihood \citep{Csillery-etal-2010, Fasiolo-etal-2016}. Instead, these methods generate samples from the model and transform them into a vector of summary statistics that describe the data in the simplest manner possible \citep{Csillery-etal-2010, Fasiolo-etal-2016}. The simulated summary statistics are then compared to observed summary statistics using a predefined distance measure \citep{Fasiolo-etal-2016}. Information reduction approaches smooth the likelihood, reducing some of the common implementation problems encountered with other fitting methods. However, the results from information reduction approaches are often imprecise and, thus, may be most useful in the model development phase \citep{Fasiolo-etal-2016, Fasiolo-Wood-2018}. Interested readers are referred to \citet{Csillery-etal-2010}, \citet{Fasiolo-etal-2016}, and \citet{ Fasiolo-Wood-2018}.

\subsection{Fitting models with discrete states}
\label{S.discrete.states}

Depending on the complexity of the SSM and ones favored inferential approach, having discrete states can either facilitate or complicate the fitting process. A SSM with a single time series of categorical states, generally referred as an HMM (Section \ref{S.HMM}), can be relatively easily fitted with a frequentist approach. The key advantage of these HMMs is their mathematical simplicity: what
would be a high-dimensional integration in a SSM with continuous
state values (see Section \ref{S.Fitting}) is now a simple sum. As
such, having a finite number of possible state values (i.e., discrete
states) significantly simplifies the analysis \citep{Langrock-etal-2012}. The mathematical simplicity of HMMs makes
them highly attractive, and various efficient tools and \texttt{R} packages have been developed to fit HMMs to data. We refer readers interested in HMMs to \citet{McClintock-etal-2020} and \citet{Zucchini-etal-2016}.

While one can use Metropolis-Hastings samplers (Section \ref{S.Metropolis-Hastings}) to fit HMMs with a Bayesian approach \citep[e.g.,][]{Zucchini-etal-2016}, these algorithms are far less efficient than those used to fit HMMs in a frequentist framework. In addition, HMC algorithms (Section \ref{S.HMC}) do not generally allow to sample discrete states. However, recent work has demonstrated the gain in speed that can be made by marginalizing the latent states and how this can be implemented with Gibbs sampling (e.g., \texttt{JAGS}) and HMC software  \citep[e.g., \texttt{Stan};][]{Leos-barajas-Michelot-2018, Betancourt-etal-2020, Yackulic-etal-2020}. Marginalizing the states means that when we estimate the parameter values, we do not sample the hidden states at each iteration, but rather track the likelihood of being in any given state \citep{Yackulic-etal-2020}. One can then estimate the states values using the conditional distribution (Eq. \ref{E.cond.dist.state}) or approximations of it \citep[see][for more detail]{Yackulic-etal-2020}, or algorithms that are commonly used with frequentist HMMs, such as the Viterbi algorithm \citep{Zucchini-etal-2016, Leos-barajas-Michelot-2018}. This two-step approach used when marginalizing the states has many parallels with the frequentist approach described in Section \ref{S.Freq}, where we first estimate the parameters using the marginal likelihood and subsequently estimate the hidden states based on the estimated parameter values.

While there are many efficient tools to fit simple HMMs with a frequentist approach, it can be more challenging to fit SSMs that combined both continuous and discrete states. Just as for Bayesian methods, some of the computationally efficient methods (e.g., Laplace approximation method described in Section \ref{S.Laplace}) do not allow for discrete states. One can use instead frequentist methods that rely on sampling the states (e.g., Sequential Monte Carlo methods described in Section \ref{S.Iterative.Filtering}). One could potentially develop algorithms that marginalize the discrete and continuous states with different approaches. 

For Bayesian SSMs with discrete states, one additional consideration is label-switching \citep{Jonsen-etal-2013}. The labels given to the $N$ discrete states are arbitrary, and thus there are $N!$ potential label assignments \citep{Zucchini-etal-2016}. The different label permutations result in the same model. Thus, when the MCMC chains have reached convergence, all possible labels will have been assigned to each state and inference on the states will be difficult. For example, we will no longer be able to take the mean of the posterior distribution to estimate the states because all $ \hat{z}_{t,1:T} \approx N/2$. One solution is to impose constraints on the parameters that would be violated when labels are permuted \citep{Zucchini-etal-2016}. For example, in the two-behavior movement model described in Section \ref{S.HMM} we would constrain $\gamma_1 \leq \gamma_2$.

%---------------------------------------------------------
\subsection{When to use each method?}
\label{S.method.comp}

Choosing from this multitude of fitting methods can appear daunting, but can be guided by a choice of inference framework and the limitations of each methodology. In Table \ref{t.methods}, we list the methods discussed above, with some pros and cons. We simply state the associated inferential framework (frequentist vs. Bayesian), and we let the readers decide their favorite inferential framework. In general, there are more computationally efficient methods for simple models in the frequentist framework (e.g., Kalman filter and Laplace approximation), but such generalization cannot be made for more complex models.

Note that in some cases, it may also be easier to use one of the more specific ecological SSM \texttt{R} packages. For example, the package \texttt{MARSS} \citep[which stands for Multivariate Auto-Regressive State-Space,][]{Holmes-etal-2012, MARSS} can be useful to model multiple populations, if these can be reasonably formulated with a linear and normal SSM. Those interested in fisheries stock assessment SSMs should look at the package \texttt{stockassessment} \citep[available on GitHub at {https://github.com/fishfollower/SAM},][]{Nielsen-Berg-2014}. Those interested in SSMs for animal movement should explore \texttt{bsam} \citep{Jonsen-etal-2005, Jonsen-2016}, \texttt{crawl} \citep{Johnson-etal-2008, crawl}, and \texttt{momentuHMM} \citep{McClintock-Michelot-2018}.

% ****************************************
\section{\label{S.Estimability}Formulating an appropriate SSM for your data}

SSMs are powerful tools, but their inherent flexibility can tempt ecologists to formulate models that are far too complex for the available data. The model structure or the characteristics of the specific dataset may make it impossible to estimate every parameter reliably. In such cases, parameter estimates will no longer provide key information on the underlying biological process and state estimates may become unreliable \citep[e.g.,][]{AugerMethe-etal-2016}. Formulation of SSMs needs to be guided by the inference objectives and the available data. In this section, we discuss how to assess whether a model is adequate for your data and how one can alleviate potential estimation problems.

%---------------------------------------------------------
\subsection{Identifiability, parameter redundancy and estimability}
\label{parredidentest}

\begin{figure}
\begin{center}
\includegraphics[width=470pt]{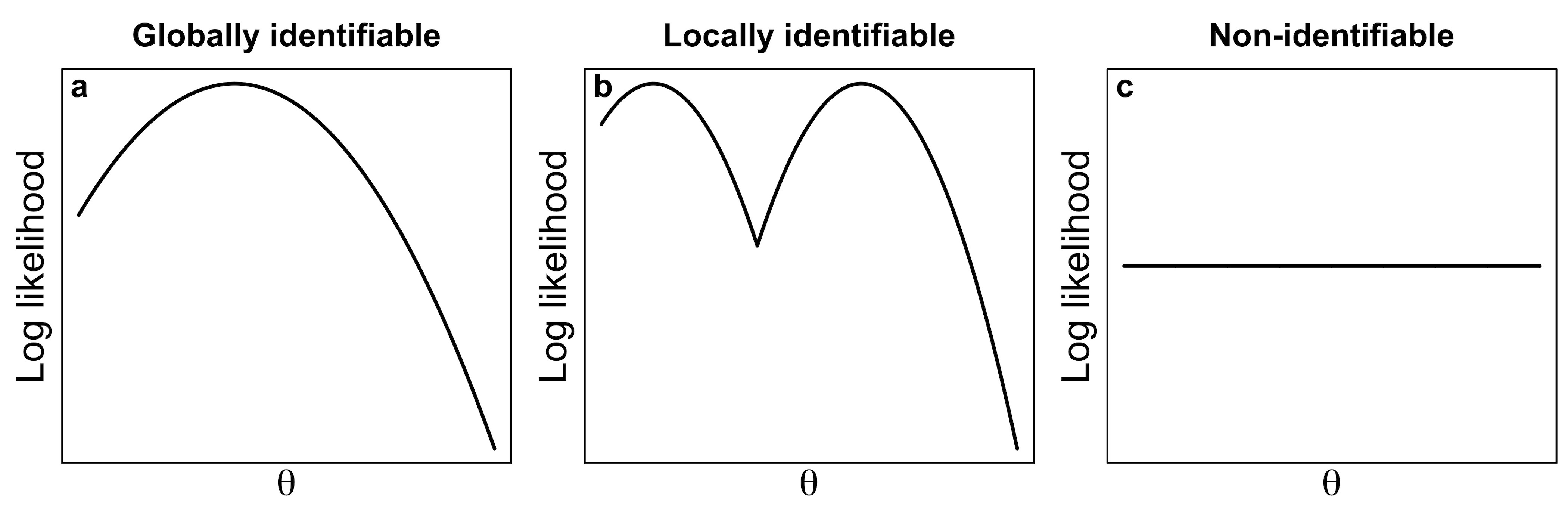}
\end{center}
\caption{Examples of log-likelihood profiles for a parameter $\theta$ under various identifiability scenarios: (a) globally identifiable, (b) locally identifiable, and (c) non-identifiable model.}
\label{fig:profile}
\end{figure}

When we estimate the parameters of a model, denoted here as $M(\boldsymbol{\theta})$, we often want to find the set of parameter values, $\boldsymbol{\theta}$, that results in the best fit to the data. For this to be possible, the model needs to be identifiable. Identifiability refers to whether or not there is a unique representation of the model. A model is globally identifiable if $M(\boldsymbol{\theta}_1) = M(\boldsymbol{\theta}_2)$ implies that $\boldsymbol{\theta}_1 = \boldsymbol{\theta}_2$. For example, in a frequentist framework, an identifiable model would have only a single $\boldsymbol{\theta}$ value that would maximise the likelihood (Fig. \ref{fig:profile}a). A model is locally identifiable if there exists a neighbourhood of $\boldsymbol{\theta}$ where this is true (Fig. \ref{fig:profile}b). Otherwise a model is non-identifiable \citep[Fig.
\ref{fig:profile}c;][]{Rothenberg1971,Coleetal2010}. 

An obvious case of non-identifiability is when a model is overparameterised and can be reparameterised with a smaller set of parameters. For example, if two parameters only appear as a product in a model (e.g., $y = \alpha \beta x$); that model could be reparameterised with a single parameter replacing that product (e.g., $y = \gamma x$, where $\gamma = \alpha \beta$). The parameter redundancy of the original model will result in non-identifiability \citep{CatchpoleandMorgan1997} and non-identifiability caused by the inherent structure of a model is referred to as intrinsic parameter redundancy \citep{Gimenez-etal-2004} or structural non-identifiability \citep{CobelliDiStefano1980}. Regardless of the amount or quality of data, it is impossible to estimate all the parameters in such a model.

Having a structurally identifiable model does not guarantee that one can estimate its parameters with the data at hand. Non-identifiability can be caused by a specific dataset with, for example, missing or sparse data \citep{Gimenez-etal-2004}. This problem is known as extrinsic parameter redundancy \citep{Gimenez-etal-2004} or practical non-identifiability \citep{Raue-etal-2009}. A parameter is defined as practically non-identifiable if it has a confidence interval that is infinite \citep{Raue-etal-2009}. It is also possible for a dataset to create estimation problems with an otherwise structurally and practically identifiable model, a phenomenon referred to as statistical inestimability  \citep{Campbellele2014}. If a model is statistical inestimable, a confidence interval for a parameter will be extremely large but not infinite. This often occurs because the model is very similar to a submodel that is parameter redundant for a particular dataset, which is known as near redundancy \citep{Catchpoleetal2001,Coleetal2010}. 

Having a non-identifiable model (either structurally or practically) leads to several problems. First, there will be a flat ridge in the likelihood of a parameter redundant model \citep{CatchpoleandMorgan1997}, resulting in more than one set of MLEs. However, despite the parameter redundancy, numerical methods for parameter estimation usually converge to a single set of MLEs. Therefore, without further diagnostics, one may not realise that
the MLEs are not unique.  Second, the Fisher information matrix will be singular \citep{Rothenberg1971} and therefore the standard errors will be undefined in a non-identifiable model. However, the exact Fisher information matrix is rarely known and standard errors are typically approximated using a Hessian matrix. The Hessian describes the local curvature of a multi-parameter likelihood surface. The Hessian is generally evaluated numerically, which can lead to explicit (but incorrect) estimates of standard errors. Third, many model selection methods (see Section \ref{S.Model.Comparison}) are based on the assumption that a model is identifiable and that the penalty for complexity is a function of the number of unique and estimable parameters \citep{Gimenez-etal-2004}. If a model is statistically inestimable or near redundant, these three problems may also occur, as the model is close to being non-identifiable. For example the log-likelihood profile will be almost flat.

Checking for identifiability and estimability should become part of the model fitting process and several methods are available to do so. A clear sign of problems is a flat log-likelihood profile (Fig. \ref{fig:profile}c), and plotting the log-likelihood profile for each parameter can serve as a diagnostic for this \citep[Fig. \ref{fig:profile};][]{Dennis-etal-2006, Raue-etal-2009, AugerMethe-etal-2016}. Correlation between parameters can also be indicative of estimation problems, and it may be useful to inspect the log-likelihood or posterior surface of pairs of parameters \citep{Campbellele2014, AugerMethe-etal-2016}. Depending on model complexity and computation time, simulations can be an easy way to investigate the estimability of SSMs \citep{AugerMethe-etal-2016}. For a specified SSM and a known set of parameters, one simulates the state process and observation time
series, and then estimates the parameters and states. One  then compares  estimated parameter and state values with the known
true values. Parameter estimates from non-identifiable models will usually be biased with large variances. 

In addition to these simple checks, three advanced methods to assess estimability and identifiability problems exist. First, data cloning has been shown to be useful with ecological models \citep{Peacock-etal-2016}. Data cloning involves using Bayesian methodology with a likelihood based on $K$ copies of the data (clones).
The posterior variance of a parameter will tend towards $K$ times the asymptotic variance of the parameter, so that if a parameter is identifiable the posterior variance will tend to zero as $K$ tends to infinity. If a parameter is not identifiable, the posterior variance will tend to a fixed (non-zero) value \citep{Leleetal2010}. \cite{Campbellele2014} show how this method can be extended to find estimable parameter combinations in non-identifiable models.

Second, one can use the fact that the Hessian matrix in a non-identifiable model will be singular at the MLE. As a singular matrix has at least one zero eigenvalue, the Hessian method involves finding the eigenvalues of the Hessian matrix. If the Hessian matrix is found numerically, the eigenvalues for a singular matrix may be close to zero rather than exactly zero. Therefore, if any of the eigenvalues are zero or close to zero, the model is deemed non-identifiable or parameter redundant, at least for that particular dataset \citep{Viallefontetal1998}. The Hessian matrix will also have eigenvalues close to zero if the model is statistically inestimable or near redundant \citep{Catchpoleetal2001}.

Third, one can use the symbolic method. This method uses the concept that a model can be represented by an exhaustive summary, which is a vector of parameter combinations that uniquely define the model. For example, this vector could be $\mathbf{k} = (L_{\textsc{m}}(\boldsymbol{\theta} | y_1), L_{\textsc{m}}(\boldsymbol{\theta} | \mathbf{y}_{1:2}), \hdots, L_{\textsc{m}}(\boldsymbol{\theta} | \mathbf{y}_{1:T}))'$, where the first element is the marginal likelihood (Eq. \ref{E.Marginal.Likelihood}) for the first observation ($y_1$), the second element is the marginal likelihood for the first two observations ($\mathbf{y}_{1:2}$), etc. This straightforward exhaustive summary works well for HMMs \citep{Cole-2019}, but can be impractical for SSMs with continuous states as it involves integration. Suitable, but more complex to derive, exhaustive summaries for SSMs are given in \cite{ColeandMcCrea2016}. To investigate identifiability, we form a derivative matrix by differentiating each term of the vector with respect to each parameter. Then, we find the rank of this matrix. The rank of a matrix is the number of columns that are linearly independent. Since each column of the derivative matrix is associated with one of the parameters, the rank is the number of estimable parameters (or parameter combinations). If the rank is less than the number of parameters, then the model is non-identifiable or parameter redundant \citep{CatchpoleandMorgan1997,Coleetal2010}. This method can be used to investigate  practical identifiability as well as structural identifiability by choosing an exhaustive summary that includes the specific dataset \citep{Coleetal2012}. In some more complex models, the computer can run out of memory calculating the rank of the derivative matrix. \cite{Coleetal2010} and \cite{ColeandMcCrea2016} provide symbolic algebra methods for overcoming this issue. The alternative is a hybrid symbolic-numerical method, which involves finding the derivative matrix using symbolic algebra, but then finding the rank at five random points in the parameter space \citep{ChoquetandCole2012}.

Each of the numerical methods (log-likelihood profile, simulation, data cloning, Hessian method) can be inaccurate. They are also not able to distinguish between estimability, practical identifiability and structural identifiability when applied to a specific dataset, although in some cases a large simulated dataset could be used to test structural identifiability. Being able to distinguish between these problems is useful as it can help us assess whether gathering more data will help. The symbolic method is accurate, but is more complicated to use as it involves using a symbolic algebra package. Code for assessing estimability using simulations and the Hessian method is given in Appendix S1: Section S1 1.4. Code for the symbolic algebra method is given in Appendix S3.

In Bayesian analysis, identifiability and estimability issues have a different focus because priors can affect our capacity to differentiate between parameters \citep{Cressie-etal-2009}. In general, parameters are said to be weakly indentifiable when the posterior distribution significantly overlaps with the prior \citep{Garrett-Zeger-2000, Gimenez-etal-2009-indentifiability}. If priors are well informed by previous data or expert knowledge, their strong influence on the posterior distribution is no longer an identifiability/estimability issue but one of the benefits of Bayesian analysis. However, misusing informed priors (e.g., when the information is not reliable) may hide identifiability issue or cause the estimability problems \citep{Yin-etal-2019}. Thus, one should choose priors with great care. To help ensure that the data inform the model and that the posterior is well behaved, \citet{Gelfand-Sahu-1999} suggested to use informative priors that are not too precise (see Section \ref{S.priors} for other considerations). Weak identifiability can result in multiple implementation issues, including slow convergence \citep{Gimenez-etal-2009-indentifiability}. Diagnostics for parameter identifiability in the Bayesian framework include some of the tools described above and the visual or numerical assessment of the overlap between priors and posterior distributions \citep{Garrett-Zeger-2000, Gimenez-etal-2009-indentifiability}.

%---------------------------------------------------------
\subsection{Remedies for identifiability issues}

When we fit a SSM to our data, we hope that it will provide accurate and precise estimates of our parameters and states. But how can we achieve these goals? First, we need to have a structurally identifiable model. Second, one needs a dataset appropriate for the model and vice-versa, otherwise one can face estimation problems even with structurally identifiable models. Generally, we assume that having more data will allow us to better estimate parameters and states. However, as discussed below, increasing the length of the time series may not be the best way to improve estimation.

\subsubsection{Reformulate the SSM}

To create a structurally identifiable model, one should start by avoiding overparametrization. As mentioned above, models where some
parameters only appear as products of each other should be simplified. The same holds  for models where parameters only appear as sums (e.g., $y = (\alpha + \beta)x$), or differences, or fractions. Models where the magnitude of two sources of error are simply additive are also problematic (e.g., $Y \sim N(X, \sigma^2)$ and $X \sim N(\mu, \tau^2)$, will result in $Y \sim N(\mu, \sigma^2 + \tau^2)$ where $\sigma$ and $\tau$ cannot be uniquely identified). As such, one needs to check that none of the parameters are confounded and carefully inspect the combination of the sources of variability in all hierarchical models, including SSMs (see below). Some of the tools discussed above can help construct structurally identifiable models. In particular, the symbolic method can be used to identify the parameters that are confounded in a non-identifiable model, and thus can be used to select estimable parameter combinations. This involves solving a set of partial differential equations formed from the same derivative matrix used to check identifiability \citep{Catchpoleetal1998,Coleetal2010}.

\subsubsection{Make simplifying assumptions when data are limited}

A full model may be too complex for the data-at-hand and it may be advantageous to make simplifying assumptions. For example, when the data available for older age classes are limited, researchers can have difficulties fitting the fisheries stock assessment model presented in Section \ref{S.M.stock.assessment}. To help the estimation process, one can create a cumulative age class, $\text{A}^+$, that accounts for all fish older than a certain age \citep{Nielsen-Berg-2014}. To allow fish to remain in the cumulative age class, we need to add the following equation to the model:
\begin{linenomath*}
\begin{align}
    \log (N_{\textsc{a}^+,t}) &=\log( N_{\textsc{a}^+-1,t-1}e^{- F_{\textsc{a}^+-1,t-1} - M_{\textsc{a}^+-1,t-1}} + N_{\textsc{a}^+,t-1}e^{- F_{\textsc{a}^+,t-1} - M_{\textsc{a}^+,t-1}})  + \epsilon_{N_{\textsc{a}^+, t}}.
    \end{align}
\end{linenomath*}
The similar size of these older fish makes them more likely to be caught by the same type of fishing gear, and thus their catchability and fishing mortality can be further assumed to equal that of the previous age class ($Q_{\textsc{a}^+,s} = Q_{\textsc{a}^+-1,s}$ and $F_{\textsc{a}^+,t} = F_{\textsc{a}^+-1,t}$). While this appears to add complexity, creating this cumulative age class and equating some terms reduces the number of states and parameters to estimate. However, some simplifying assumptions may result in estimation problems. For example, the original DCRW model of \citet{Jonsen-etal-2005} has a single correction factor rather than one per coordinate ($\psi = \psi_{lon} = \psi_{lat}$, see Section \ref{S.M.movement.model}). The common correction factor can results in estimation problems because longitude and latitude often differ in the degree of correction they need \citep{AugerMethe-etal-2017}. As long as they are biologically reasonable, such simplifying assumptions can be useful in a wide range of fields, including in community ecology where SSMs can link multiple species to common latent variables and thus reduce the dimension of the model \citep{Thorson-etal-2016}.

\subsubsection{Estimate of measurement errors externally}

SSMs can be associated with significant estimability problems, particularly when trying to estimate the two main sources of variability \citep{Knape-2008, AugerMethe-etal-2016}. As a result, researchers often fix some of the parameters to known values, or use informed priors if they are working in a Bayesian framework. In particular, many use fixed values for the measurement errors and use for them independent estimates of measurement errors \citep[e.g.,][]{Jonsen-etal-2005}. While such method can alleviate estimation problems \citep{Knape-2008}, one must be careful not to use biased or misspecified values.

\subsubsection{Integrate additional data}

Covariates that provide additional information about a state or a process (e.g., survival) may be a means of overcoming identifiability problems. \citet{Polansky-etal-2019} showed that non-identifiability in the estimation of a fecundity and observation correction parameter could be overcome by including a covariate in the model for fecundity.

Similarly, identifiability issues can be overcome by combining a SSM with a model for another data set that has parameters in common with the SSM. For example, in integrated population models, SSMs for time series of census data are combined with capture-recapture data \citep{Besbeas-etal-2002,Abadi-etal-2010}. Adding additional data sources can be extremely useful but may not remove all identifiability issues. Methods for checking identifiability in integrated models are discussed in \citet{ColeandMcCrea2016}.

\subsubsection{Use replicated observations}

Having replicated observations through time (e.g., two independent population surveys) can help differentiate process variation from observation error, improve the parameter estimates accuracy, and improve the capacity of model selection methods to identify the correct observation distribution \citep{Dennis-etal-2010, Knape-etal-2011}. In many instances, such replicated observations have already been collected, but are aggregated. For example, in population monitoring studies, subsamples (e.g., transect portions) are often aggregated into one overall estimate of abundance. \citet{Dennis-etal-2010} demonstrated that using these as replicates, rather than aggregating them, can improve the estimates. One can also take advantages of time series with multiple data sources to estimate the errors of each data source \citep[e.g., double tagged individuals in movement SSMs,][]{Winship-eta-2012}. For animal movement models, individuals can be also seen as replicates of the same process, but often the SSMs are fitted separately to each individual track. To improve inference, one can create a population model, where each individual track is linked to a distinct state time series but all share the same parameters \citep{Jonsen-2016}. While the gains that can be made with replications are significant, one must understand the assumptions of models for replicated data. Simple population models for replicated datasets may assume that the replicates are independent \citep{Dennis-etal-2010}. However, many temporally varying factors (e.g., weather) may affect the sampling conditions and/or the behaviour of animals and result in correlations between replicates. \citet{Knape-etal-2011} demonstrated how to account for such dependence in population dynamics models. For animal movement, one may want to consider whether it is appropriate to assume that the behavioral mechanism driving movement is identical across individuals and, if not, may want to modify the model accordingly. However, as the gains that can be made with replications far surpasses those that could be made with longer time-series \citep{Dennis-etal-2010}, one should consider using replication in their models and when designing their studies. For example, \citet{Knape-etal-2011} suggested that in some cases managers may want to sample a population twice every second year rather than once a year. As SSMs are becoming the prime method to fit ecological time series, such study design issues should be explored further.

\subsubsection{Match temporal resolution for states and observations}

The temporal resolution of the data can affect the parameter and state estimates and it is important to define a model at a resolution that is appropriate for the data. In many cases, adequate temporal span or resolution is more important than increased data quantity. For example, if a model describes a long-term cycle, then collecting data from more individuals is unlikely to make parameters estimable if the dataset is not long enough to span the cycle being described \citep{Peacock-etal-2016}. If developing a model to classify a movement path into distinct behavioral modes, one must sample the movement track at a high enough frequency so that multiple locations are recorded in each movement bout \citep{Postlethwaite-Dennis-2013}. If one has a dataset with locations every 8 hrs, it would be challenging to estimate behavioral states lasting less than 16-24 hrs. One can use pilot data, simulations, and data cloning to identify the temporal (and spatial) scale of sampling appropriate for the model, in something akin to a power analysis \citep{Peacock-etal-2016}. Overall, finding an appropriate model for your data, or collecting the appropriate data for your questions, can be an iterative process where one assess the estimability of different models under different data conditions.

% ***************************************
\section{\label{S.Model.Comparison}Computationally-efficient model comparison methods}

Model comparison (or selection) can be used to compare the relative fit of models representing multiple working hypotheses, and to identify the model amongst these that best describes the data (see Section \ref{S.Diagnostics} for methods to evaluate the absolute fit of a model). Because different model structures can affect the estimated states and parameters \citep{Knape-etal-2011}, model comparison can be extremely useful in helping to refine state estimates \citep{AugerMethe-etal-2017}. Model comparison is common in ecology and has been used to compare SSMs \citep[e.g.,][]{Siple-Francis-2016}. However, it is not uncommon for users to fit only a single SSM, likely due to the computational burden of fitting complex SSMs and some of the known limitations of applying model selection methods to SSMs \citep{Jonsen-etal-2013}. With the improved efficiency of fitting algorithms and advancements in model selection measures, model comparison of SSMs is becoming more attainable. 

One common view is that ecological systems are so complex that it  is impossible to develop a model that truly describes them, and that the goal of model selection is to find the best approximation of the truth \citep{Burnham-Anderson-2002}. Under this paradigm, a useful way to compare models is to assess how well they can be used to predict new data. Comparing the out-of-sample predictive accuracy of models can be done with cross-validation. However, it is rarely done with ecological SSMs because it requires fitting the same model multiple times, and thus can add significant computational burden to the analysis. Many advocate cross-validation as the best method for model selection \citep{Gelman-etal-2014, Link-etal-2017}, and gains in efficiency of fitting algorithms are making its use increasingly feasible. We discuss cross-validation as a model selection , and validation, method in Section \ref{S.Diagnostics}. Here, we focus on what can be considered approximations of predictive accuracy. In particular, we discuss information criteria measures used with frequentist and Bayesian approaches.

%---------------------------------------------------------
\subsection{\label{S.Model.Comparison.ML}Frequentist approach}

The most common model comparison measure in ecology is Akaike's Information Criterion \citep[AIC;][]{Aho-etal-2014}. AIC was derived to estimate the expected and relative distance between the fitted model and the unknown true data-generating mechanism \citep{Burnham-Anderson-2002}, and can be viewed as $-2$ times an approximation of the predictive accuracy of the model \citep{Gelman-etal-2014}:
\begin{linenomath*}
\begin{equation}
    \text{AIC} = - 2 \log L(\boldsymbol{\hat{\theta}}_\textsc{mle} | \mathbf{y}) + 2k,
    \label{E.AIC}
\end{equation}
\end{linenomath*}
where $L(\boldsymbol{\hat{\theta}}_\textsc{mle} | \mathbf{y})$ is the likelihood of the model at the MLE (i.e., the probability of the observed data given the model) and $k$ is the number of parameters estimated. The model with the lowest AIC, thus the shortest distance from the truth, is considered the best model. Models with more parameters will be more flexible and will tend to fit the existing data better by chance alone. Thus, AIC penalizes a model for its number of estimated parameters to compensate for overfitting.

There are many issues related to using AIC with SSMs, and some have cautioned against this practice \citep[e.g.,][]{Jonsen-etal-2013}. We identified five different concerns. The first three concerns are related to the fact that the states of a SSM can be considered as random effects. First, using AIC to understand whether including random effects improves the model is difficult because some of the models may have parameters at the boundary of parameter space
\citep{Bolker-etal-2009}. For example, testing whether or not there is process variance in SSMs (e.g., comparing our toy model to a model with no process variance, where $\sigma_p = 0$) could result in boundary problems, and is not recommended. Second, when you have random effects it is difficult to quantify the effective number of parameters \citep{Bolker-etal-2009}. For SSMs, it is difficult to know to what extent the states should be counted as estimated parameters and contribute to $k$. However, if all of the compared SSMs have the same number of states and no additional random effects, these two issues should be less problematic. In such cases, we would expect any bias in the penalty $k$ to be the same across models and thus have little effect on the difference in AIC across models. Third, one must decide whether the marginal likelihood or the conditional likelihood should be used when calculating AIC of a model with random effects \citep{Muller-etal-2013}. In contrast to the marginal likelihood, where we integrate out the states (Eq. \ref{E.Marginal.Likelihood}), the conditional likelihood considers the states as known: $L_{\textsc{C}} (\boldsymbol{\theta}_o| \mathbf{z}_{1:T}, \mathbf{y}_{1:T}) = \prod_{t=1}^T g(y_t | z_t, \boldsymbol{\theta}_o)$. When the conditional likelihood is used in the AIC framework, both the parameter and state estimates are plugged in and different approaches can be used to account for the number of states \citep{Vaida-Blanchard-2005, Muller-etal-2013}. This conditional AIC is a measure of the model's ability to predict new observations that share the same latent states, while the marginal AIC does not assume that the latent states are shared with the new observations and measures the model's ability to predict new observations from the same process \citep{Vaida-Blanchard-2005}. For example, for a SSM describing the population dynamics of a fish species, we would interpret the conditional AIC as assessing the ability to predict another survey of the same population during the same time period. The marginal AIC would be assessing the ability of the model to predict a survey from a similar population of the same species. To our knowledge the marginal likelihood has always been used with SSMs fitted in a frequentist framework. In most SSMs, the number of states increases with the sample size (i.e., with the length of the time series). Because frequentist model selection methods rely on asymptotic properties, which can be attained when the sample size is large compared to the number of quantities estimated, conditional AIC may be unreliable for most SSMs. This characteristic may explain why potential advantages of using the conditional likelihood remain uninvestigated in the frequentist SSM literature (the conditional likelihood is used in Bayesian information criteria, see Section \ref{S.Model.Comparison.B} for a discussion). The fourth source of concern is related to the problems associated with using AIC to choose the number of components in mixture models, which are particularly relevant for choosing the number of states in HMMs \citep{Jonsen-etal-2013}. \citet{Pohle-etal-2017} outline solutions to this HMM-specific problem.

The final concern, which is specific to cases with small sample size, is one that has been studied in the SSM literature. When the sample size, $n$, is small and the number of parameters, $k$, is relatively large (e.g., when $k \approx n/2$), the $2k$ penalty is inadequate and AIC has a tendency to favor more complex SSMs \citep{Cavanaugh-Shumway-1997}. Many use the corrected AIC (AICc) for small sample size \citep{Burnham-Anderson-2002}. However, \citet{Cavanaugh-Shumway-1997} noted that AICc may be inadequate for many SSMs, and suggested an alternative: the bootstrap-corrected measure, AICb. AICb has been used for ecological SSMs \citep{Ward-etal-2010, Siple-Francis-2016}, especially by users of the \texttt{R} package \texttt{MARSS} \citep{Holmes-etal-2012}. This package for estimating the parameters of linear multivariate auto-regressive SSMs with Gaussian errors (i.e., multivariate dynamic linear models) has a function that calculates various versions of AICb. AICb was developed in the context of linear Gaussian SSMs, but is thought to be relatively robust to violations to normality \citep{Cavanaugh-Shumway-1997}. We can describe AICb as:
\begin{linenomath*}
\begin{equation}
    \text{AICb} = - 2 \log L_{\textsc{m}}(\boldsymbol{\hat{\theta}_\textsc{mle}} | \mathbf{y}_{1:T}) + 2\left(\frac{1}{N} \sum_{i=1}^{N} - 2\ \log \frac{L_{\textsc{m}}(\boldsymbol{\hat{\theta}}^i | \mathbf{y}_{1:T})}{L_{\textsc{m}}(\boldsymbol{\hat{\theta}_\textsc{mle}} | \mathbf{y}_{1:T})} \right),
\end{equation}
\end{linenomath*}
where $\boldsymbol{\hat{\theta}}^i$ is the $i^{th}$ bootstrap replicate of $\boldsymbol{\hat{\theta}}$, $N$ is the number of replicates, and $L_{\textsc{m}}(\boldsymbol{\hat{\theta}}^i | \mathbf{y}_{1:T})$ is the marginal likelihood of the model with the bootstrapped parameter sets given the original data. This bootstrap replicate can be achieved by simulating a time series from our model with $\boldsymbol{\hat{\theta}_{\textsc{mle}}}$ and estimating the parameters using this new time series. AICb was shown to outperform AIC and AICc when used with SSMs that had relatively small sample size for the number of parameter estimated \citep{Cavanaugh-Shumway-1997}. The disadvantage of AICb is that it requires fitting the model $N$ times. In the case of models that are computationally demanding to fit, one may need to continue to rely on AICc when sample sizes are small. While AICc tends to erroneously choose more complex models compared to AICb, it is better than AIC and many other metrics for SSMs with small sample size \citep{Cavanaugh-Shumway-1997}. Another similar computationally-intensive AIC variant for SSMs fitted to small samples has been developed by \citet{Bengtsson-Cavanaugh-2006}, but its use in ecology has been limited by some of its constraints \citep[e.g.,][]{Ward-etal-2010}. For large datasets, some ecologists prefer to use BIC over AIC because AIC tends to choose more complex models as sample size increases. However, these two measures are used to achieve different inferential goals, and choosing between them is largely a philosophical question \citep[see][]{Aho-etal-2014, Hooten-Hobbs-2015}.

Overall, AIC and its small-sample alternatives can be used with SSMs in many instances, especially when the number of states and random effects are the same. AIC has been used for decades with SSMs \citep{Harvey-1990}, and simple simulation studies have shown that AIC can be used to reliably select between SSMs \citep{AugerMethe-etal-2017}. Further research on the capacity of AIC to compare the predictive abilities of SSMs when the number of states or random effects vary, and research on how to account for the number of states in the penalty term, would be useful. In the meantime, one should be aware of the limitations outlined above, and interpret the results accordingly.

Other frequentist methods may be used to select between SSMs. For example, likelihood ratio tests can be used to select between nested models, especially when conducting planned hypothesis testing \citep[e.g.,][]{Karban-and-deValpine-2010}. However, likelihood ratio tests will suffer from some of the same issues as the those highlighted for AIC. \citet{Newman-etal-2014} also highlighted the potential use of score tests, transdimensional simulated annealing, and other methods. To our knowledge, these alternative methods have not been used in the SSM literature, but may be the focus of future research.

%---------------------------------------------------------
\subsection{\label{S.Model.Comparison.B}Bayesian approaches}

Two Bayesian information criteria, the Deviance Information Criteria (DIC, see Appendix S4) and the Watanabe-Akaike information criterion (WAIC), are popular with hierarchical models, and have been used with SSMs. They replace the information criteria based on MLEs, such as AIC, which do not have a clear interpretation for Bayesians \citep{Hooten-Hobbs-2015}. DIC and WAIC are similar to AIC, but they both use information from the posterior and estimate the effective number of parameters using data-based bias correction rather than a fixed rule. These data-based methods attempt to account for the effects of priors and the hierarchical structure (e.g., the characteristics of the random effects) on the flexibility of the model. 

While DIC has been used to select ecological SSMs \citep[e.g.,][]{Michielsens-etal-2006}, and MCMC sampler software \citep[e.g., JAGS,][]{Plummer:2003} and \texttt{R} packages like \texttt{rjags} \citep{rjags} have functions that compute it easily,  this information criterion is known to have many drawbacks that hinder its suitability for SSMs. DIC performs better when the number of effective parameters is much smaller than the sample size \citep{Hooten-Hobbs-2015}, a condition likely uncommon with SSMs because the number of latent states scales with the sample size. In addition, DIC is known to be problematic for mixture models, can poorly estimate the effective number of parameters (e.g., can return negative numbers), relies on approximate posterior normality, and is not fully Bayesian because its measure of fit relies on the posterior mean
of $\boldsymbol{\theta}$ (i.e., a point estimate, see Appendix S4) instead of the entire posterior distribution  \citep{Gelman-etal-2014, Hooten-Hobbs-2015, Kai-Yokoi-2019}. These limitations may explain why \citet{Chang-etal-2015}, in contrast to \citet{Wilberg-Bence-2008}, showed that DIC had difficulties selecting amongst ecological SSMs.

\sloppy Many now favor WAIC, a recently developed Bayesian information criterion  \citep{Gelman-etal-2014, Hooten-Hobbs-2015}:
\begin{linenomath*}
\begin{equation}
    \text{WAIC} = -2 \sum_{i=1}^T \log \int p(y_i | \boldsymbol{\theta}) p(\boldsymbol{\theta} | \mathbf{y}) \text{d} \boldsymbol{\theta} + 2 p_\textsc{waic}.
    \label{E.WAIC}
\end{equation}
\end{linenomath*}
The first component of WAIC is also a measure of fit, but unlike DIC it uses the entire posterior distribution
for $\boldsymbol{\theta}$ rather than a point estimate. As such, we can consider this measure of fit as truly Bayesian. There are different ways to estimate the effective number of parameters, $p_\textsc{waic}$. \citet{Gelman-etal-2014} recommend using $\sum_{i=1}^T \Var_{\textsc{post}}(\log p(y_i | \boldsymbol{\theta}))$ as it gives results closer to the leave-one-out cross validation. In our formulation of WAIC (Eq. \ref{E.WAIC}), we used a $-2$ multiplier as it helps highlight the similarity to AIC (Eq. \ref{E.AIC}). However, this multiplier may obscure how WAIC is a measure of the predictive accuracy of the model, and some researchers prefer not using it \citep[e.g.,][]{Vehtari-etal-2017}. See Appendix S4 for how Eq. \ref{E.WAIC} is calculated in practice.

WAIC has been used to compare ecological SSMs \citep[e.g.,][]{Baldwin-etal-2018, Ferretti-etal-2018} and can be computed using the R package \texttt{loo}  \citep{Vehtari-etal-2017}. Recent reviews of Bayesian model comparison methods favor WAIC over DIC \citep{Gelman-etal-2014, Hooten-Hobbs-2015} because it is a fully Bayesian metric, it is not affected by parametrization, and will not return negative values for the effective number of parameters. However, WAIC has a few shortcomings, and new approximations of predictive accuracy have been recently proposed \citep[e.g., Pareto-smoothed importance sampling leave-one-out cross validation,][]{Vehtari-etal-2017}. Both parts of WAIC are computed by using the sum over each data point $i$, and thus rely on partitioning the data into disjoint, ideally conditionally independent, pieces \citep{Gelman-etal-2014}. Naively partitioning can be problematic with SSMs since the time-series nature of the data generally results in dependence structures (see Appendix S4 for a potential solution). While AIC and DIC rely on a point estimate rather than summing over each data point, they also assume conditional independence. 

Just as for AIC, we could use either the conditional or marginal likelihood with DIC and WAIC \citep{Kai-Yokoi-2019,Merkle-etal-2019}. With the Bayesian approach, the likelihood is generally defined as fully conditional on both parameters and latent states and both are generally sampled when sampling the posterior. Thus, the conditional likelihood is usually used with Bayesian metrics even though this is rarely specified \citep{Millar-2018, Merkle-etal-2019}. While computing the marginal likelihood version of these Bayesian metrics is more computationally expensive, their conditional counterparts are often unreliable \citep{Millar-2009, Millar-2018, Merkle-etal-2019}. In particular, DIC and WAIC were shown to more reliably select the true underlying SSM when the marginal likelihood is used \citep{Kai-Yokoi-2019}.

As \citet{Gelman-etal-2014} noted, we are asking close to the impossible from these information criteria measures: an unbiased estimate of out-of-sample prediction error based on data used to fit the model that works for all model classes and requires minimum computation. As such, metrics such as WAIC can be unreliable estimates of the predictive ability of ecological models \citep{Link-etal-2017}. While further research is needed to assess when WAIC is appropriate for SSMs and to identify data partitioning schemes that resolve some of the potential biases, WAIC based on the marginal likelihood is likely the best information criterion for Bayesian SSMs at this point. Future work should explore how promising new approximation methods \citep[see][]{Vehtari-etal-2017,Burkner-etal-2020} perform with ecological SSMs. If the models are relatively inexpensive to fit, then one can bypass many of the shortcomings of WAIC, and other approximations of predictive ability, by comparing models using more computer intensive cross-validation methods \citep{Gelman-etal-2014, Link-etal-2017, Vehtari-etal-2017}. Cross-validation will also require one to partition data intelligently, but this may be more easily implemented with blocking \citep{Gelman-etal-2014, Roberts-etal-2017}.

Other methods could be used to compare models in a Bayesian framework \citep[e.g.][]{Newman-etal-2014}. For example, reversible-jump MCMC has been used to compare SSMs \citep{McClintock-etal-2012}, but is known to be difficult to implement \citep{Hooten-Hobbs-2015}. The importance of multiple covariates in a model (e.g., the effect of temperature and precipitation on bird survival) can be assessed
by multiplying coefficients in a model by indicator variables which
when equal to one include the covariate and when equal to zero exclude the covariate \citep{OHara-Sillanpaa-2009}. Such techniques have
been used to compare ecological SSMs \citep{Sanderlin-etal-2019}, but such an
approach is designed for nested models only. Posterior predictive loss approaches appear to be suitable for time-series data  \citep{Hooten-Hobbs-2015} and have been used to compare ecological SSMs \citep{MillsFlemming-etal-2010}. While these alternative approaches may not be as commonly used to compare ecological SSMs, and will have drawbacks, many of them warrant further exploration.

%---------------------------------------------------
\subsection{Model averaging}

Model averaging can combine the strength of several models and account for model uncertainty, something model selection cannot offer \citep{Buckland-etal_1997, Hooten-Hobbs-2015}. \citet{Wintle-etal-2003} argued against using a single model to make predictions because uncertainty about model structure is often high in ecology, and alternative models can have prediction differences with important repercussions for management decisions. When one selects
a single model, and presents the parameter and state estimates based on this best model, one implicitly assumes that the model is true and that the uncertainty is only in the estimation process \citep{Buckland-etal_1997, Wintle-etal-2003}. One can instead use model averaging, where, for example, each model is weighted and the predictions are a weighted sum across the plausible models \citep{Wintle-etal-2003}. Both the parameters and the predictions could be averaged, but this must be done with care and we would generally caution against averaging parameters. In many cases, differences in model structure result in changing the meaning of parameters, thus making their average nonsensical \citep{Dormann-etal-2018}. Model averaging has been used in a few studies applying SSMs to ecological data \citep[e.g.,][]{Maunder-Deriso-2011, Moore-and-Barlow-2011} and was shown to provide unbiased estimates \citep{Wilberg-Bence-2008}. However, simulations studies have shown that model averaging may not always provide more accurate point estimates than the best SSMs \citep{Wilberg-Bence-2008, Chang-etal-2015}. In addition, while model averaging generally reduces prediction errors compared to each of the contributing models, these gains can be counteracted by factors such as uncertainty in the model weights and covariance between models \citep{Dormann-etal-2018}. In addition, calculating weights using parametric methods such as AIC can perform poorly \citep{Dormann-etal-2018}. We refer  interested readers to a recent review by \citet{Dormann-etal-2018}, which provides an in-depth discussion of model averaging in ecology.

% ***************************************
\section{\label{S.Diagnostics}Diagnostics and model validation for SSMs}

While model selection can help us identify which of the fitted models best describes the data, it rarely provides an assessment of the absolute fit of that model. As such, the selected model could be a poor representation of the data generating process (i.e., could poorly describe the ecological process and/or measurement process) and relative measures of fit, such as AIC, do not quantify how closely the model matches the data. Thus, before interpreting model results, it is crucial to carry out some of the following model diagnostics. First, it is essential to examine whether estimated parameters
seem biologically reasonable. For example, our understanding
of the system may stipulate that a response variable should increase with a covariate.
A model with parameter estimates inconsistent with such \textit{a priori}
understanding or with unrealistic effect sizes will be suspect. Second, it is important to assess the influence
of individual observations on estimated parameters. For example, outliers can have a strong influence on parameter estimates. Third, one should examine whether the model assumptions
are reasonable. For example, with SSMs, assumptions
are made about the probability distributions for states and observations (e.g., Eqs. \ref{E.state.NDLM}-\ref{E.obs.NDLM} assume both are normal). Fourth, it is important to examine the goodness of fit, which defines how well the model describes the data used to fit the model.  At the individual observation level, goodness of fit measures how far an observation is from its predicted value (e.g., $|y_t-\hat{y}_t|$, $t$=1,$\ldots$, $T$). At the model level, it
summarizes the overall fit of a model to all observations (e.g., the
average squared errors). Fifth, one ought to assess the model's predictive
accuracy, or how well the model predicts an outcome for an observation that was not used to fit the model (e.g., via cross-validation). With time-series models, including SSMs, one can use the first $t$ observations
to fit the model, and then use the model to predict the $t+1$ observation, or fit the model to all $T$ observations
and see how well future observations are predicted.

%--------------------------------------------------------------
\subsection{Challenges with SSMs}

For simple statistical
models, such as linear regression, diagnostics 
for most of the above features are well established. Diagnostics for SSMs, however, can be challenging for two reasons. First, observations are temporally dependent. 
Many diagnoses rely on response or conventional residuals, which we define as follows for our toy model:
\begin{linenomath*}
\begin{align}
     \label{E.naive.SSM.residual}
    e_{t|1:T} &= y_t - \hat{y}_{t|1:T},
\end{align}
\end{linenomath*}
where  $\hat{y}_{t|1:T}$ is the predicted observation at time $t$ given all observations. This predicted value depends on the smoothed state estimate at time $t$, $\hat{z}_{t|1:T}$, and the observation equation. For example, for our toy model (Eqs. \ref{E.state.NDLM}-\ref{E.obs.NDLM}), $\hat{y}_{t|1:T} = E[y_t|y_{1:T}] = \alpha\hat{z}_{t|1:T}$.  \citet{Harvey-1990}
notes that these response residuals are not serially
independent. Their use can
impair one's capacity to identify model misspecification
\citep{Harvey-1990}, and can have negative consequences for model inference and further model diagnosis \citep[e.g., inflated goodness of fit, ][]{Thygesen-etal-2017}.
Second, as for most hierarchical
models, we generally do 
not have direct observations of the hidden states, $z_t$, thus
one cannot directly compare predicted states with their ``true'' values.

Because of these challenges, researchers often fail to check the absolute
fit of SSMs, and thus risk making conclusions based on
a misspecified model or risk having biased parameter and state estimates. Here, we
provide a list of tools to help researchers perform this essential
model-checking step. We start with the tools commonly used to assess Bayesian hierarchical models. These tools can be easily used with frequentist and Bayesian SSMs alike, but have important limitations. We then discuss the tools that have been the focus of model validation developments for SSMs, which specifically address the issue of temporal dependence in the residuals. We end with methods relying on out-of-sample validation (e.g., cross-validation), which we believe is the gold standard for assessing the predictive ability of a model, and we hope will become the focus of future developments for SSMs. This order also reflects an increased division between the data used to estimate the model parameters and hidden states and the data used to perform the diagnostics.

%---------------------------------------------------
\subsection{Posterior predictive measures}

Posterior predictive checking is a common Bayesian method to quantify the discrepancies between the data and the model \citep{Gelman-etal-2013, Conn-etal-2018}. It has been used to verify the fit of SSMs to ecological data \citep[e.g.,][]{Hobbs-etal-2015}. The idea behind posterior predictive checking is that if the model fits the data well, then data generated from the model should have characteristics similar to those of the observed data \citep{Gelman-etal-2013}. These posterior predictive checks often involve calculating a posterior predictive p-value, $p_B$:
\begin{linenomath*}
\begin{equation}
    p_B = Pr(T(\mathbf{y}^i, \boldsymbol{\theta}) \geq T(\mathbf{y}, \boldsymbol{\theta}) |\mathbf{y}),
\end{equation}
\end{linenomath*}
where each $\mathbf{y}^i$
is a time series that has been simulated from the fitted model (i.e., representing a replicate time series that could have been observed from the model), $\mathbf{y}$ is the observed data, $\boldsymbol{\theta}$ contains the model parameters, and $T(\mathbf{y}, \boldsymbol{\theta})$ 
is a test quantity summarizing the data (e.g., the mean) or a discrepancy function (e.g., $\chi^2$
measure). This p-value is similar to the one used in frequentist inference. It measures the probability, under the model of interest, of finding a test quantity as extreme as that associated with the data. Posterior predictive checks use three steps: 1)  sample a set of posterior $\boldsymbol{\theta}$ values, 2) simulate one $\mathbf{y}^i$ from each, and 3) calculate the test quantity for each $\mathbf{y}^i$. We estimate the p-value with the proportion of the replicates that have a test quantity value greater or equal to that of the real data. Posterior predictive p-values near 0 or 1 indicate that the pattern observed with the data would be unlikely if the model were true. Thus, unlike p-values associated with classic statistical tests used to reject null hypotheses (e.g., t-test), we are seeking a posterior predictive p-value close to 0.5 not smaller than 0.05. The relevance of the p-value largely depends on the choice of test quantity. \citet{Hobbs-etal-2015} used the mean and standard deviation of the observed data, as well as a discrepancy function ($T(\mathbf{y,\boldsymbol{\theta}}) = \sum_{t=1}^T (y_t - \hat{y}_{t|1:T})^2 = \sum_{t=1}^T e_{t|1:T}^2$) that measures the disagreement between the SSM and the data. \citet{Newman-etal-2014} and \citet{Conn-etal-2018} provide lists of important alternative functions. Although we described posterior predictive checks in a Bayesian framework and have defined the test quantity as a function of $\mathbf{y}$, \citet{King-etal-2015} have applied similar concepts in a frequentist framework, using test quantities that describe characteristics of the estimated hidden states, $\mathbf{z}$ (e.g.,  autocorrelation function at lag 1 of the states).

\begin{figure}
\begin{center}
\includegraphics[width=470pt]{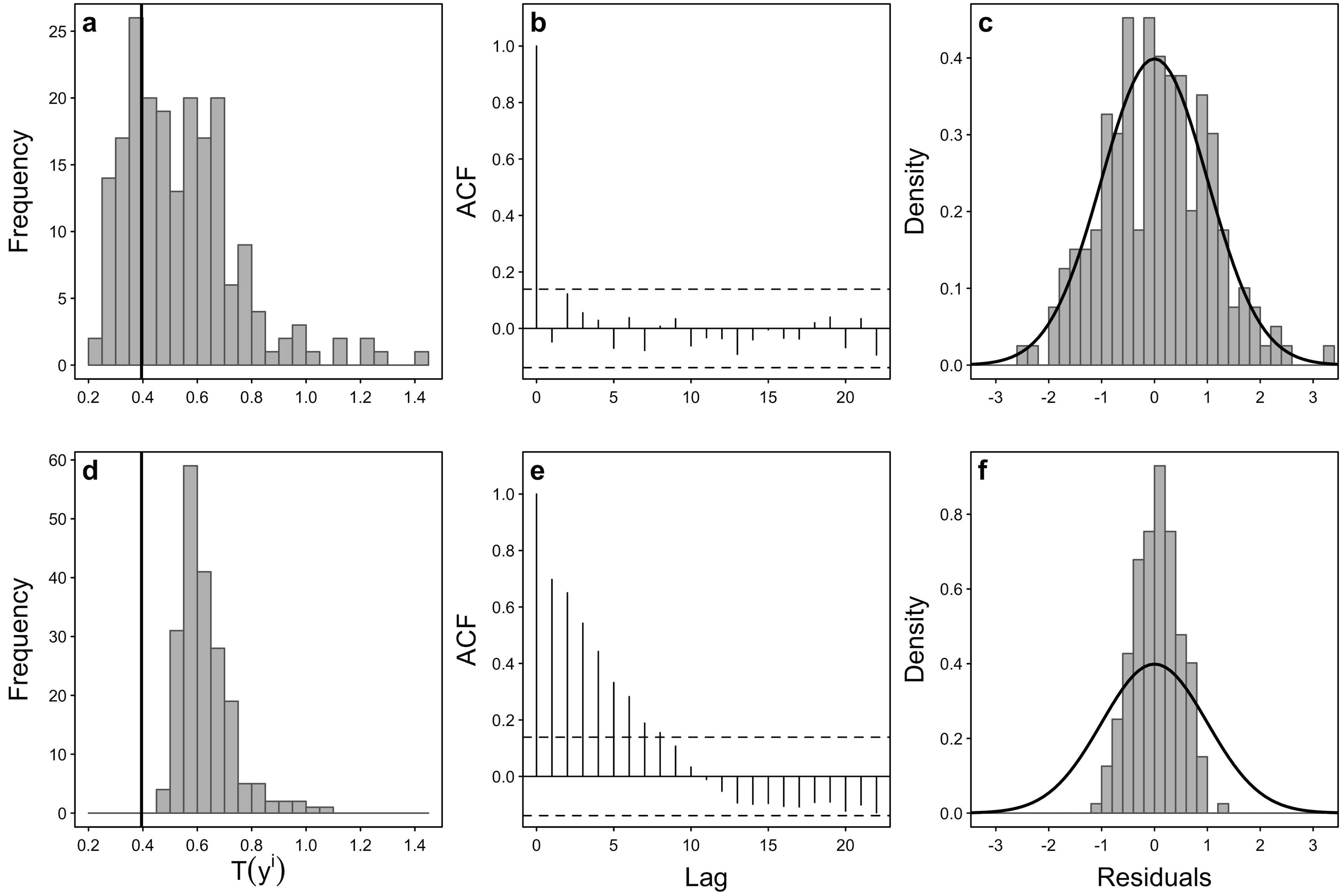}
\end{center}
\caption{Examples of diagnostic plots for a well specified (panels a, b, c) and a misspecified model (panels d, e, f). The data for all plots were simulated from the toy model (Eqs. \ref{E.state.NDLM}-\ref{E.obs.NDLM}) with $\alpha=\beta=1$ and the $\sigma_p = \sigma_o = 0.1$. The correctly specified model used the correct known values of $\alpha$ and $\beta$, and estimated $\sigma_p$ and $\sigma_a$. While the misspecified model also used the correct known values of $\alpha$ and $\beta$, it wrongly assumed a value of $\sigma_o = 0.5$ and then estimated $\sigma_p$. Panels a and d represent a frequentist version of the posterior predictive check, where the test quantity is the standard deviation of the observations, $T(\mathbf{y}) = \sqrt{\sum_{t=1}^{T} { (y_t - \bar{y})^2/(T-1)}}$. The histograms represent the frequency of test quantity for 200 datasets simulated using the estimated parameters with the original dataset. The vertical bar is the test quantity for the original dataset. Panels b and e represent the autocorrelation function of the one-step-ahead residuals. Panels c and f compare the distribution of the observed one-step-ahead residuals (histograms) to a standard normal probability density function (curves).}
\label{fig:diag}
\end{figure}

Although common, posterior predictive p-values have important limitations \citep{Conn-etal-2018}. Because they use the data twice, once to fit the model and once to test the model fit, they tend to be conservative (i.e., tend to return value closer to 0.5 than
to 0 or 1), and often have insufficient power to detect lack of fit. One can alter the method described above and generate all the observation replicates using only a single sample from  the posterior parameter distribution. This method was shown to have better theoretical properties (e.g., better Type I error rate control), and to detect lack of fit more reliably for some ecological hierarchical models \citep{Conn-etal-2018}. Following \citet{King-etal-2015}, we recommend assessing discrepancies between the SSM and the data by looking at where $T(\mathbf{y},\boldsymbol{\theta})$ falls in the frequency distribution of $T(\mathbf{y}^i,\boldsymbol{\theta})$ (Fig. \ref{fig:diag}a,d). This graphical method is also more useful in assessing the ecological importance of the discrepancies than looking at the p-value, and can provide a better sense of why the model may be inappropriate for the data \citep{Conn-etal-2018}.

Posterior predictive checks are also useful to assess the validity of model assumptions \citep{Gelman-etal-2013}. We can use a single sample from the posterior distribution of the hidden states to assess the assumptions associated with the process equation \citep{Thygesen-etal-2017}. For example, we can sample a time series of state, $\mathbf{z}^i$, from the posterior state distribution of our toy model to calculate the process variation as $\epsilon_t^i = z_t^i - \beta z_{t-1}^i$, and verify whether the $\epsilon_t^i$ are normally distributed with a mean of 0 as assumed by Eq. \ref{E.state.NDLM}. Departures from the assumed distribution (e.g., if the mean of the process variation is far from 0), indicate that the model is not adequate for the data. This method is generally recommended for assessing the assumptions of Bayesian hierarchical models \citep{Gelman-etal-2013}, but \citet{Thygesen-etal-2017} used the Laplace approximation implemented in \texttt{TMB} to create a posterior distribution of the states for non-Bayesian models.

%-------------------------------------------------------------
\subsection{One-step-ahead residuals and their extensions}
\label{S.One.step.ahead}

The model diagnostic that has received the most attention in the SSM literature is the one-step-ahead residuals \citep{Harvey-1990, Thygesen-etal-2017}, also known
as recursive residuals \citep{Fruhwirth-Schnatter-1996}. Unlike the response residuals
(Eq. \ref{E.naive.SSM.residual}), the one-step-ahead residuals should not have temporal dependence when the model is adequate because the residual 
for the $t^{th}$ observation uses the expected observation at time $t$ given
observations only up to time $t-1$:
\begin{linenomath*}
\begin{align}
    \label{E.one.step.ahead.residual}
    e_{t|1:t-1} &= y_t - \hat{y}_{t|1:t-1}.
\end{align}
\end{linenomath*}
Effectively, for response residuals (Eq. \ref{E.naive.SSM.residual}) we use the smoothed estimates of states, $\hat{z}_{t|1:T}$, to predict the observation at time $t$, while for one-step-ahead residuals, we use the prediction of the states, $\hat{z}_{t|1:t-1}$. In the context of a Kalman filter, we can calculate $\hat{y}_{t|1:t-1}$ using the one-step-ahead forecast prediction that is already calculated as part of
the recursive algorithm. As more information is available for fitting the model as $t$ increases, the variance of prediction residuals will tend to decrease with $t$. To account for this change in variance, it is useful to scale
the prediction residuals by their standard deviations (a procedure equivalent to calculating standardized Pearson residuals):
\begin{linenomath*}
\begin{align}
    \tilde{e}_{t|1:t-1} = 
    \frac{e_{t|1:t-1}}{sd(e_{t|1:t-1})}.
\end{align}
\end{linenomath*}
For the special case of SSMs with normally
distributed states and observations, such standardized residuals are independent
and identically distributed with a standard normal distribution and can be used to test a variety of assumptions. Diagnostic procedures include
qq-normal plots to check for normality, auto-correlation function plots
to see if the residuals are independent, and plots of the
residuals against observed values to check for non-constant
variance.

For non-normal SSMs, the probability distribution of these standardized residuals are not standard normal, making 
the exploration of residuals harder. Probability scores (P-scores), and their transformed version, prediction quantile residuals, are useful alternatives \citep{Fruhwirth-Schnatter-1996, Thygesen-etal-2017}. A P-score, $u_t$, is the
cumulative distribution function for the predicted
observations evaluated at the $t^{th}$ observed value:
\begin{linenomath*}
\begin{align}
 u_t = F_{Y_t|y_{1:t-1}}(y_t) &= \text{Pr}(Y_t \leq y_t | Y_{1:t-1} = y_{1:t-1}).
\end{align}
\end{linenomath*}
If $F_{Y_t|y_{1:t-1}}$ describes the cumulative distribution function of the true model, then the resulting
$u_t$ are uniformly distributed \citep{Conn-etal-2018}. Deviations from uniformity suggest model misspecification. As this is simply an
application of the probability integral transformation  \citep[i.e., 
if $Y$ has the cumulative density function $F_Y$, then $F_Y(Y)$ is distributed with Uniform(0,1),][]{Smith-1985}, these are a specific
case of probability integral transform (PIT) residuals \citep{Warton-etal-2017}. To get normally distributed residuals, we can transform the P-scores to
prediction quantile residuals, $v_t$,
as follows:
\begin{linenomath*}
\begin{equation}
  v_t = \Phi^{-1}(u_t),
\end{equation}
\end{linenomath*}
where $\Phi^{-1}$ is the inverse of the standard normal
cumulative distribution function (also known as the standard normal
quantile function). When the model is true,
$v_t$ should 
be an independent sample from a standard normal. Thus, we can assess whether the data fits the 
model assumptions using the same 
diagnostic procedures available for standardized one-step-ahead
prediction residuals in the case of normally distributed SSMs \citep[see Fig. \ref{fig:diag}b-c,e-f; and][]{Newman-etal-2014,Thygesen-etal-2017}.

P-scores and prediction quantile residuals can be difficult to estimate for non-normal SSMs because their  calculation requires knowledge of the cumulative distribution function (cdf) for $Y_t|y_{1:t-1}$, which
in many cases will not be known nor have an analytical
form. However, \citet{Thygesen-etal-2017} developed methods for approximating the cdf based on the Laplace approximation that can be implemented easily in \texttt{TMB}. Because this method depends on the Laplace approximation, it is important to assess the accuracy of this approximation (see Appendix S1: Section S1 1.2.3). The quantile residuals of \citet{Thygesen-etal-2017} are applicable to a broad range of frequentist SSMs, although there are some limitations in using them with multivariate time series. We are not aware of equivalent methods for as broad a range of Bayesian SSMs, although some exist for a limited class \citep{Fruhwirth-Schnatter-1996, Gamerman-etal-2013}.

%---------------------------------------
\subsection{Cross-validation}

While one-step-ahead residuals and their extensions remove data when calculating the expected value of the observation at time $t$, they use the complete dataset to estimate the model parameters. Thus, these residuals cannot be used to fully assess the predictive ability of the model. Assessing the predictive ability of a model is thought to be best achieved with out-of-sample data, where two independent datasets are used: one to fit (or train) the model and one to validate (or test or evaluate) it \citep{Hooten-Hobbs-2015}. While rarely done with SSMs, there are examples where independent information on the true values of the hidden states was collected \citep[e.g., ][and Appendix S1: Section S1 2.2]{AugerMethe-etal-2017}, a data stream was used as validation data \citep[e.g.,][]{Hobbs-etal-2015}, or part of a time series was selected as a validation time period \citep[e.g.,][]{Holdo-etal-2009}.

When using a single subset of the data as validation, we can only assess the predictive ability for those specific observations. Instead, one can use cross-validation methods that look at the predictive ability of all data points by sequentially leaving out small subsets of the data \citep{Hooten-Hobbs-2015}. $k$-fold cross validation is a ubiquitous statistical method, where $k$ groups of similar size sequentially serve as the validation dataset while the remaining $k-1$ groups are collectively used as the training set. Leave-one-out is a common version that 
leaves each of the data points out sequentially. To assess the predictive ability of the model, we can use score or discrepancy
functions, such as the mean squared prediction error (MSPE) for group $k$:
\begin{linenomath*}
\begin{align}
    \text{MSPE}_k &= \sum_{i=1}^{n} (\mathbf{y}_{i,k,\text{oos}}- \hat{\mathbf{y}}_{i,k,\text{oos}})^2/n,
\end{align}
\end{linenomath*}
where we assume that $T/k$ is an integer $n$, oos means out of sample, $\mathbf{y}_{i,k,\text{oos}}$ is the $i^{th}$ observation in subsample $k$, and $\hat{\mathbf{y}}_{i,k,\text{oos}}$ is the expected observation based on the model fitted to the dataset without sample $k$. As an overall value, we can then average the $k$ $\text{MSPE}_k$. Such functions directly assess the predictive ability of the model and thus are intuitive measures of how good a model is.

As mentioned in Section \ref{S.Model.Comparison}, cross-validation is also often deemed the preferred method for model comparison \citep{Gelman-etal-2014, Hooten-Hobbs-2015}. The model set can be ranked based on their predictive accuracy, with better models having lower prediction error \citep[e.g., lower MSPE or its square root, RMSPE,][]{Hooten-Hobbs-2015}. While cross-validation can be implemented relatively easily, it can be computationally demanding \citep{Link-Sauer-2016, Vehtari-etal-2017}. Cross-validation generally requires refitting the models $k$ times, which can be a daunting task with Bayesian models \citep[but see][for suggested solutions]{Hooten-Hobbs-2015}. In addition, cross-validation assumes that the training and evaluation datasets are independent \citep{Roberts-etal-2017}. The main challenge with using cross-validation with SSMs is that, due to the temporal dependency in the data, removing only a few data points will underestimate the prediction error and removing many will lead to propagation of error \citep{Newman-etal-2014}. 

Despite these drawbacks cross-validation is a powerful tool, which has been promoted for use with complex ecological models \citep{Link-etal-2017}. At present, there are few cross-validation methods specifically designed to handle the dependency structure of SSMs \citep{Ansley-Kohn-1987, DeJong-1988}. These are appropriate for only a restricted set of SSMs and appear to be rarely used. However, the time-series literature \citep[e.g.,][]{Tashman-2000, Bergmeir-Benitez-2012, Burkner-etal-2020} and the suggestions of \citet{Roberts-etal-2017} on block cross-validation methods to account for dependence structure in ecological data are useful starting points for the development and evaluation of such methods for SSMs. Cross-validation methods for time series include procedures analogous to the one-step-ahead residuals (Section \ref{S.One.step.ahead}), but where model parameters differ across the $k$ folds and are estimated using only observations prior to the expected values \citep{Hyndman-Athanasopoulos-2018}. One may need to consider additional modifications, such as whether one should use a rolling window for the training dataset \citep{Tashman-2000}.

The topic of model validation for SSM is one that has been relatively poorly studied, with a few notable exceptions \citep[e.g.,][]{Fruhwirth-Schnatter-1996,King-etal-2015, Thygesen-etal-2017}. Because of the additional parameter identifiability and estimability problems discussed in Section \ref{S.Estimability}, we believe this topic deserves more attention. Beyond the tools we have outlined above, SSM developers and users can gain inspiration from the tools developed for hierarchical models \citep[e.g., PIT-trap residuals,][]{Warton-etal-2017}. For researchers using Bayesian SSMs, we point readers towards the review of \citet{Conn-etal-2018} on model checking methods for Bayesian hierarchical models. Finally, we would like to remind readers that, while it is crucial to perform a model validation step, passing this step does not mean that the model is representing the truth. It simply means that one could not find difference between the data generating system and the model. This could be due to a low sample size or the conservative nature of some of the methods described above.

% **************************************************
\section{\label{S.Conclusion}Conclusion}

Through a diverse set of examples, we have demonstrated that SSMs are flexible models for time series that can be used to
answer a broad range of ecological questions. They can be used to model univariate or multivariate time series. SSMs can be linear or nonlinear, and have discrete or continuous time steps. They can have normal or non-normal sources of stochasticity, and thus can model continuous, count, binary, or categorical data. They are particularly useful when one
has significant process variation and observation error. Accounting for these sources of uncertainty can substantially affect management decisions, making SSMs the perfect modeling tool in many contexts \citep[e.g.,][]{Jamieson-Brooks-2004, Hobbs-etal-2015}.

As we have outlined, a variety of tools to fit SSMs to data exist. Historically, many researchers wrote SSMs so they could be fitted with the Kalman filter and its extensions. However, the diversity of fitting procedures available now allows researchers to create models that are more representative of the structure of their data and the ecological processes they are interested in. In addition, flexible fitting tools now exist in both the frequentist and Bayesian frameworks, allowing researchers to choose their preferred inferential framework rather than have their model dictate the framework they can use. Within each inferential framework, the choice of a fitting procedure will be a compromise between flexibility and efficiency. In particular, highly efficient fitting methods (e.g., Laplace approximation and Hamiltonian Monte Carlo) have more restrictions than their slower alternatives (e.g., particle filter and Gibbs).

While these tools provide the means to fit complex SSMs, it is crucial to appropriately formulate the model. As discussed, SSMs can suffer from parameter estimability problems, but various tools exist to assess whether this is the case and to identify the type of study design or model simplification that will resolve these problems. In general, making use of replication or including covariates can help reduce some of the common estimation problems.

Researchers often forgo doing model selection and validation with SSMs, but we advocate that these should become part of every SSM user's workflow. Model mispecification can affect ecological inferences and the accuracy of state estimates. While no model selection measure is perfect for SSMs, AIC and WAIC, can be useful. While model validation is also difficult with SSMs, posterior predictive measures, and one-step-ahead residuals and their extensions are relatively easy ways to assess whether the model describes the data well and whether some of the model assumptions are met. Cross-validation methods are often computationally expensive, but provides one of the best ways to select and evaluate models when correlation is handled appropriately.

While there are many tools already available to fit, compare, and validate SSMs, five topics warrant further research. First, while we advocate that SSMs be a default framework to model many ecological time series, it is important to pinpoint the conditions under which simpler alternatives perform adequately (e.g., when do models without observation error provide reliable parameter and state estimates?). Such research should account for the additional identifiability and estimability issues that comes with fitting SSMs and the types of datasets that allow SSMs to return reliable estimates. Second, as SSMs are often the primary tools used to analyze time series, it is important to explore the data collection designs that optimize the estimation of SSMs, so that the best data possible are collected. Third, there is a need for further developments of computationally efficient model selection procedures for SSMs. Using the marginal likelihood with AIC and WAIC appears most adequate for SSMs, especially if one has a single observation time series. However, we should explore when the conditional likelihood can be used and whether it affects the predictive accuracy of the states and parameters differently. To facilitate the uptake of WAIC based on the marginal likelihood, new \texttt{R} functions that automatically calculate this information criterion should be written. In addition, we it would be helpful to explore how newer tools to approximate predictive ability \citep[e.g.,][]{Vehtari-etal-2017, Burkner-etal-2020} perform with SSMs. Fourth, while there have been a few important advances in model validation methods for SSMs, this remains a relatively untouched research area. Given the increasing use of SSMs in management, it is crucial that a broader range of validation methods be developed for these complex models. Fifth, with the increasing efficiency of fitting procedures, cross-validation is becoming a feasible procedure to assess predictive accuracy and compare models. As such, the time is ripe to start developing proper cross-validation procedures that will account for dependencies in the data.

Overall, we provided a review of the topics needed to formulate and fit SSMs to ecological data, and Appendix S1 provide an extensive set of examples of methods to facilitate this process. We hope this guide will help researchers develop and apply SSMs to
their own data, and foster the development of SSMs in multiple fields of ecology.

% ***************************************************
\section{Acknowledgments}

This paper was instigated during a Banff International Research Station (BIRS) workshop hosted at the Casa Matem\'atica Oaxaca (CMO) entitled \textit{New perspectives on state-space models}. We thank BIRS and CMO for their support and the lead organizer of the workshop, David Campbell, and all participants for their insights. This effort was also supported by the Canadian Statistical Sciences Institute through a Collaborative Research Team Project led by JMF. MAM thanks the Natural Sciences and Engineering Research Council of Canada and the Canada Research Chairs Program. We thank Andrew Derocher for the polar bear track used in Fig. \ref{fig:polarbear} and Appendix S1. We also thank Devin Lyons, Perry de Valpine, Aki Vehtari, and three anonymous reviewers for their insightful comments on previous versions of the manuscript.

% ----------------------------------------

\bibliographystyle{apalike}

\bibliography{references_ssm}

\begin{thebibliography}{}

\bibitem[Abadi et~al., 2010]{Abadi-etal-2010}
Abadi, F., Gimenez, O., Ullrich, B., Arlettaz, R., and Schaub, M. (2010).
\newblock Estimation of immigration rate using integrated population models.
\newblock {\em Journal of Applied Ecology}, 47:393--400.

\bibitem[Aeberhard et~al., 2018]{Aeberhard-etal-2018}
Aeberhard, W.~H., {Mills Flemming}, J., and Nielsen, A. (2018).
\newblock {Review of state-space models for fisheries Science}.
\newblock {\em Annual Review of Statistics and Its Application}, 5:215--235.

\bibitem[Aho et~al., 2014]{Aho-etal-2014}
Aho, K., Derryberry, D., and Peterson, T. (2014).
\newblock {Model selection for ecologists: the worldviews of AIC and BIC}.
\newblock {\em Ecology}, 95:631--636.

\bibitem[Albertsen et~al., 2015]{Albertsen-etal-2015}
Albertsen, C.~M., Whoriskey, K., Yurkowski, D., Nielsen, A., and {Mills
  Flemming}, J. (2015).
\newblock {Fast fitting of non-Gaussian state-space models to animal movement
  data via Template Model Builder}.
\newblock {\em Ecology}, 96:2598--2604.

\bibitem[Anderson-Sprecher and Ledolter, 1991]{Anderson-Sprecher-1991}
Anderson-Sprecher, R. and Ledolter, J. (1991).
\newblock State-space analysis of wildlife telemetry data.
\newblock {\em Journal of the American Statistical Association}, 86:596--602.

\bibitem[Andrieu et~al., 2010]{Andrieu-etal-2010}
Andrieu, C., Doucet, A., and Holenstein, R. (2010).
\newblock {Particle Markov chain Monte Carlo methods}.
\newblock {\em Journal of the Royal Statistical Society. Series B: Statistical
  Methodology}, 72:269--342.

\bibitem[Ansley and Kohn, 1987]{Ansley-Kohn-1987}
Ansley, G.~F. and Kohn, R. (1987).
\newblock {Efficient Generalized Cross-Validation for State Space Models}.
\newblock {\em Biometrika}, 74:139--148.

\bibitem[Appling et~al., 2018]{Appling-etal-2018}
Appling, A.~P., Hall, R. O.~J., Yackulic, C.~B., and Arroita, M. (2018).
\newblock {Overcoming equifinality: leveraging long time series for stream
  metabolism estimation}.
\newblock {\em Journal of Geophysical Research: Biogeosciences}, 123:624--645.

\bibitem[Auger-M\'eth\'e et~al., 2017]{AugerMethe-etal-2017}
Auger-M\'eth\'e, M., Albertsen, C.~M., Jonsen, I.~D., Derocher, A.~E., Lidgard,
  D.~C., Studholme, K.~R., Bowen, W.~D., Crossin, G.~T., and {Mills Flemming},
  J. (2017).
\newblock {Spatiotemporal modelling of marine movement data using Template
  Model Builder (TMB)}.
\newblock {\em Marine Ecology Progress Series}, 565:237--249.

\bibitem[Auger-M{\'{e}}th{\'{e}} et~al., 2016]{AugerMethe-etal-2016}
Auger-M{\'{e}}th{\'{e}}, M., Field, C., Albertsen, C.~M., Derocher, A.~E.,
  Lewis, M.~a., Jonsen, I.~D., and Flemming, J.~M. (2016).
\newblock {State-space models' dirty little secrets: even simple linear
  Gaussian models can have estimation problems}.
\newblock {\em Scientific Reports}, 6:26677.

\bibitem[Baldwin et~al., 2018]{Baldwin-etal-2018}
Baldwin, J.~W., Leap, K., Finn, J.~T., and Smetzer, J.~R. (2018).
\newblock {Bayesian state-space models reveal unobserved off-shore nocturnal
  migration from Motus data}.
\newblock {\em Ecological Modelling}, 386:38--46.

\bibitem[Bell et~al., 2015]{Bell-etal-2015}
Bell, D.~M., Ward, E.~J., Oishi, A.~C., Oren, R., Flikkema, P.~G., and Clark,
  J.~S. (2015).
\newblock {A state-space modeling approach to estimating canopy conductance and
  associated uncertainties from sap flux density data}.
\newblock {\em Tree Physiology}, 35:792--802.

\bibitem[Bengtsson and Cavanaugh, 2006]{Bengtsson-Cavanaugh-2006}
Bengtsson, T. and Cavanaugh, J.~E. (2006).
\newblock {An improved Akaike information criterion for state-space model
  selection}.
\newblock {\em Computational Statistics and Data Analysis}, 50:2635--2654.

\bibitem[Berg and Nielsen, 2016]{Berg-Nielsen-2016}
Berg, C.~W. and Nielsen, A. (2016).
\newblock {Accounting for correlated observations in an age-based state-space
  stock assessment model.}
\newblock {\em ICES Journal of Marine Science: Journal du Conseil},
  73:1788--1797.

\bibitem[Bergmeir and Ben{\'{i}}tez, 2012]{Bergmeir-Benitez-2012}
Bergmeir, C. and Ben{\'{i}}tez, J.~M. (2012).
\newblock {On the use of cross-validation for time series predictor
  evaluation}.
\newblock {\em Information Sciences}, 191:192--213.

\bibitem[Besbeas et~al., 2002]{Besbeas-etal-2002}
Besbeas, P., Freeman, S.~N., Morgan, B. J.~T., and Catchpole, E.~A. (2002).
\newblock Integrating mark-recapture-recovery and census data to estimate
  animal abundance and demographic parameters.
\newblock {\em Biometrics}, 58:540--547.

\bibitem[Besbeas and Morgan, 2019]{Besbeas-Morgan-2019}
Besbeas, P. and Morgan, B. J.~T. (2019).
\newblock {Exact inference for integrated population modelling}.
\newblock {\em Biometrics}, 75:475--484.

\bibitem[Best and Punt, 2020]{Best-Punt-2020}
Best, J.~K. and Punt, A.~E. (2020).
\newblock {Parameterizations for Bayesian state-space surplus production
  models}.
\newblock {\em Fisheries Research}, 222:105411.

\bibitem[{Betancourt}, 2017]{Betancourt:2017}
{Betancourt}, M. (2017).
\newblock {A conceptual introduction to Hamiltonian Monte Carlo}.
\newblock {\em arXiv e-prints}, page arXiv:1701.02434.

\bibitem[Betancourt et~al., 2020]{Betancourt-etal-2020}
Betancourt, M., Margossian, C.~C., and Leos-Barajas, V. (2020).
\newblock {The discrete adjoint method: efficient derivatives for functions of
  discrete sequences}.
\newblock {\em arXiv e-prints}, page arXiv:2002.00326.

\bibitem[Boero et~al., 2015]{Boero-etal-2015}
Boero, F., Kraberg, A.~C., Krause, G., and Wiltshire, K.~H. (2015).
\newblock {Time is an affliction: why ecology cannot be as predictive as
  physics and why it needs time series}.
\newblock {\em Journal of Sea Research}, 101:12--18.

\bibitem[Bolker, 2008]{Bolker-2008}
Bolker, B.~M. (2008).
\newblock {\em Ecological models and data in R}.
\newblock Princeton University Press, Princeton, NJ.

\bibitem[Bolker et~al., 2009]{Bolker-etal-2009}
Bolker, B.~M., Brooks, M.~E., Clark, C.~J., Geange, S.~W., Poulsen, J.~R.,
  Stevens, M. H.~H., and White, J. S.~S. (2009).
\newblock {Generalized linear mixed models: a practical guide for ecology and
  evolution}.
\newblock {\em Trends in Ecology \& Evolution}, 24:127--135.

\bibitem[Boone et~al., 2014]{Boone-etal-2014}
Boone, E.~L., Merrick, J.~R., and Krachey, M.~J. (2014).
\newblock {A Hellinger distance approach to MCMC diagnostics}.
\newblock {\em Journal of Statistical Computation and Simulation}, 84:833--849.

\bibitem[Breed et~al., 2012]{Breed-etal-2012}
Breed, G.~A., Costa, D.~P., Jonsen, I.~D., Robinson, P.~W., and {Mills
  Flemming}, J. (2012).
\newblock {State-space methods for more completely capturing behavioral
  dynamics from animal tracks}.
\newblock {\em Ecological Modelling}, 235-236:49--58.

\bibitem[Brooks and Gelman, 1998]{Brooks-Gelman-1998}
Brooks, S.~P. and Gelman, A. (1998).
\newblock {General methods for monitoring convergence of iterative
  simulations}.
\newblock {\em Journal of Computational and Graphical Statistics}, 7:434--455.

\bibitem[Buckland et~al., 1997]{Buckland-etal_1997}
Buckland, S.~T., Burnham, K.~P., and Augustin, N.~H. (1997).
\newblock {Model selection: an integral part of inference}.
\newblock {\em Biometrics}, 53:603--618.

\bibitem[Buckland et~al., 2004]{Buckland-etal-2004}
Buckland, S.~T., Newman, K.~B., Thomas, L., and Koesters, N.~B. (2004).
\newblock {State-space models for the dynamics of wild animal populations}.
\newblock {\em Ecological Modelling}, 171:157--175.

\bibitem[B{\"{u}}rkner et~al., 2020]{Burkner-etal-2020}
B{\"{u}}rkner, P.-C., Gabry, J., and Vehtari, A. (2020).
\newblock {Approximate leave-future-out cross-validation for Bayesian time
  series models}.
\newblock {\em Journal of Statistical Computation and Simulation},
  90:2499--2523.

\bibitem[Burnham and Anderson, 2002]{Burnham-Anderson-2002}
Burnham, K.~P. and Anderson, D.~R. (2002).
\newblock {\em Model selection and multimodel inference: a practical
  information-theoretic approach}.
\newblock Springer-Verlag, New York, NY, 2nd edition.

\bibitem[Campbell and Lele, 2014]{Campbellele2014}
Campbell, D. and Lele, S. (2014).
\newblock An anova test for parameter estimability using data cloning with
  application to statistical inference for dynamic systems.
\newblock {\em Computational Statistics and Data Analysis}, 70:257--267.

\bibitem[Carter and Kohn, 1994]{Carter+Kohn:1994}
Carter, C.~K. and Kohn, R. (1994).
\newblock On {G}ibbs sampling for state space models.
\newblock {\em Biometrika}, 81:541--553.

\bibitem[Catchpole et~al., 2001]{Catchpoleetal2001}
Catchpole, E.~A., Kgosi, P.~M., and Morgan, B. J.~T. (2001).
\newblock On the near-singularity of models for animal recovery data.
\newblock {\em Biometrics}, 57:720--726.

\bibitem[Catchpole and Morgan, 1997]{CatchpoleandMorgan1997}
Catchpole, E.~A. and Morgan, B. J.~T. (1997).
\newblock Detecting parameter redundancy.
\newblock {\em Biometrika}, 84:187--196.

\bibitem[Catchpole et~al., 1998]{Catchpoleetal1998}
Catchpole, E.~A., Morgan, B. J.~T., and Freeman, S.~N. (1998).
\newblock Estimation in parameter-redundant models.
\newblock {\em Biometrika}, 85:462--468.

\bibitem[Cavanaugh and Shumway, 1997]{Cavanaugh-Shumway-1997}
Cavanaugh, J. and Shumway, R. (1997).
\newblock {A bootstrap variant of the AIC for state-space model selection}.
\newblock {\em Statistica Sinica}, 7:473--496.

\bibitem[Chaloupka and Balazs, 2007]{Chaloupka-Balazs-2007}
Chaloupka, M. and Balazs, G. (2007).
\newblock {Using Bayesian state-space modelling to assess the recovery and
  harvest potential of the Hawaiian green sea turtle stock}.
\newblock {\em Ecological Modelling}, 205:93--109.

\bibitem[Chang et~al., 2015]{Chang-etal-2015}
Chang, Y.-J., Brodziak, J., O'Malley, J., Lee, H.~H., DiNardo, G., and Sun,
  C.-L. (2015).
\newblock {Model selection and multi-model inference for Bayesian surplus
  production models: a case study for Pacific blue and striped marlin}.
\newblock {\em Fisheries Research}, 166:129--139.

\bibitem[Choquet and Cole, 2012]{ChoquetandCole2012}
Choquet, R. and Cole, D.~J. (2012).
\newblock A hybrid symbolic-numerical method for determining model structure.
\newblock {\em Mathematical Biosciences}, 236:117--125.

\bibitem[Choquet and Gimenez, 2012]{Choquet-Gimenez-2012}
Choquet, R. and Gimenez, O. (2012).
\newblock {Towards built-in capture – recapture mixed models in program
  E-SURGE}.
\newblock {\em Journal of Ornithology}, 152:S625--S639.

\bibitem[Clark et~al., 2011]{Clark-etal-2011}
Clark, J.~S., Bell, D.~M., Hersh, M.~H., and Nichols, L. (2011).
\newblock {Climate change vulnerability of forest biodiversity: climate and
  competition tracking of demographic rates}.
\newblock {\em Global Change Biology}, 17:1834--1849.

\bibitem[Cobelli and DiStefano~III, 1980]{CobelliDiStefano1980}
Cobelli, C. and DiStefano~III, J. (1980).
\newblock Parameter and structural identifiability concepts and ambiguities: a
  critical review and analysis.
\newblock {\em American Journal of Physiology - Regulatory, Integrative and
  Comparative Physiology}, 239:7--24.

\bibitem[Cole, 2019]{Cole-2019}
Cole, D.~J. (2019).
\newblock {Parameter redundancy and identifiability in hidden Markov models}.
\newblock {\em Metron}, 77:105--118.

\bibitem[Cole and McCrea, 2016]{ColeandMcCrea2016}
Cole, D.~J. and McCrea, R.~S. (2016).
\newblock Parameter redundancy in discrete state-space and integrated models.
\newblock {\em Biometrical Journal}, 58:1071--1090.

\bibitem[Cole et~al., 2012]{Coleetal2012}
Cole, D.~J., Morgan, B. J.~T., Catchpole, E.~A., and Hubbard, B.~A. (2012).
\newblock Parameter redundancy in mark-reovery models.
\newblock {\em Biometrical Journal}, 54:507--523.

\bibitem[Cole et~al., 2010]{Coleetal2010}
Cole, D.~J., Morgan, B. J.~T., and Titterington, D.~M. (2010).
\newblock Determining the parametric structure of models.
\newblock {\em Mathematical Biosciences}, 228:16--30.

\bibitem[Conn et~al., 2018]{Conn-etal-2018}
Conn, P.~B., Johnson, D.~S., Williams, P.~J., Melin, S.~R., and Hooten, M.~B.
  (2018).
\newblock {A guide to Bayesian model checking for ecologists}.
\newblock {\em Ecological Monographs}, 88:526--542.

\bibitem[Costa et~al., 2010]{Costa-etal-2010}
Costa, D.~P., Robinson, P.~W., Arnould, J. P.~Y., Harrison, A.~L., Simmons,
  S.~E., Hassrick, J.~L., Hoskins, A.~J., Kirkman, S.~P., Oosthuizen, H.,
  Villegas-Amtmann, S., and Crocker, D.~E. (2010).
\newblock {Accuracy of ARGOS locations of pinnipeds at-sea estimated using
  fastloc GPS}.
\newblock {\em PLoS ONE}, 5:e8677.

\bibitem[Cowles and Carlin, 1996]{Cowles-Carlin-1996}
Cowles, M.~K. and Carlin, B.~P. (1996).
\newblock {Markov Chain Monte Carlo convergence diagnostics: a comparative
  review}.
\newblock {\em Journal of the American Statistical Association}, 91:883--904.

\bibitem[Cressie et~al., 2009]{Cressie-etal-2009}
Cressie, N., Calder, C.~A., Clark, J.~S., {Ver Hoef}, J.~M., and Wikle, C.~K.
  (2009).
\newblock {Accounting for uncertainty in ecological analysis: the strengths and
  limitations of hierarchical statistical modeling}.
\newblock {\em Ecological Applications}, 19:553--570.

\bibitem[Csill{\'{e}}ry et~al., 2010]{Csillery-etal-2010}
Csill{\'{e}}ry, K., Blum, M. G.~B., Gaggiotti, O.~E., and Fran{\c{c}}ois, O.
  (2010).
\newblock {Approximate Bayesian Computation (ABC) in practice.}
\newblock {\em Trends in Ecology \& Evolution}, 25:410--418.

\bibitem[Damgaard, 2012]{Damgaard-2012}
Damgaard, C. (2012).
\newblock {Trend analyses of hierarchical pin-point cover data}.
\newblock {\em Ecology}, 93(6):1269--1274.

\bibitem[Damgaard, 2019]{Damgaard-2019}
Damgaard, C. (2019).
\newblock {A critique of the space-for-time substitution practice in community
  ecology}.
\newblock {\em Trends in Ecology \& Evolution}, 34:416--421.

\bibitem[Damgaard et~al., 2011]{Damgaard-etal-2011}
Damgaard, C., Nygaard, B., Ejrn{\ae}s, R., and Kollmann, J. (2011).
\newblock {State-space modeling indicates rapid invasion of an alien shrub in
  coastal dunes}.
\newblock {\em Journal of Coastal Research}, 27:595--599.

\bibitem[de~Jong, 1988]{DeJong-1988}
de~Jong, P. (1988).
\newblock {A cross-validation filter for time series models}.
\newblock {\em Biometrika}, 75:594--600.

\bibitem[{de Valpine}, 2002]{DeValpine-2002}
{de Valpine}, P. (2002).
\newblock {Review of methods for fitting time-series models with process and
  observation error and likelihood calculations for nonlinear, non-Gaussian
  state-space models}.
\newblock {\em Bulletin of Marine Science}, 70:455--471.

\bibitem[de~Valpine, 2004]{deValpine-2004}
de~Valpine, P. (2004).
\newblock Monte carlo state-space likelihoods by weighted posterior kernel
  density estimation.
\newblock {\em Journal of the American Statistical Association}, 99:523--536.

\bibitem[de~Valpine, 2012]{DeValpine-2012}
de~Valpine, P. (2012).
\newblock {Frequentist analysis of hierarchical models for population dynamics
  and demographic data}.
\newblock {\em Journal of Ornithology}, 152:S393--S408.

\bibitem[de~Valpine and Hastings, 2002]{deValpine-Hastings-2002}
de~Valpine, P. and Hastings, A. (2002).
\newblock {Fitting population models incorporating process noise and
  observation error}.
\newblock {\em Ecological Monographs}, 72:57--76.

\bibitem[{de Valpine} et~al., 2017]{deValpine_et_al:2017}
{de Valpine}, P., Turek, D., Paciorek, C., Anderson-Bergman, C., {Temple Lang},
  D., and Bodik, R. (2017).
\newblock Programming with models: writing statistical algorithms for general
  model structures with nimble.
\newblock {\em Journal of Computational and Graphical Statistics}, 26:403--413.

\bibitem[Dennis and Ponciano, 2014]{Dennis-etal-2014}
Dennis, B. and Ponciano, J.~M. (2014).
\newblock {Density-dependent state-space model for population-abundance data
  with unequal time intervals}.
\newblock {\em Ecology}, 95:2069--2076.

\bibitem[Dennis et~al., 2006]{Dennis-etal-2006}
Dennis, B., Ponciano, J.~M., Lele, S.~R., Taper, M.~L., and Staples, D.~F.
  (2006).
\newblock {Estimating density dependence, process noise, and observation
  error}.
\newblock {\em Ecological Monographs}, 76:323--341.

\bibitem[Dennis et~al., 2010]{Dennis-etal-2010}
Dennis, B., Ponciano, J.~M., and Taper, M.~L. (2010).
\newblock {Replicated sampling increases efficiency in monitoring biological
  populations}.
\newblock {\em Ecology}, 91:610--620.

\bibitem[Dennis and Taper, 1994]{dennis1994}
Dennis, B. and Taper, M.~L. (1994).
\newblock Density dependence in time series observations of natural
  populations: Estimation and testing.
\newblock {\em Ecological Monographs}, 64:205--224.

\bibitem[Dormann et~al., 2018]{Dormann-etal-2018}
Dormann, C.~F., Calabrese, J.~M., Guillera-Arroita, G., Matechou, E., Bahn, V.,
  Barto{\'{n}}, K., Beale, C.~M., Ciuti, S., Elith, J., Gerstner, K., Guelat,
  J., Keil, P., Lahoz-Monfort, J.~J., Pollock, L.~J., Reineking, B., Roberts,
  D.~R., Schr{\"{o}}der, B., Thuiller, W., Warton, D.~I., Wintle, B.~A., Wood,
  S.~N., W{\"{u}}est, R.~O., and Hartig, F. (2018).
\newblock {Model averaging in ecology: a review of Bayesian,
  information-theoretic, and tactical approaches for predictive inference}.
\newblock {\em Ecological Monographs}, 88:485--504.

\bibitem[Douc et~al., 2011]{Douc-etal-2011}
Douc, R., Moulines, E., Olsson, J., and van Handel, R. (2011).
\newblock {Consistency of the maximum likelihood estimator for general hidden
  Markov models}.
\newblock {\em The Annals of Statistics}, 39:474--513.

\bibitem[Doucet et~al., 2001]{Doucet-etal-2001}
Doucet, A., De~Freitas, N., and Gordon, N. (2001).
\newblock An introduction to sequential {Monte Carlo} methods.
\newblock In {\em Sequential Monte Carlo methods in practice}, pages 3--14.
  Springer.

\bibitem[Dukic et~al., 2012]{Dukic-etal-2012}
Dukic, V., Lopes, H.~F., and Polson, N.~G. (2012).
\newblock Tracking epidemics with {G}oogle flu trends data and a state-space
  {SEIR} model.
\newblock {\em Journal of the American Statistical Association},
  107:1410--1426.

\bibitem[Dunham and Grand, 2016]{Dunham-Grand-2016}
Dunham, K. and Grand, J.~B. (2016).
\newblock {Effects of model complexity and priors on estimation using
  sequential importance sampling/resampling for species conservation}.
\newblock {\em Ecological Modelling}, 340:28--36.

\bibitem[Dupuis, 1995]{Dupuis-1995}
Dupuis, J.~A. (1995).
\newblock {Bayesian estimation of movement and survival probabilities from
  capture-recapture data}.
\newblock {\em Biometrika}, 82:761--772.

\bibitem[Durbin and Koopman, 2012]{Durbin-and-Koopman-2012}
Durbin, J. and Koopman, S.~J. (2012).
\newblock {\em Time series analysis by state space methods}.
\newblock Oxford University Press, Oxford, UK, 2nd edition.

\bibitem[Einarsson et~al., 2016]{Einarsson-etal-2016}
Einarsson, {\'A}., Hauptfleisch, U., Leavitt, P.~R., and Ives, A.~R. (2016).
\newblock {Identifying consumer-resource population dynamics using ­
  paleoecological data}.
\newblock {\em Ecology}, 97:361--371.

\bibitem[Fasiolo et~al., 2016]{Fasiolo-etal-2016}
Fasiolo, M., Pya, N., and Wood, S.~N. (2016).
\newblock {A comparison of inferential methods for highly nonlinear state space
  models in ecology and epidemiology}.
\newblock {\em Statistical Science}, 31:96--118.

\bibitem[Fasiolo and Wood, 2018]{Fasiolo-Wood-2018}
Fasiolo, M. and Wood, S.~N. (2018).
\newblock {ABC in Ecological Modelling}.
\newblock In Sisson, S.~A., Fan, Y., and Beaumont, M., editors, {\em Handbook
  of approximate Bayesian computation}, pages 597--622. Chapman and Hall/CRC,
  New York, NY.

\bibitem[Ferretti et~al., 2018]{Ferretti-etal-2018}
Ferretti, F., Curnick, D., Liu, K., Romanov, E.~V., and Block, B.~A. (2018).
\newblock {Shark baselines and the conservation role of remote coral reef
  ecosystems}.
\newblock {\em Science Advances}, 4:eaaq033.

\bibitem[Fournier et~al., 2012]{Fournier-etal-2012}
Fournier, D.~A., Skaug, H.~J., Ancheta, J., Ianelli, J., Magnusson, A.,
  Maunder, M.~N., Nielsen, A., and Sibert, J. (2012).
\newblock {AD Model Builder: using automatic differentiation for statistical
  inference of highly parameterized complex nonlinear models}.
\newblock {\em Optimization Methods and Software}, 27:233--249.

\bibitem[Fr{\"{u}}hwirth-Schnatter, 1996]{Fruhwirth-Schnatter-1996}
Fr{\"{u}}hwirth-Schnatter, S. (1996).
\newblock {Recursive residuals and model diagnostics for normal and non-normal
  state space models}.
\newblock {\em Environmental and Ecological Statistics}, 3:291--309.

\bibitem[Fujiwara et~al., 2005]{Fujiwara-etal-2005}
Fujiwara, M., Kendall, B.~E., Nisbet, R.~M., and Bennett, W.~A. (2005).
\newblock {Analysis of size trajectory data using an energetic-based growth
  model}.
\newblock {\em Ecology}, 86:1441--1451.

\bibitem[Gamerman et~al., 2013]{Gamerman-etal-2013}
Gamerman, D., {Rezende dos Santos}, T., and Franco, G.~C. (2013).
\newblock {A non-{Gaussian} family of state-space models with exact marginal
  likelihood}.
\newblock {\em Journal of Time Series Analysis}, 34:625--645.

\bibitem[Garrett and Zeger, 2000]{Garrett-Zeger-2000}
Garrett, E.~S. and Zeger, S.~L. (2000).
\newblock Latent class model diagnosis.
\newblock {\em Biometrics}, 56:1055--1067.

\bibitem[Gelfand and Sahu, 1999]{Gelfand-Sahu-1999}
Gelfand, A.~E. and Sahu, S.~K. (1999).
\newblock {Identifiability, improper priors, and Gibbs sampling for generalized
  linear models}.
\newblock {\em Journal of the American Statistical Association}, 94:247--253.

\bibitem[Gelman et~al., 2013]{Gelman-etal-2013}
Gelman, A., Carlin, J.~B., Stern, H.~S., Dunson, D.~B., Vehtari, A., and Rubin,
  D.~B. (2013).
\newblock {\em {Bayesian data analysis}}.
\newblock CRC Press, Boca Raton, FL, 3rd edition.

\bibitem[Gelman et~al., 2014]{Gelman-etal-2014}
Gelman, A., Hwang, J., and Vehtari, A. (2014).
\newblock {Understanding predictive information criteria for Bayesian models}.
\newblock {\em Statistics and Computing}, 24:997--1016.

\bibitem[Gelman and Rubin, 1992]{Gelman-Rubin-1992}
Gelman, A. and Rubin, D.~B. (1992).
\newblock {Inference from iterative simulation using multiple sequences}.
\newblock {\em Statistical Science}, 7:457--511.

\bibitem[Gelman and Shirley, 2011]{Gelman-Shirley-2011}
Gelman, A. and Shirley, K. (2011).
\newblock {Inference from simulations and monitoring convergence}.
\newblock In Brooks, S., Gelman, A., Jones, G., and Meng, X.-L., editors, {\em
  Handbook of Markov chain Monte Carlo}, chapter~6, pages 163--174. Chapman and
  Hall/CRC, New York, NY.

\bibitem[Gelman et~al., 2017]{Gelman-etal-2017}
Gelman, A., Simpson, D., and Betancourt, M. (2017).
\newblock {The prior can often only be understood in the context of the
  likelihood}.
\newblock {\em Entropy}, 19:555.

\bibitem[Geyer, 2011]{Geyer:2011}
Geyer, C. (2011).
\newblock Introduction to {M}arkov {C}hain {M}onte {C}arlo.
\newblock In Brooks, S., Gelman, A., Jones, G., and Meng, X.-L., editors, {\em
  Handbook of {M}arkov {C}hain {M}onte {C}arlo}, chapter~1, pages 3--48.
  Chapman and Hall/CRC, New York, NY.

\bibitem[Gilks et~al., 1995]{Gilks-etal-1995}
Gilks, W.~R., Richardson, S., and Spiegelhalter, D. (1995).
\newblock {\em Markov chain Monte Carlo in practice}.
\newblock Chapman and Hall/CRC, New York, NY.

\bibitem[Gimenez et~al., 2009a]{gimenez2009}
Gimenez, O., Bonner, S., King, R., A.~Parker, R., Brooks, S., Jamieson, L.,
  Grosbois, V., Morgan, B., and Thomas, L. (2009a).
\newblock Winbugs for population ecologists: Bayesian modeling using markov
  chain monte carlo methods.
\newblock {\em Environmental and Ecological Statistics}, 3:883--915.

\bibitem[Gimenez et~al., 2009b]{Gimenez-etal-2009-indentifiability}
Gimenez, O., Morgan, B. J.~T., and Brooks, S.~P. (2009b).
\newblock {Weak identifiability in models for mark-recapture-recovery data}.
\newblock In Thomson, D.~L., Cooch, E.~G., and Conroy, M.~J., editors, {\em
  Modeling demographic processes in marked populations}, pages 1055--1067.
  Environmental and Ecological Statistics Series.

\bibitem[Gimenez et~al., 2007]{Gimenez-etal-2007}
Gimenez, O., Rossi, V., Choquet, R., Dehais, C., Doris, B., Varella, H., Vila,
  J.~P., and Pradel, R. (2007).
\newblock {State-space modelling of data on marked individuals}.
\newblock {\em Ecological Modelling}, 206:431--438.

\bibitem[Gimenez et~al., 2004]{Gimenez-etal-2004}
Gimenez, O., Viallefont, A., Catchpole, E.~A., Choquet, R., and Morgan, B.
  J.~T. (2004).
\newblock Methods for investigating parameter redundancy.
\newblock {\em Animal Biodiversity and Conservation}, 27:1--12.

\bibitem[{Gordon} et~al., 1993]{Gordon-etal-1993}
{Gordon}, N.~J., {Salmond}, D.~J., and {Smith}, A. F.~M. (1993).
\newblock Novel approach to nonlinear/non-gaussian bayesian state estimation.
\newblock {\em IEE Proceedings F - Radar and Signal Processing}, 140:107--113.

\bibitem[{Goudie} et~al., 2017]{Goudie-etal-2017}
{Goudie}, R. J.~B., {Turner}, R.~M., {De Angelis}, D., and {Thomas}, A. (2017).
\newblock {MultiBUGS: A parallel implementation of the BUGS modelling framework
  for faster Bayesian inference}.
\newblock {\em arXiv e-prints}, page arXiv:1704.03216.

\bibitem[Grewal and Andrews, 2010]{Grewal-Mohinder-2010}
Grewal, M.~S. and Andrews, A.~P. (2010).
\newblock {Applications of Kalman filtering in aerospace 1960 to the present}.
\newblock {\em IEEE Control Systems Magazine}, 30:69--78.

\bibitem[Harvey, 1990]{Harvey-1990}
Harvey, A.~C. (1990).
\newblock {\em Forecasting, structural time series models and the Kalman
  filter}.
\newblock Cambridge university press, Cambridge, UK.

\bibitem[Hobbs et~al., 2015]{Hobbs-etal-2015}
Hobbs, N.~T., Geremia, C., Treanor, J., Wallen, R., White, P.~J., Hooten,
  M.~B., and Rhyan, J.~C. (2015).
\newblock {State-space modeling to support management of brucellosis in the
  Yellowstone bison population}.
\newblock {\em Ecological Monographs}, 85:525--556.

\bibitem[Holdo et~al., 2009]{Holdo-etal-2009}
Holdo, R.~M., Sinclair, A. R.~E., Dobson, A.~P., Metzger, K.~L., Bolker, B.~M.,
  Ritchie, M.~E., and Holt, R.~D. (2009).
\newblock {A disease-mediated trophic cascade in the Serengeti and its
  implications for ecosystem C}.
\newblock {\em PLoS Biology}, 7:e1000210.

\bibitem[Holmes et~al., 2018]{MARSS}
Holmes, E., Ward, E., Scheuerell, M., and Wills, K. (2018).
\newblock {\em MARSS: Multivariate Autoregressive State-Space Modeling}.
\newblock R package version 3.10.10.

\bibitem[Holmes et~al., 2012]{Holmes-etal-2012}
Holmes, E.~E., Ward, E.~J., and Wills, K. (2012).
\newblock {MARSS}: Multivariate autoregressive state-space models for analyzing
  time-series data.
\newblock {\em The R Journal}, 4:11--19.

\bibitem[Hooten and Hefley, 2019]{Hooten-Hefley-2019}
Hooten, M.~B. and Hefley, T.~J. (2019).
\newblock {\em Bringing Bayesian models to life}.
\newblock CRC Press, Boca Raton, FL.

\bibitem[Hooten and Hobbs, 2015]{Hooten-Hobbs-2015}
Hooten, M.~B. and Hobbs, N.~T. (2015).
\newblock A guide to bayesian model selection for ecologists.
\newblock {\em Ecological Monographs}, 85:3--28.

\bibitem[Hooten et~al., 2009]{Hooten-etal-2009}
Hooten, M.~B., Wikle, C.~K., Sheriff, S.~L., and Rushin, J.~W. (2009).
\newblock {Optimal spatio-temporal hybrid sampling designs for ecological
  monitoring}.
\newblock {\em Journal of Vegetation Science}, 20:639--649.

\bibitem[Hyndman and Athanasopoulos, 2018]{Hyndman-Athanasopoulos-2018}
Hyndman, R.~J. and Athanasopoulos, G. (2018).
\newblock {\em Forecasting: principles and practice, 2nd Ed.}
\newblock OTexts, Melbourne, Australia. OTexts.com/fpp2. Accessed on January
  14, 2021.

\bibitem[Ionides et~al., 2015]{Ionides-etal-2015}
Ionides, E.~L., Nguyen, D., Atchad{\'{e}}, Y., Stoev, S., and King, A.~A.
  (2015).
\newblock {Inference for dynamic and latent variable models via iterated,
  perturbed Bayes maps}.
\newblock {\em Proceedings of the National Academy of Sciences}, 112:719--724.

\bibitem[Ives et~al., 2003]{Ives-etal-2003}
Ives, A.~R., Dennis, B., Cottingham, K.~L., and Carpenter, S.~R. (2003).
\newblock {Estimating community stability and ecological interactions from
  time-series data}.
\newblock {\em Ecological Monographs}, 73:301--330.

\bibitem[Jamieson and Brooks, 2004]{Jamieson-Brooks-2004}
Jamieson, L.~E. and Brooks, S.~P. (2004).
\newblock {Density dependence in North American ducks}.
\newblock {\em Animal Biodiversity and Conservation}, 27.1:113--128.

\bibitem[Jia et~al., 2011]{Jia-etal-2011}
Jia, X., Shao, M., Wei, X., Horton, R., and Li, X. (2011).
\newblock {Estimating total net primary productivity of managed grasslands by a
  state-space modeling approach in a small catchment on the Loess Plateau,
  China}.
\newblock {\em Geoderma}, 160:281--291.

\bibitem[Johnson et~al., 2016]{Johnson-etal-2016}
Johnson, D.~S., Laake, J.~L., Melin, S.~R., Delong, R.~L., Johnson, D.~S.,
  Laake, J.~L., Melin, S.~R., and Delong, R.~L. (2016).
\newblock {Multivariate state hidden Markov models for Mark-Recapture Data}.
\newblock {\em Statistical Science}, 31:233--244.

\bibitem[Johnson and London, 2018]{crawl}
Johnson, D.~S. and London, J.~M. (2018).
\newblock crawl: an r package for fitting continuous-time correlated random
  walk models to animal movement data.

\bibitem[Johnson et~al., 2008]{Johnson-etal-2008}
Johnson, D.~S., London, J.~M., Lea, M.-A., and Durban, J.~W. (2008).
\newblock Continuous-time correlated random walk model for animal telemetry
  data.
\newblock {\em Ecology}, 89:1208--1215.

\bibitem[Jonsen, 2016]{Jonsen-2016}
Jonsen, I. (2016).
\newblock {Joint estimation over multiple individuals improves behavioural
  state inference from animal movement data}.
\newblock {\em Scientific Reports}, 6:20625.

\bibitem[Jonsen et~al., 2013]{Jonsen-etal-2013}
Jonsen, I., Basson, M., Bestley, S., Bravington, M., Patterson, T., Pedersen,
  M., Thomson, R., Thygesen, U., and Wotherspoon, S. (2013).
\newblock {State-space models for bio-loggers: a methodological road map}.
\newblock {\em Deep Sea Research Part II: Topical Studies in Oceanography},
  88-89:34--46.

\bibitem[Jonsen et~al., 2005]{Jonsen-etal-2005}
Jonsen, I.~D., {Mills Flemming}, J., and Myers, R.~A. (2005).
\newblock Robust state-space modeling of animal movement data.
\newblock {\em Ecology}, 86:2874--2880.

\bibitem[Kai and Yokoi, 2019]{Kai-Yokoi-2019}
Kai, M. and Yokoi, H. (2019).
\newblock {Performance evaluation of information criteria for estimating a
  shape parameter in a Bayesian state-space biomass dynamics model}.
\newblock {\em Fisheries Research}, 219:105326.

\bibitem[Kalman, 1960]{Kalman-1960}
Kalman, R.~E. (1960).
\newblock A new approach to linear filtering and prediction problems.
\newblock {\em Journal of basic Engineering}, 82:35--45.

\bibitem[Kalman and Bucy, 1961]{Kalman-and-Bucy-1961}
Kalman, R.~E. and Bucy, R.~S. (1961).
\newblock New results in linear filtering and prediction theory.
\newblock {\em Journal of basic engineering}, 83:95--108.

\bibitem[Karban and de~Valpine, 2010]{Karban-and-deValpine-2010}
Karban, R. and de~Valpine, P. (2010).
\newblock {Population dynamics of an Arctiid caterpillar-tachinid parasitoid
  system using state-space models}.
\newblock {\em Journal of Animal Ecology}, 79:650--661.

\bibitem[Kass and Wasserman, 1996]{Kass-Wasserman-1996}
Kass, R.~E. and Wasserman, L. (1996).
\newblock {The selection of prior distributions by formal rules stable}.
\newblock {\em Journal of the American Statistical Association}, 91:1343--1370.

\bibitem[K\'ery et~al., 2009]{Kery-etal-2009}
K\'ery, M., Royle, J.~A., Plattner, M., and Dorazio, R.~M. (2009).
\newblock {Species richness and occupancy estimation in communities subject to
  temporary emigration}.
\newblock {\em Ecology}, 90:1279--1290.

\bibitem[Kindsvater et~al., 2018]{Kindsvater-etal-2018}
Kindsvater, H.~K., Dulvy, N.~K., Horswill, C., Juan-Jord{\`{a}}, M.-J., Mangel,
  M., and Matthiopoulos, J. (2018).
\newblock {Overcoming the data crisis in biodiversity conservation}.
\newblock {\em Trends in Ecology \& Evolution}, 33:676--688.

\bibitem[King et~al., 2015]{King-etal-2015}
King, A.~A., Domenech~de Cell{\`e}s, M., Magpantay, F. M.~G., and Rohani, P.
  (2015).
\newblock Avoidable errors in the modelling of outbreaks of emerging pathogens,
  with special reference to {Ebola}.
\newblock {\em Proceedings of the Royal Society of London, Series B},
  282:20150347.

\bibitem[King et~al., 2016]{King-etal-2016}
King, A.~A., Nguyen, D., and Ionides, E.~L. (2016).
\newblock Statistical inference for partially observed {Markov} processes via
  the {R} package {pomp}.
\newblock {\em Journal of Statistical Software}, 69:1--43.

\bibitem[King, 2012]{King-2012}
King, R. (2012).
\newblock {A review of Bayesian state-space modelling of capture – recapture
  – recovery data}.
\newblock {\em Interface focus}, 2:190--204.

\bibitem[Kitagawa, 1987]{Kitagawa-1987}
Kitagawa, G. (1987).
\newblock {Non-Gaussian state-space modeling of nonstationary time series}.
\newblock {\em Journal of the American Statistical Association}, 82:1032--1041.

\bibitem[Knape, 2008]{Knape-2008}
Knape, J. (2008).
\newblock {Estimability of density dependence in models of time series data}.
\newblock {\em Ecology}, 89:2994--3000.

\bibitem[Knape and de~Valpine, 2012]{Knape-deValpine-2012}
Knape, J. and de~Valpine, P. (2012).
\newblock {Fitting complex population models by combining particle filters with
  Markov chain Monte Carlo}.
\newblock {\em Ecology}, 93:256--263.

\bibitem[Knape et~al., 2011]{Knape-etal-2011}
Knape, J., Jonz{\'{e}}n, N., and Sk{\"{o}}ld, M. (2011).
\newblock {On observation distributions for state space models of population
  survey data.}
\newblock {\em The Journal of animal ecology}, 80:1269--1277.

\bibitem[Kristensen et~al., 2016]{Kristensen-etal-2016}
Kristensen, K., Nielsen, A., Berg, C.~W., Skaug, H., and Bell, B. (2016).
\newblock {TMB: automatic differentiation and Laplace approximation}.
\newblock {\em Journal of Statistical Software}, 70:1--21.

\bibitem[Langrock et~al., 2012]{Langrock-etal-2012}
Langrock, R., King, R., Matthiopoulos, J., Thomas, L., Fortin, D., and Morales,
  J.~M. (2012).
\newblock {Flexible and practical modeling of animal telemetry data: hidden
  Markov models and extensions}.
\newblock {\em Ecology}, 93:2336--2342.

\bibitem[Lele, 2020]{Lele-2020}
Lele, S.~R. (2020).
\newblock {Consequences of lack of parameterization invariance of
  non-informative Bayesian analysis for wildlife management: survival of San
  Joaquin kit fox and declines in amphibian populations}.
\newblock {\em Frontiers in Ecology and Evolution}, 7:501.

\bibitem[Lele et~al., 2010]{Leleetal2010}
Lele, S.~R., Nadeem, K., and Schmuland, B. (2010).
\newblock Estimability and likelihood inference for generalized linear mixed
  models using data cloning.
\newblock {\em Journal of the American Statistical Association}, 10:1617--1625.

\bibitem[Lemoine, 2019]{Lemoine-2019}
Lemoine, N.~P. (2019).
\newblock {Moving beyond noninformative priors: why and how to choose weakly
  informative priors in Bayesian analyses}.
\newblock {\em Oikos}, 128:912--928.

\bibitem[Leos-Barajas and Michelot, 2018]{Leos-barajas-Michelot-2018}
Leos-Barajas, V. and Michelot, T. (2018).
\newblock {An Introduction to Animal Movement Modeling with Hidden Markov
  Models using Stan for Bayesian Inference}.
\newblock {\em arXiv e-prints}, page arXiv:1806.10639.

\bibitem[Lind{\'{e}}n and Knape, 2009]{Linden-Knape-2009}
Lind{\'{e}}n, A. and Knape, J. (2009).
\newblock {Estimating environmental effects on population dynamics:
  consequences of observation error}.
\newblock {\em Oikos}, 118:675--680.

\bibitem[Link and Sauer, 2016]{Link-Sauer-2016}
Link, W.~A. and Sauer, J.~R. (2016).
\newblock {Bayesian cross-validation for model evaluation and selection, with
  application to the North American Breeding Bird Survey}.
\newblock {\em Ecology}, 97:1746--1758.

\bibitem[Link et~al., 2017]{Link-etal-2017}
Link, W.~A., Sauer, J.~R., and Niven, D.~K. (2017).
\newblock {Model selection for the North American Breeding Bird Survey: a
  comparison of methods}.
\newblock {\em The Condor}, 119:546--556.

\bibitem[Liu and Chen, 1998]{Liu-Chen-1998}
Liu, J.~S. and Chen, R. (1998).
\newblock Sequential monte carlo methods for dynamic systems.
\newblock {\em Journal of the American statistical association}, 93:1032--1044.

\bibitem[Louca and Doebeli, 2015]{Louca-Doebeli-2015}
Louca, S. and Doebeli, M. (2015).
\newblock {Detecting cyclicity in ecological time series}.
\newblock {\em Ecology}, 96:1724--1732.

\bibitem[Lunn et~al., 2013]{Lunn_et_al:2013}
Lunn, D., Jackson, C., Best, N., Thomas, A., and Spiegelhalter, D. (2013).
\newblock {\em The {BUGS} book}.
\newblock Chapman \& Hall/CRC, Boca Raton, FL.

\bibitem[Lunn et~al., 2009]{Lunn-etal-2009}
Lunn, D., Spiegelhalter, D., Thomas, A., and Best, N. (2009).
\newblock The {BUGS} project: evolution, critique and future directions.
\newblock {\em Statistics in medicine}, 28:3049--3067.

\bibitem[Lunn et~al., 2000]{Lunn-et-al-2000}
Lunn, D.~J., Thomas, A., Best, N., and Spiegelhalter, D. (2000).
\newblock {WinBUGS - A Bayesian modelling framework: concepts, structure, and
  extensibility}.
\newblock {\em Statistics and Computing}, 10:325--337.

\bibitem[Maunder and Deriso, 2011]{Maunder-Deriso-2011}
Maunder, M.~N. and Deriso, R.~B. (2011).
\newblock { A state–space multistage life cycle model to evaluate population
  impacts in the presence of density dependence: illustrated with application
  to delta smelt (\textit{Hyposmesus transpacificus})}.
\newblock {\em Canadian Journal of Fisheries and Aquatic Sciences},
  68:1285--1306.

\bibitem[McClintock et~al., 2014]{McClintock-etal-2014}
McClintock, B.~T., Johnson, D.~S., Hooten, M.~B., {Ver Hoef}, J.~M., and
  Morales, J.~M. (2014).
\newblock {When to be discrete: the importance of time formulation in
  understanding animal movement}.
\newblock {\em Movement Ecology}, 2:21.

\bibitem[McClintock et~al., 2012]{McClintock-etal-2012}
McClintock, B.~T., King, R., Thomas, L., Matthiopoulos, J., McConnell, B.~J.,
  and Morales, J.~M. (2012).
\newblock {A general discrete-time modeling framework for animal movement using
  multistate random walks}.
\newblock {\em Ecological Monographs}, 82:335--349.

\bibitem[McClintock et~al., 2020]{McClintock-etal-2020}
McClintock, B.~T., Langrock, R., Gimenez, O., Cam, E., Borchers, D.~L.,
  Glennie, R., and Patterson, T.~A. (2020).
\newblock Uncovering ecological state dynamics with hidden {M}arkov models.
\newblock {\em Ecology Letters}, 23:1878--1903.

\bibitem[McClintock et~al., 2017]{Mcclintock-etal-2017}
McClintock, B.~T., London, J.~M., Cameron, M.~F., and Boveng, P.~L. (2017).
\newblock {Bridging the gaps in animal movement: hidden behaviors and
  ecological relationships revealed by integrated data streams}.
\newblock {\em Ecosphere}, 8:e01751.

\bibitem[McClintock and Michelot, 2018]{McClintock-Michelot-2018}
McClintock, B.~T. and Michelot, T. (2018).
\newblock momentu{HMM}: {R} package for generalized hidden {M}arkov models of
  animal movement.
\newblock {\em Methods in Ecology and Evolution}, 9:1518--1530.

\bibitem[Mendelssohn, 1988]{Mendelssohn-1988}
Mendelssohn, R. (1988).
\newblock Some problems in estimating population sizes from catch-at-age data.
\newblock {\em Fishery Bulletin}, 86:617--630.

\bibitem[Merkle et~al., 2019]{Merkle-etal-2019}
Merkle, E.~C., Furr, D., and Rabe-Hesketh, S. (2019).
\newblock {Bayesian Comparison of latent variable models: conditional versus
  marginal likelihoods}.
\newblock {\em Psychometrika}, 84:802--829.

\bibitem[Meyer and Millar, 1999]{Meyer-Millar-1999}
Meyer, R. and Millar, R.~B. (1999).
\newblock Bugs in bayesian stock assessments.
\newblock {\em Canadian Journal of Fisheries and Aquatic Sciences},
  56:1078--1087.

\bibitem[Michaud et~al., 2020]{Michaud-etal-2020}
Michaud, N., de~Valpine, P., Turek, D., Paciorek, C.~J., and Nguyen, D. (2020).
\newblock {Sequential Monte Carlo methods in the nimble R package}.
\newblock {\em arXiv e-prints}, page arXiv:1703.06206.

\bibitem[Michielsens et~al., 2006]{Michielsens-etal-2006}
Michielsens, C. G.~J., McAllister, M.~K., Kuikka, S., Pakarinen, T., Karlsson,
  L., Romakkaniemi, A., Per{\"{a}}, I., and M{\"{a}}ntyniemi, S. (2006).
\newblock {A Bayesian state-space mark-recapture model to estimate exploitation
  rates in mixed-stock fisheries}.
\newblock {\em Canadian Journal of Fisheries and Aquatic Sciences},
  63:321--334.

\bibitem[Millar, 2002]{Millar-2002}
Millar, R.~B. (2002).
\newblock {Reference priors for Bayesian fisheries models}.
\newblock {\em Canadian Journal of Fisheries and Aquatic Sciences},
  59:1492--1502.

\bibitem[Millar, 2009]{Millar-2009}
Millar, R.~B. (2009).
\newblock {Comparison of hierarchical Bayesian models for overdispersed count
  data using DIC and Bayes' factors}.
\newblock {\em Biometrics}, 65:962--969.

\bibitem[Millar, 2018]{Millar-2018}
Millar, R.~B. (2018).
\newblock {Conditional vs marginal estimation of the predictive loss of
  hierarchical models using WAIC and cross-validation}.
\newblock {\em Statistics and Computing}, 28:375--385.

\bibitem[{Mills Flemming} et~al., 2010]{MillsFlemming-etal-2010}
{Mills Flemming}, J., Jonsen, I.~D., Myers, R.~A., and Field, C.~A. (2010).
\newblock {Hierarchical state-space estimation of leatherback turtle navigation
  ability}.
\newblock {\em PLoS ONE}, 5:e14245.

\bibitem[Monnahan et~al., 2017]{Monnahan-etal-2017}
Monnahan, C.~C., Thorson, J.~T., and Branch, T.~A. (2017).
\newblock {Faster estimation of Bayesian models in ecology using Hamiltonian
  Monte Carlo}.
\newblock {\em Methods in Ecology and Evolution}, 8:339--348.

\bibitem[Moore and Barlow, 2011]{Moore-and-Barlow-2011}
Moore, J.~E. and Barlow, J. (2011).
\newblock {Bayesian state-space model of fin whale abundance trends from a
  1991-2008 time series of line-transect surveys in the California Current}.
\newblock {\em Journal of Applied Ecology}, 48:1195--1205.

\bibitem[Mordecai et~al., 2011]{Mordecai-etal-2011}
Mordecai, R.~S., Mattsson, B.~J., Tzilkowski, C.~J., and Cooper, R.~J. (2011).
\newblock {Addressing challenges when studying mobile or episodic species:
  hierarchical Bayes estimation of occupancy and use}.
\newblock {\em Journal of Applied Ecology}, 48:56--66.

\bibitem[M{\"{u}}ller et~al., 2013]{Muller-etal-2013}
M{\"{u}}ller, S., Scealy, J.~L., and Welsh, A.~H. (2013).
\newblock {Model selection in linear mixed models}.
\newblock {\em Statistical Science}, 28:135--167.

\bibitem[Neal, 2011]{Neal:2011}
Neal, R. (2011).
\newblock {MCMC} using {H}amiltonian dynamics.
\newblock In Brooks, S., Gelman, A., Jones, G., and Meng, X., editors, {\em
  Handbook of {M}arkov {C}hain {M}onte {C}arlo}. Chapman \& Hall/CRC, Boca
  Raton, FL.

\bibitem[Newman et~al., 2014]{Newman-etal-2014}
Newman, K.~B., Buckland, S.~T., Morgan, B. J.~T., King, R., Borchers, D.~L.,
  Cole, D.~J., Besbeas, P., Gimenez, O., and Thomas, L. (2014).
\newblock {\em Modelling population dynamics: model formulation, fitting and
  assessment using state-space methods}.
\newblock Springer, New York, NY.

\bibitem[Nielsen and Berg, 2014]{Nielsen-Berg-2014}
Nielsen, A. and Berg, C.~W. (2014).
\newblock {Estimation of time-varying selectivity in stock assessments using
  state-space models}.
\newblock {\em Fisheries Research}, 158:96--101.

\bibitem[O'Hara et~al., 2009]{OHara-Sillanpaa-2009}
O'Hara, R.~B., Sillanp{\"a}{\"a}, M.~J., et~al. (2009).
\newblock A review of bayesian variable selection methods: what, how and which.
\newblock {\em Bayesian analysis}, 4:85--117.

\bibitem[Ong et~al., 2018]{Ong-etal-2018}
Ong, V. M.~H., Nott, D.~J., Tran, M.-N., Sisson, S.~A., and Drovandi, C.~C.
  (2018).
\newblock {Variational Bayes with synthetic likelihood}.
\newblock {\em Statistics and Computing}, 28:971--988.

\bibitem[Osada et~al., 2019]{Osada-etal-2019}
Osada, Y., Kuriyama, T., Asada, M., Yokomizo, H., and Miyashita, T. (2019).
\newblock {Estimating range expansion of wildlife in heterogeneous landscapes:
  a spatially explicit state-space matrix model coupled with an improved
  numerical integration technique}.
\newblock {\em Ecology and Evolution}, 9:318--327.

\bibitem[Patterson et~al., 2008]{Patterson-etal-2008}
Patterson, T.~A., Thomas, L., Wilcox, C., Ovaskainen, O., and Matthiopoulos, J.
  (2008).
\newblock {State-space models of individual animal movement}.
\newblock {\em Trends in Ecology and Evolution}, 23:87--94.

\bibitem[Peacock et~al., 2017]{Peacock-etal-2016}
Peacock, S.~J., Krko{\v{s}}ek, M., Lewis, M.~A., and Lele, S. (2017).
\newblock Study design and parameter estimability for spatial and temporal
  ecological models.
\newblock {\em Ecology and Evolution}, 7:762--770.

\bibitem[Pedersen et~al., 2011]{Pedersen-etal-2011}
Pedersen, M.~W., Berg, C.~W., Thygesen, U.~H., Nielsen, A., and Madsen, H.
  (2011).
\newblock {Estimation methods for nonlinear state-space models in ecology}.
\newblock {\em Ecological Modelling}, 222:1394--1400.

\bibitem[Petris, 2010]{petris2010}
Petris, G. (2010).
\newblock An {R} package for {D}ynamic {L}inear {M}odels.
\newblock {\em Journal of Statistical Software}, 36:1--16.

\bibitem[Petris et~al., 2009]{DLMwR}
Petris, G., Petrone, S., and Campagnoli, P. (2009).
\newblock {\em Dynamic Linear Models with {R}}.
\newblock Springer.

\bibitem[Pitt and Shephard, 1999]{Pitt-Shephard-1999}
Pitt, M.~K. and Shephard, N. (1999).
\newblock Filtering via simulation: auxiliary particle filters.
\newblock {\em Journal of the American Statistical Association}, 94:590--599.

\bibitem[Plummer, 2003]{Plummer:2003}
Plummer, M. (2003).
\newblock {JAGS}: a program for analysis of {B}ayesian graphical models using
  {G}ibbs sampling.
\newblock {\em Proceedings of the 3rd international workshop on distributed
  statistical computing}, 124:1--10.

\bibitem[Plummer, 2018]{rjags}
Plummer, M. (2018).
\newblock {\em rjags: Bayesian Graphical Models using MCMC}.
\newblock R package version 4-8.

\bibitem[Plummer et~al., 2006]{coda}
Plummer, M., Best, N., Cowles, K., and Vines, K. (2006).
\newblock Coda: convergence diagnosis and output analysis for {MCMC}.
\newblock {\em R News}, 6:7--11.

\bibitem[Pohle et~al., 2017]{Pohle-etal-2017}
Pohle, J., Langrock, R., van Beest, F.~M., and Schmidt, N.~M. (2017).
\newblock {Selecting the number of states in hidden Markov models: pragmatic
  solutions illustrated using animal movement}.
\newblock {\em Journal of Agricultural, Biological, and Environmental
  Statistics}, 22:270--293.

\bibitem[Polansky et~al., 2019]{Polansky-etal-2019}
Polansky, L., Newman, K.~B., and Mitchell, L. (2019).
\newblock Improving inference for nonlinear state-space models of animal
  population dynamics given biased sequential life stage data.
\newblock {\em arXiv e-prints}, page arXiv:1909.09111.

\bibitem[Postlethwaite and Dennis, 2013]{Postlethwaite-Dennis-2013}
Postlethwaite, C.~M. and Dennis, T.~E. (2013).
\newblock {Effects of temporal resolution on an inferential model of animal
  movement}.
\newblock {\em PLoS ONE}, 8:e57640.

\bibitem[Prado and West, 2010]{Prado-West-2010}
Prado, R. and West, M. (2010).
\newblock {\em Time series: modeling, computation, and inference}.
\newblock Chapman \& Hall/CRC, Boca Raton, FL.

\bibitem[{R Core Team}, 2019]{R}
{R Core Team} (2019).
\newblock {\em R: A Language and Environment for Statistical Computing}.
\newblock R Foundation for Statistical Computing, Vienna, Austria.

\bibitem[Raue et~al., 2009]{Raue-etal-2009}
Raue, A., Kreutz, C., Maiwald, T., Bachmann, J., Schilling, M.,
  {Klingm\"uller}, U., and Timmer, J. (2009).
\newblock Structural and practical identifiability analysis of partially
  observed dynamical models by exploiting the profile likelihood.
\newblock {\em Bioinformatics}, 25:1923--1929.

\bibitem[Robert, 2007]{Robert-2007}
Robert, C.~P. (2007).
\newblock {\em The Bayesian choice: from decision theoretic foundations to
  computation implementation}.
\newblock Springer Sciences + Business Media, LLC, New York, NY, 2nd edition.

\bibitem[Roberts et~al., 2017]{Roberts-etal-2017}
Roberts, D.~R., Bahn, V., Ciuti, S., Boyce, M.~S., Elith, J., Guillera-arroita,
  G., Hauenstein, S., Lahoz-monfort, J.~J., Schr{\"{o}}der, B., Thuiller, W.,
  Warton, D.~I., Wintle, B.~A., Hartig, F., and Dormann, C.~F. (2017).
\newblock {Cross-validation strategies for data with temporal, spatial,
  hierarchical ,or phylogenetic structure}.
\newblock {\em Ecography}, 49:913--929.

\bibitem[Rothenberg, 1971]{Rothenberg1971}
Rothenberg, T.~J. (1971).
\newblock Identification in parametric models.
\newblock {\em Econometrica}, 39:577--591.

\bibitem[Royle, 2008]{Royle-2008}
Royle, J.~A. (2008).
\newblock {Modeling individual effects in the Cormack-Jolly-Seber model: a
  state-space formulation}.
\newblock {\em Biometrics}, 64:364--370.

\bibitem[S{\ae}ther et~al., 2008]{Saether-etal-2008}
S{\ae}ther, B.-e., Lilleg{\aa}rd, M., Gr{\o}tan, V., Drever, M.~C., Engen, S.,
  Nudds, T.~D., and Podruzny, K.~M. (2008).
\newblock {Geographical gradients in the population dynamics of North American
  prairie ducks}.
\newblock {\em Journal of Animal Ecology}, 77:869--882.

\bibitem[Sanderlin et~al., 2019]{Sanderlin-etal-2019}
Sanderlin, J.~S., Block, W.~M., Strohmeyer, B.~E., Saab, V.~A., and Ganey,
  J.~L. (2019).
\newblock {Precision gain versus effort with joint models using
  detection/non-detection and banding data}.
\newblock {\em Ecology and Evolution}, 9:804--817.

\bibitem[Schick et~al., 2013]{Schick-etal-2013}
Schick, R.~S., Kraus, S.~D., Rolland, R.~M., Knowlton, A.~R., Hamilton, P.~K.,
  Pettis, H.~M., Kenney, R.~D., and Clark, J.~S. (2013).
\newblock {Using hierarchical Bayes to understand movement, health, and
  survival in the endangered North Atlantic right whale}.
\newblock {\em PLoS ONE}, 8:e64166.

\bibitem[Shumway and Stoffer, 2016]{Shumway-Stoffer-2016}
Shumway, R.~H. and Stoffer, D.~S. (2016).
\newblock {\em Time series analysis and its applications: with R examples.}
\newblock Springer International Publishing, New York, NY, 4th edition.

\bibitem[Siple and Francis, 2016]{Siple-Francis-2016}
Siple, M.~C. and Francis, T.~B. (2016).
\newblock {Population diversity in Pacific herring of the Puget Sound, USA}.
\newblock {\em Oecologia}, 180:111--125.

\bibitem[Smith, 1985]{Smith-1985}
Smith, J. (1985).
\newblock Diagnostic checks of non-standard time series models.
\newblock {\em Journal of Forecasting}, 4:283--291.

\bibitem[{Stan Development Team}, 2012]{STAN:2012}
{Stan Development Team} (2012).
\newblock {\em Stan: A {C++} library for probability and sampling}.
\newblock http://mc-stan.org.

\bibitem[{Stan Development Team}, 2018]{rstan:2018}
{Stan Development Team} (2018).
\newblock {RStan}: the {R} interface to {Stan}.
\newblock R package version 2.18.2.

\bibitem[Sullivan, 1992]{Sullivan-1992}
Sullivan, P.~J. (1992).
\newblock A {K}alman filter approach to catch-at-length analysis.
\newblock {\em Biometrics}, 48:237--257.

\bibitem[Tashman, 2000]{Tashman-2000}
Tashman, L.~J. (2000).
\newblock {Out-of-sample tests of forecasting accuracy: an analysis and
  review}.
\newblock {\em International Journal of Forecasting}, 16:437--450.

\bibitem[Thorson et~al., 2016]{Thorson-etal-2016}
Thorson, J.~T., Ianelli, J.~N., Larsen, E.~A., Ries, L., Scheuerell, M.~D.,
  Szuwalski, C., and Zipkin, E.~F. (2016).
\newblock {Joint dynamic species distribution models: a tool for community
  ordination and spatio-temporal monitoring}.
\newblock {\em Global Ecology and Biogeography}, 25:1144--1158.

\bibitem[Thygesen et~al., 2017]{Thygesen-etal-2017}
Thygesen, U.~H., Albertsen, C.~M., Berg, C.~W., Kasper, K., and Nielsen, A.
  (2017).
\newblock {Validation of ecological state space models using the Laplace
  approximation}.
\newblock {\em Environmental and Ecological Statistics}, 24:317--339.

\bibitem[Tom\'e et~al., 2020]{Tome-etal-2020}
Tom\'e, C.~P., Smith, E. A.~E., Lyons, S.~K., Newsome, S.~D., and Smith, F.~A.
  (2020).
\newblock {Changes in the diet and body size of a small herbivorous mammal
  (hispid cotton rat, \textit{Sigmodon hispidus}) following the late
  Pleistocene megafauna extinction}.
\newblock {\em Ecography}, 43:604--619.

\bibitem[Vaida and Blanchard, 2005]{Vaida-Blanchard-2005}
Vaida, F. and Blanchard, S. (2005).
\newblock {Conditional Akaike information for mixed-effects models}.
\newblock {\em Biometrika}, 92:351--370.

\bibitem[VanDerwerken and Schmidler, 2017]{VanDerwerken-Schmidler-2017}
VanDerwerken, D. and Schmidler, S.~C. (2017).
\newblock {Monitoring joint convergence of MCMC samplers}.
\newblock {\em Journal of Computational and Graphical Statistics}, 26:558--568.

\bibitem[{Vats} and {Knudson}, 2018]{Vats-Knudson-2018}
{Vats}, D. and {Knudson}, C. (2018).
\newblock {Revisiting the Gelman-Rubin diagnostic}.
\newblock {\em arXiv e-prints}, page arXiv:1812.09384.

\bibitem[Vehtari et~al., 2017]{Vehtari-etal-2017}
Vehtari, A., Gelman, A., and Gabry, J. (2017).
\newblock {Practical Bayesian model evaluation using leave-one-out
  cross-validation and WAIC}.
\newblock {\em Statistics and Computing}, 27:1413--1432.

\bibitem[Viallefont et~al., 1998]{Viallefontetal1998}
Viallefont, A., Lebreton, J.-D., and Reboulet, A.-M. (1998).
\newblock Parameter identifiability and model selection in capture-recapture
  models: A numerical approach.
\newblock {\em Biometrical Journal}, 40:313--325.

\bibitem[Viljugrein et~al., 2005]{viljugrein2005}
Viljugrein, H., Stenseth, N.~C., Smith, G.~W., and Steinbakk, G.~H. (2005).
\newblock {Density dependence in North American ducks}.
\newblock {\em Ecology}, 86:245--254.

\bibitem[Ward et~al., 2010]{Ward-etal-2010}
Ward, E.~J., Chirakkal, H., Gonz{\'a}lez-Su{\'a}rez, M., Aurioles-Gamboa, D.,
  Holmes, E.~E., and Gerber, L. (2010).
\newblock {Inferring spatial structure from time-series data: using
  multivariate state-space models to detect metapopulation structure of
  California sea lions in the Gulf of California, Mexico}.
\newblock {\em Journal of Applied Ecology}, 47:47--56.

\bibitem[Warton et~al., 2017]{Warton-etal-2017}
Warton, D.~I., Thibault, L., and Wang, Y.~A. (2017).
\newblock {The PIT-trap - a ``model-free" bootstrap procedure for inference
  about regression models with discrete, multivariate responses}.
\newblock {\em PLoS ONE}, 12:e0181790.

\bibitem[White et~al., 2016]{White-etal-2016}
White, J.~W., Nickols, J.~K., Malone, D., Carr, M.~H., Starr, R.~M.,
  Cordoleani, F., Baskett, M.~L., Hasting, A., and Botsford, L.~W. (2016).
\newblock {Fitting state-space integral projection models to size-structured
  time series data to estimate unknown parameters}.
\newblock {\em Ecological Applications}, 26:2677--2694.

\bibitem[Wilberg and Bence, 2008]{Wilberg-Bence-2008}
Wilberg, M.~J. and Bence, J.~R. (2008).
\newblock {Performance of deviance information criterion model selection in
  statistical catch-at-age analysis}.
\newblock {\em Fisheries Research}, 93:212--221.

\bibitem[Winship et~al., 2012]{Winship-eta-2012}
Winship, A.~J., Jorgensen, S.~J., Shaffer, S.~A., Jonsen, I.~D., Robinson,
  P.~W., Costa, D.~P., and Block, B.~A. (2012).
\newblock {State-space framework for estimating measurement error from
  double-tagging telemetry experiments}.
\newblock {\em Methods in Ecology and Evolution}, 3:291--302.

\bibitem[Wintle et~al., 2003]{Wintle-etal-2003}
Wintle, B.~A., McCarthy, M.~A., Volinsky, C.~T., and Kavanagh, R.~P. (2003).
\newblock {The use of Bayesian model averaging to better represent uncertainty
  in ecological models}.
\newblock {\em Conservation Biology}, 17:1579--1590.

\bibitem[Wolkovich et~al., 2014]{Wolkovich-etal-2014}
Wolkovich, E.~M., Cook, B.~I., McLauchlan, K.~K., and Davies, T.~J. (2014).
\newblock {Temporal ecology in the Anthropocene}.
\newblock {\em Ecology Letters}, 17:1365--1379.

\bibitem[Yackulic et~al., 2020]{Yackulic-etal-2020}
Yackulic, C.~B., Dodrill, M., Dzul, M., Sanderlin, J.~S., and Reid, J.~A.
  (2020).
\newblock {A need for speed in Bayesian population models: a practical guide to
  marginalizing and recovering discrete latent states}.
\newblock {\em Ecological Applications}, 30:e02112.

\bibitem[Yin et~al., 2019]{Yin-etal-2019}
Yin, Y., Aeberhard, W.~H., Smith, S.~J., and {Mills Flemming}, J. (2019).
\newblock {Identifiable state-space models: a case study of the Bay of Fundy
  sea scallop fishery}.
\newblock {\em Canadian Journal of Statistics}, 47:27--45.

\bibitem[Zucchini et~al., 2016]{Zucchini-etal-2016}
Zucchini, W., Macdonald, I.~L., and Langrock, R. (2016).
\newblock {\em {Hidden Markov} models for time series: an introduction using
  {R}}.
\newblock Taylor \& Francis, Boca Raton, FL, 2nd edition.

\end{thebibliography}

\section*{Data Availability}

Data are available from the Dryad Digital Repository (Auger-Méthé \& Derocher 2021): doi:10.5061/dryad.4qrfj6q96

\newpage

%\begin{table}
\footnotesize
\begin{longtable}{ p{3.15cm}  p{12.5cm} } 
    
    \caption{Our definitions of important terms in the context of SSMs.}\label{t.terms} \\ 
    %\begin{tabular}{ll}
    \hline
    \hline
        Term & Definition \\
        \hline
        Conditional likelihood & \multirow{4}{12.5cm}{Likelihood function of the parameters of the SSM conditional on the states. In contrast to the joint and marginal likelihoods, the function only includes the probability distribution of the observations (not that of the states) and use sufficient statistics for the state values (e.g., states are fixed to their estimated values).}\\
        $(L_{\textsc{c}})$ & \\
        & \\
        & \\
        \hline
        Data stream & \multirow{3}{12.5cm}{Distinct set of observations. The term is generally used when more than one source of data is used in a single model (e.g., a SSM that jointly models data from systematic survey and citizen science).}\\
        & \\
        & \\
        \hline
        Hidden Markov model & \multirow{4}{12.5cm}{Class of SSMs with a finite number of discrete hidden states. For example, these discrete states could be categorical variables representing the behavioral modes of an animal (e.g., foraging, resting, traveling) or binary variables representing whether the individual is alive or dead.}\\
        (HMM) & \\
        & \\
        & \\
        \hline
        Hierarchical model & \multirow{4}{12.5cm}{Class of statistical models that has multiple levels of stochasticity. They model randomness in the data and in the process. Linear mixed effects models (i.e., linear models with fixed and random effects) are a commonly used type of hierarchical models in ecology. SSMs are another type of hierarchical model.}\\
        & \\
        & \\
        & \\
        \hline
        Joint likelihood & \multirow{4}{12.5cm}{Likelihood function of both hidden states and parameters, which summarizes the complete SSM. While it is common to call this function a likelihood, this term can cause confusion because states, but not parameters, are often viewed as random variables and often cannot be jointly estimated with parameters by maximizing this function.}\\
        ($L_\textsc{j}$) & \\
        & \\
        & \\
        \hline
        Marginal likelihood & \multirow{2}{12.5cm}{Likelihood function of the parameters of the SSM, where the states have been marginalized (i.e., integrated out or summed over all possible state values).}\\
        ($L_\textsc{m}$) & \\
        \hline
        Observation equation & \multirow{3}{12.5cm}{Equation from a SSM that models how observations depend on hidden states. Synonymous, or closely related, terms used in the literature include: observation model, and measurement equation or model.}\\
        & \\
        & \\
        \hline
        Observation error & \multirow{2}{12.5cm}{Variation associated with the discrepancy between hidden states and observations. The observation error will often reflect the imprecision of the sampling methodology.}\\
        & \\
        \hline
        Process equation & \multirow{3}{12.5cm}{Equation from a SSM that models how unobserved states at a given time depend on past states. Synonymous, or closely related, terms used in the literature include: process model, state equation, state model, and transition equation.}\\
        & \\
        & \\ 
        \hline
        Process variation & \multirow{2}{12.5cm}{Variation associated with the underlying process and hidden states. In ecology, process variation often represents biological variability.}\\
        & \\
        \hline
        State & \multirow{4}{12.5cm}{Unobserved random variable that generally represents a true attribute of the system. A SSM has at least one time series of hidden states (e.g., true population size through time or whether the individual is alive or dead during each sampling period). Synonymous, or closely related, terms used in the literature include: latent state and latent variable.}\\
        & \\
        & \\
        & \\
        \hline
        State-space model & \multirow{2}{12.5cm}{Class of hierarchical models for time series that specifies the dynamic of the hidden states and their link to the observations (see Fig. \ref{fig:structure}a).}\\
        (SSM) & \\
        \hline
   % \end{tabular}

\end{longtable}
%\end{table}

\begin{table}
    \centering
        \caption{Comparison of the fitting methods discussed in Section \ref{S.Fitting}.}
    \begin{tabular}{lllll}
    \hline
    \hline
        Method & Framework & Pros & Cons & \texttt{R} package \\
        \hline
        Kalman filter \& MLE & Frequentist & \multirow{2}{2.5cm}{Efficient \& exact} & \multirow{2}{4.7cm}{Only applicable to linear Gaussian SSMs} & \texttt{dlm, MARSS} \\
        & & & & \\
        Laplace approximation & Frequentist & \multirow{2}{2.5cm}{Efficient \& flexible} & \multirow{5}{5.3cm}{States need to be approximable with a continuous unimodal distribution (e.g., no discrete states)} & \texttt{TMB} \\
        & & & & \\
        & & & & \\
        & & & & \\
        & & & & \\
         \multirow{2}{4.1cm}{Particle filter \& iterative filtering} & Frequentist & Flexible & \multirow{2}{4.7cm}{Can be slow and sensitive to starting values} & \texttt{pomp}, \texttt{nimble} \\
        & & & & \\
        MCMC-MH & Bayesian & Flexible & \multirow{2}{4.7cm}{Can be slow and sensitive to convergence problems} & \multirow{4}{1.3cm}{\texttt{rjags}, \texttt{nimble}, \texttt{R2WinBUGS},
        \texttt{BRugs}} \\
        & & & & \\
        & & & & \\
        & & & & \\
        MCMC-HMC & Bayesian & \multirow{2}{2.5cm}{Efficient \& flexible} & \multirow{3}{4.7cm}{Require continuous parameters and states or marginalization} & \texttt{rstan} \\
        & & & & \\
        & & & & \\
        Information reduction & Bayesian & \multirow{3}{2.5cm}{Flexible \& fewer estimation problems} & Can be slow and imprecise & \texttt{EasyABC}\\
          & & & & \\
          & & & & \\
           & & & & \\
        \hline
    \end{tabular}
    \label{t.methods}
\end{table}

\newpage

\clearpage
\renewcommand{\listfigurename}{Figure captions}
\listoffigures

\clearpage

\clearpage

\clearpage

\end{document}